# Neural mechanisms of predictive processing: A collaborative community experiment through the OpenScope program

Ido Aizenbud[1], Nicholas Audette[2], Ryszard Auksztulewicz[3], Krzysztof Basiński[4], André M. Bastos[5], Michael Berry[6], Andres Canales-Johnson[8,9,10], Hannah Choi[7], Claudia Clopath[11], Uri Cohen[12], Rui Ponte Costa[13], Roberto De Filippo[14], Roman Doronin[15], Séverine Durand[23], Steven P. Errington[16], Jeffrey P. Gavornik[17], Colleen J. Gillon[11], Arno Granier[18], Jordan P. Hamm[19,20], Loreen Hertäg[21], Henry Kennedy[22], Sandeep Kumar[6], Alexander Ladd[24], Hugo Ladret[15], Jérôme A. Lecoq[23, *], Alexander Maier[25], Patrick McCarthy[26], Jie Mei[27,28], Jorge Mejias[29], John Hongyu Meng[30], Fabian Mikulasch[31], Noga Mudrik[32], Farzaneh Najafi[33], Kevin Nejad[13], Hamed Nejat[25], Karim Oweiss[34], Mihai A. Petrovici[18], Viola Priesemann[31], Lucas Rudelt[31], Sarah Ruediger[35], Simone Russo[36], Alessandro Salatiello[37], Walter Senn[18], Eli Sennesh[25], Sepehr Sima[38], Cem Uran[39,40], Anna Vasilevskaya[15], Julien Vezoli[22], Martin Vinck[39,40], Xiao-Jing Wang[30], Jacob A. Westerberg[41,25], Katharina Wilmes[18], and Yihan Sophy Xiong[25]

1. Edmond and Lily Safra Center for Brain Sciences (ELSC), The Hebrew University of Jerusalem, Jerusalem, Israel
2. Department of Psychological Sciences, University of Connecticut, Storrs, CT, USA
3. Maastricht University, Maastricht, Netherlands
4. Auditory Neuroscience Laboratory, Department of Psychology, Medical University of Gdańsk, Gdańsk, Poland
5. Department of Psychology and Vanderbilt Brain Institute, Vanderbilt University, Nashville, TN, USA
6. Princeton Neuroscience Institute, Princeton University, NJ, USA
7. School of Mathematics, Georgia Institute of Technology, USA
8. CINPSI Neurocog, Facultad de Ciencias de la Salud, Universidad Catolica del Maule, Chile
9. Department of Psychology, University of Cambridge, United Kingdom
10. Neuroscience Center, Helsinki Institute of Life Science, University of Helsinki, Finland
11. Imperial College London, London, United Kingdom
12. Computational and Biological Learning Lab, University of Cambridge, Cambridge, United Kingdom
13. Department of Physiology, Anatomy & Genetics (DPAG), University of Oxford, United Kingdom
14. Humboldt-Universität zu Berlin, Berlin, Germany
15. Friedrich Miescher Institute for Biomedical Research, Basel, Switzerland
16. Newcastle University, Newcastle upon Tyne, United Kingdom
17. Boston University, Boston, MA, USA
18. Department of Physiology, University of Bern, Switzerland
19. Department of Psychiatry, New York University Grossman School of Medicine, New York, NY, USA
20. Nathan S. Kline Institute for Psychiatric Research, Orangeburg, New York, USA
21. Modeling of Cognitive Processes, Institute of Software Engineering and Theoretical Computer Science, Berlin Institute of Technology, Berlin, Germany
22. University of Lyon, Université Claude Bernard Lyon 1, INSERM, Stem Cell and Brain Research Institute U1208, Bron, France
23. Allen Institute, Neural dynamics program, Seattle, WA, USA
24. University of Washington, Seattle, WA, USA
25. Department of Psychology, Vanderbilt University, Nashville, TN, USA
26. Mathematical, Physical and Life Sciences Doctoral Training Centre, University of Oxford, United Kingdom
27. IT:U– Interdisciplinary Transformation University Austria, Austria

28. Department of Anatomy, University of Quebec in Trois-Rivières, QC, Canada
29. Swammerdam Institute for Life Sciences, University of Amsterdam, Amsterdam, Netherlands
30. Center of Neural Science, New York University, New York, NYU, USA
31. Max Planck Institute for Dynamics and Self-organization, Goettingen, Germany
32. Department of Biomedical Engineering, Johns Hopkins University, Baltimore, MD, USA
33. Georgia Institute of Technology, School of Biological Sciences, Atlanta, GA, USA
34. Departments of Electrical & Computer Engineering, Neuroscience and Neurology, University of Florida, Gainesville, FL, USA
35. University College London, London, United Kingdom
36. Wallace H Coulter Department of Biomedical Engineering, Georgia Institute of Technology and Emory University, Atlanta, GA, USA
37. University of Tübingen, Tübingen, Germany
38. Zuckerman Mind Brain Behavior Institute, Department of Neuroscience, Columbia University, New York, NY, USA
39. Ernst Strüngmann Institute (ESI) for Neuroscience, Frankfurt am Main, Germany
40. Donders Centre for Neuroscience, Department of Neurophysics, Radboud University Nijmegen, Netherlands
41. Department of Vision and Cognition, Netherlands Institute for Neuroscience, Royal Netherlands Academy of Arts and Sciences, Amsterdam, Netherlands

* Correspondence
jeromel@alleninstitute.org

# Graphical abstract

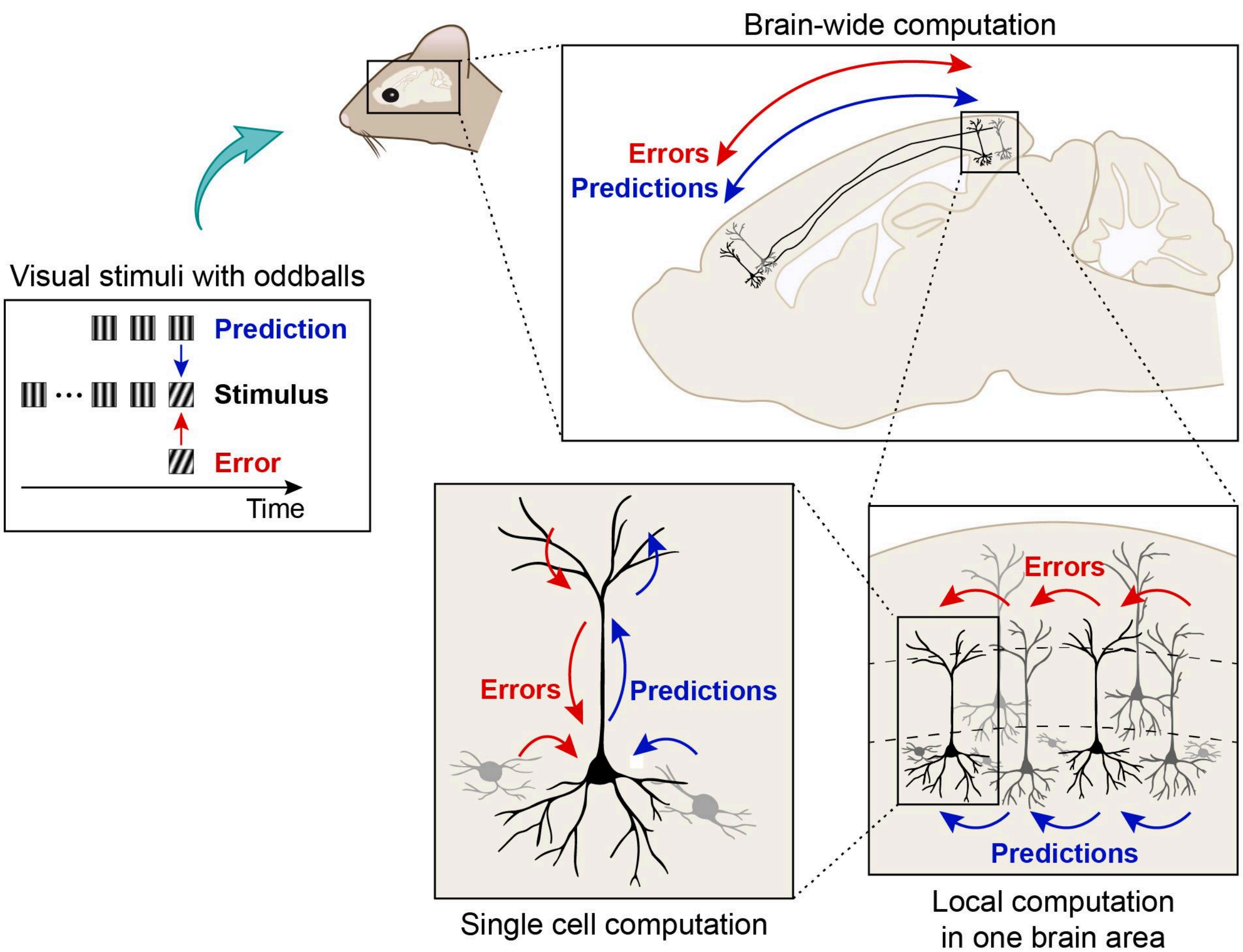

# Abstract

This review synthesizes advances in predictive processing within the sensory cortex. Predictive processing theorizes that the brain continuously predicts sensory inputs, refining neuronal responses by highlighting prediction errors. We identify key computational primitives, such as stimulus adaptation, dendritic computation, excitatory/inhibitory balance and hierarchical processing, as central to this framework. Our review highlights convergences, such as top-down inputs and inhibitory interneurons shaping mismatch signals, and divergences, including species-specific hierarchies and modality-dependent layer roles. To address these conflicts, we propose experiments in mice and primates using in-vivo two-photon imaging and electrophysiological recordings to test whether temporal, motor, and omission mismatch stimuli engage shared or distinct mechanisms. The resulting dataset, collected and shared via the OpenScope program, will enable model validation and community analysis, fostering iterative refinement and refutability to decode the neural circuits of predictive processing.

# Glossary

Predictive Processing theories frequently use colloquial terms, such as “prediction”, “belief”, or “learning” as specialized jargon with narrow definitions. It is therefore important to understand how these terms are specifically used within this context. We define the following terms, based on common definitions in the predictive processing literature, drawing heavily from past publications (Jordan and Rumelhart, 1992; Rao and Ballard, 1999). All definitions are self-contained and therefore do not directly reference the sources they are based upon.

**Computation**: The process by which neural circuits transform and integrate sensory inputs, internal states, and prior knowledge into representations and predictions. For the purpose of this review, we conceptualize sensory inputs as teaching signals (see definition below), internal states as internal models (see definition below), and prior knowledge as belief (see definition below) This can take place across several brain areas (*distributed computation*) or within a single brain area *(local computation*).

**Predictive processing:** Broad family of theories postulating that the brain uses an internal model of its environment to predict a set of ground truth inputs, e.g., incoming sensory inputs. It is important to note that the existing literature sometimes uses "predictive coding" and "predictive processing" as loose synonyms. Here, predictive processing models include predictive coding models.

**Predictive coding (dendritic and cellular predictive coding):** For this review, we will equate predictive coding with the hierarchical variant of predictive coding. In hierarchical predictive coding, information is assumed to flow across brain regions in a hierarchy (e.g., from primary sensory to higher-order sensory to associative/integrative and motor brain areas). Each level in the hierarchy receives predictions from higher-level areas and computes prediction errors by comparing such predictions with the relevant bottom-up signal to that area. The resulting prediction errors, and/or the modulated bottom-up signal, are sent to higher levels in the hierarchy. For clarity, we will introduce alternative variants of predictive coding using specific terms: "dendritic" and "cellular" variants of predictive coding refer to proposed models that differentiate where the error is computed.

**Learning:** A process by which neural networks alter their structure and function in response to experience. These changes can occur at multiple levels, from molecular changes to adjustments in network connectivity or synapses. In most predictive processing models, this process involves adjusting synaptic weights between neurons via prediction error minimization to update the internal model.

**Habituation:** Period of time during which an experimental subject is exposed to consistently repeated stimuli or constant stimulus patterns. The expected outcome is an updated internal model that predicts the continuation of the stable stimulus statistics.

**Internal representation:** The neuronal activity pattern that reflects the inferred state of the organism's environment. In predictive processing, internal representations are used to generate predictions and are continuously updated by integrating prediction error signals.

**Teaching signal:** A neuronal signal that represents the "ground truth" (real-world data) that should be matched by a corresponding prediction signal by improving the internal model through learning. In hierarchical variants of predictive processing, the teaching signal serves as the initial input to an area.

**Prediction:** We define a prediction to be the internal model's estimate of the teaching signal. More generally, predictions are neuronal signals produced by the internal model through the transformation of internal representations. During learning, predictions are modified via updates to the internal model in order to better match their target teaching signal. In hierarchical variants of predictive processing, predictions correspond to the top-down inputs neurons receive from higher levels in the hierarchy.

**Prediction error:** A signal that represents the deviation between a teaching signal and a corresponding prediction. The prediction error signal has two functions: it updates the corresponding internal representation quickly by adjusting neuronal activity), and drives slower corrective learning in the internal model by adjusting the synaptic weights). In some hierarchical variants of predictive processing, this error signal serves as bottom-up input to higher stages of processing.

**Coding space:** Abstract multidimensional space in which neural representations or signals of a particular type are organized and interpreted. This concept describes how information is stored, processed, and communicated by neural populations, in particular for specific types of neural activity, such as motor or sensory signals.

**Internal model:** An internal model reflects processes within the brain that enable it to simulate or predict aspects of the environment. The internal model maps internal representations, patterns of neural activity that encode perceptions of the environment, from the coding space of one brain area to that of another. This mapping process enables the brain to anticipate sensory inputs, guide motor actions to achieve desired outcomes, and combine sensory information to interpret the external environment. Internal models are shaped and adjusted through synaptic connections and refined by experience-dependent plasticity, enhancing accuracy in predicting and responding to changes in the environment.

**Bottom-up input.** Input carried by projections from lower to higher areas in a hierarchical processing system, e.g., input from Lateral Geniculate Nucleus (LGN) to primary visual cortex (V1), or V1 to secondary visual cortex (V2), etc.

**Top-down input:** Input carried by projections from higher to lower areas in a hierarchical processing system, e.g., input from V2 to V1, or Anterior Cingulate Cortex (ACC) to V1, etc.

**(Prediction) error neurons:** Neurons postulated to encode the magnitude of a prediction error with their firing rate. Given a sufficiently high baseline activity, positive (Teaching signal > Prediction) and negative (Teaching signal < Prediction) prediction errors can be represented within the same neurons by an increase and a decrease in activity, respectively. If neurons are constrained by low baseline firing rates, separate neural populations may be required to represent positive and negative prediction errors, thus each can only report errors unidirectionally, through an increase of activity (see next)

**Positive prediction error signals or neurons** increase their activity when the magnitude of a sensory stimulus is larger than predicted (Teaching signal > Prediction).

**Negative prediction error signals or neurons** increase their activity relative to baseline when the magnitude of a sensory stimulus is smaller than predicted (Teaching signal < Prediction).

**Mismatch stimulus**: An experimentally induced violation of a learned association between a sensory stimulus and a predictor. The predictor can be the stimulus history, another sensory stimulus, a general sensory context (e.g., spatial location), or motor output.

**Oddball or deviant stimulus:** A specific kind of mismatch stimulus that interrupts a series of repeated and frequent stimuli. An oddball or deviant stimulus is not predicted by the stimulus history.

**Belief:** Probabilistic assumption or set of assumptions about the environment, based on an internal model and current sensory input.

**Bayesian inference:** A statistical method for updating the estimated probability of a hypothesis being true as more evidence or information becomes available. Bayesian inference is based on Bayes' theorem, which states that: $P(A \mid B) = [P(B \mid A) * P(A)] / P(B)$, where $P(A \mid B)$ is the "posterior probability" (i.e., the probability of A given B), $P(B \mid A)$ is the likelihood (i.e., the probability of B given A), $P(A)$ is the "prior probability" of A (i.e., the frequency of past events, or belief thereof), and $P(B)$ is the "marginal probability" of B (i.e., the total probability of event B, considering all possible outcomes). Many computational models of predictive processing can be interpreted as performing Bayesian inference.

**Expectation:** Whole system-level estimate of a teaching signal.

**Adaptation:** Set of mechanisms by which neurons adjust their response to constant stimulation or the repetition of a single stimulus.

**Precision:** In the context of predictive processing, precision refers to the relative reliability assigned to prediction errors. Several models implement precision weighting via a multiplicative gain control mechanism, ensuring that updates to the internal model are mostly driven by high precision prediction errors.

**Attention:** The increased allocation of neural processing resources to specific sensory inputs or internal representations. Attention is thought to enable the information most relevant to current goals or tasks to be prioritized. In predictive processing, attention may enhance the precision of prediction errors, thereby influencing their impact on learning.

**Corollary discharge/Efference copy**: When sending a motor command to the periphery, motor areas also send an **efference copy** of the motor command directly to sensory areas. This **efference copy** is transformed by the internal model into **corollary discharges**. As a corollary discharge is in the coding space of the sensory area receiving the signal, it can be directly compared to the actual sensory input caused by the movement. **Corollary discharge** is synonymous with “prediction” in a motor-to-sensory pathway. Note that **efference copy** is often used synonymously with **corollary discharge**. However, we think it is more useful to distinguish the two based on whether the signal is in the motor coding space (efference copy) or the sensory coding space (corollary discharge).

**Explaining away**: The original concept of "explaining away" in causal inference describes how competing explanations for observed data are resolved. When multiple possible causes are considered, identifying the most likely one effectively "explains away" less probable causes by reducing their relevance or influence. In predictive coding, however, the term is often used with a different focus: it describes how prediction error is minimized when sensory input aligns with a prediction. Here, the signal no longer drives updates to internal models, as it has been "explained away" by a successful prediction. While both uses involve reducing uncertainty, the predictive coding approach emphasizes error minimization through matching predictions with input, whereas the original concept centers on the competition between alternative explanations.

# Acronyms

We aimed to limit our use of acronyms. However the following are commonly used in the predictive processing literature and community.

**EEG** Electroencephalography
**MEG** Magnetoencephalography
**fMRI** functional Magnetic Resonance Imaging
**ECoG** - Electrocorticography
**LFP** - Local Field Potential
**MMN** - Mismatch Negativity
**ACC** Anterior Cingulate Cortex
**PFC** Prefrontal Cortex
**LGN** Lateral Geniculate Nucleus
**CA1** Cornu Ammonis 1
**CA3** Cornu Ammonis 3
**V1** Primary visual cortex
**V2** Secondary visual cortex
**V4** - Visual Area 4
**LM** - Lateromedial Area
**M1** - Primary Motor Cortex
**M2** - Secondary Motor Cortex
**RSP** - Retrosplenial Cortex
**CCN** Cognitive Computational Neuroscience

**DANDI** Distributed Archives for Neurophysiology Data Integration
**NWB** Neurodata Without Border
**P300** event-related potential component that peaked around 300 ms after a stimulus is presented
**L1** Layer 1
**L2/3** Layer 2/3
**L4** Layer 4
**L5** Layer 5
**L6** Layer 6
**PYR** Pyramidal cell
**IT** IntraTelencephalic neurons
**PT** Pyramidal Tract neurons
**PV** Parvalbumin
**SOM** Somatostatin
**VIP** Vasoactive Intestinal Peptide
**DA** Dopaminergic neurons
**NDNF** Neuron-Derived Neurotrophic Factor
**LAMP5** Lysosomal-Associated Membrane Protein 5
**GABA** Gamma-Aminobutyric Acid
**E/I balance** - Excitation/Inhibition balance
**LTP** Long-Term Potentiation
**LTD** Long-Term Depression
**BTSP** Behavioral Time Scale synaptic Plasticity
**STDP** Spike-timing dependent plasticity
**DSI** depolarization-induced suppression of inhibition
**MDD** Major Depressive Disorder
**ASD** Autism Spectrum Disorder
**NHP** Non-Human Primate
**BCI** Brain Computer Interface
**RPE** Reward Prediction Error
**TD learning** Temporal Difference learning
**E-E** excitatory-to-excitatory
**E-I** excitatory-to-inhibitory
**PSTHs** peri-stimulus time histograms
**PCA** principal components analysis
**t-SNE** t-distributed stochastic neighbor embedding

# Introduction

Predictive coding is a prominent theory within a larger family of predictive processing models of the brain. These theories broadly propose that the brain refers to a model of the world, possibly based on the individual's past experiences to predict incoming sensory signals. Within predictive coding, when these predictions are accurate, the brain cancels out the predicted sensory inputs, allowing it to direct its resources on processing any unexpected, or incorrectly predicted inputs, known as prediction errors (Srinivasan et al., 1982; Rao and Ballard, 1999). Alternatively, neural networks can detect the statistical regularities of stimuli, e.g., sequence of visual stimuli or places, and rapidly learn these sequences so as to guide future behaviors, as seen in several computational models and in vivo experiments in several brain areas from V1 to hippocampus (Mehta NIPS 1997, Neuroscientist 2001, Hippocampus 2015). Although predictive coding has garnered significant interest over the past two decades, at the cortical level the precise neuronal layers, subtypes, or sub-compartments that are responsible for prediction generation and prediction error calculation are still under active investigation. Researchers are working to identify the specific circuits and mechanisms involved, aiming to pinpoint how neural circuits encode and process key elements of predictive coding. Recent reviews on predictive processing in the cerebral cortex responsible for complex functions such as perception and decision-making, have highlighted potential mechanisms and circuits involved (Bastos et al., 2012; Aitchison and Lengyel, 2017; Keller and Mrsic-Flogel, 2018; Walsh et al., 2020; Millidge et al., 2021; Mikulasch et al., 2023; Phillips et al., 2024; George et al., 2025). These mechanisms describe how predictions and sensory responses can be encoded and compared within cortical circuits. While cellular theories focus on the existence of dedicated error neurons (Rao and Ballard, 1999; Keller and Mrsic-Flogel, 2018; Hertäg and Sprekeler, 2020), dendrite-based theories emphasize the role of sub-cellular dendritic compartments in pyramidal neurons as potential recipients or computational loci of error signals (Urbanczik and Senn, 2014; Sacramento et al., 2018; Payeur et al., 2021; Mikulasch et al., 2023).

In this perspective, our goal is to first provide an overview of the field by contrasting the existing experimental research with the current theoretical literature, with a particular focus on computational models of predictive processing in the sensory domain (we especially focus on sensory representations, as opposed to action and reward signaling). Next, we propose a series of experiments aimed at testing different predictive processing models at the resolution of single neurons. These experiments leverage either in-vivo two-photon imaging or electrophysiological recordings in head-restrained mice, with a subset to be carried out through the OpenScope program. The collected datasets will be made available to the broader research community for analysis as Neurodata Without Border (NWB) files (Rübel et al., 2022) shared via the DANDI archive.

This paper originated from a 2024 CCN workshop and was developed through collaborative community efforts. With this project, we aim to promote a closer dialogue between experiments and theoretical models in the field. Whenever applicable, we cite existing reviews to broaden our support from the literature. Our goal is to discuss models in practical terms, constrained by the known architecture of the mammalian cortex and informed by the latest experimental findings in mice and other species. Within the framework of the well-known Marr levels (Marr, 2010), our focus is on level 3, and thus on understanding how a potential predictive processing algorithm would be implemented at the level of neuronal circuits.

# General outline

A core tenet of predictive processing theories is the existence of an internal model of the world shaped by the weights of synaptic cortical connections where predictions are continuously compared against sensory inputs. The differences resulting from this comparison are known as prediction errors. In the past, experimental efforts have primarily focused on measuring such prediction errors, as these are one of the key signals that distinguish predictive processing models from earlier representational variants of sensory cortical processing. To do this, research laboratories have sought to generate a diversity of error types (see **Section I - A diversity of error and mismatch types**). These errors could be the result of either local computations within a cortical region or distributed computations (see **Section II - Distributed error computation**) across distinct areas (for example, resulting from the interaction between higher-order and lower-order cortical areas). Within each brain area, a repertoire of neuronal responses emerge, potentially associated with prediction, mismatch stimuli, precision or attention signals (see **Section III - A diversity of predictive neuronal responses**). Experimental and theoretical groups have examined the roles of excitatory and inhibitory sub-population of neurons in those responses (see **Section IV - Role of Inhibitory/excitatory balance and interneurons**), as well as dendritic compartments (see **Section V - Dendritic computations with apical dendrites**). To update predictions and reduce prediction errors, it is assumed that synaptic weights are modified following specific plasticity rules (**Section VI - Synaptic plasticity and learning rules**). Finally, we consider how the transmission of prediction error and predictions and their ensuing circuit interactions relate to network dynamics/interactions at a meso- and macro-scopic scale, e.g., via transients and oscillations (**Section VII - Linking single neuron activity with meso- and macro-scopic neural dynamics**).

In the next seven sections, we review these distinct axes from the largest to the finest scale, with a stronger focus on animal studies with a neuronal resolution, and then back to a larger scale to integrate these results with human studies.

Building on this review, we next outline concrete, detailed experimental proposals designed to resolve existing conflicts and knowledge gaps in predictive processing (see **Experimental proposals**). These future studies aim to bridge theoretical models with experimental work, leveraging advanced neurophysiological techniques and computational models to validate key hypotheses and refine current theories. Finally, we discuss potential outcomes of those experimental projects and review the remaining challenges (see **Discussion**).

# I. Diversity of error and mismatch types

Predictive processing experiments and theories largely rely on constructing internal models that generate predictions of sensory inputs. In experiments, such predictions are typically challenged by introducing mismatch stimuli. Various theories and models have used such experiments to investigate potential mechanisms underlying predictive processing. In this section, we review the different types of mismatch stimuli employed in experiments and introduce the models that incorporated this diversity of mismatch stimuli.

## 1. Experimental evidence

Depending on the type of prediction that is violated, a variety of different terms have been used to describe these stimuli (see **Figure 1**), including: mismatches, oddballs, omissions, unexpected stimuli, expectation violations, or deviant stimuli (see **Glossary** and **Figure 1**).

While a theoretical study can model “error” signals directly, as it controls the underlying principles of the model, in experimental studies, the underlying principles are not fully known. Hence, one cannot determine categorically whether a response is an “error”. Thus, it has been the practice to avoid confounding the stimulus name (mismatch, oddball, etc.) with the interpretation (prediction error). In this perspective, we refer to “mismatches” when discussing experimental neuronal responses and “prediction errors” when discussing their interpretation in the theoretical literature.

We should recognize that the characteristics of mismatch stimuli can differ considerably from one prediction error type to another. As introduced in later sections, the responses they elicit may therefore be supported by different biological mechanisms. For instance, predictions learned for short temporal sequences might be achieved using information readily available in a single neuron, while predictions for long sequences could require a distributed population of neurons.

For this reason, it is helpful to group different types of mismatch stimuli based on core commonalities and key differences.

### Sensory mismatches

Pure sensory occlusions or sensory mismatches are used to create unexpected stimuli. For example, in visual experiments, an image may be presented with a missing part, or part of it may be changed (see **Figure 1A**). Similarly, local image features can be designed to conflict with global patterns within an image (Bair et al., 2003; Keller et al., 2020b; Kirchberger et al., 2023; Cuevas et al., 2025).
Although neurons in the visual cortex have visual receptive fields within which they typically respond to stimuli (Hubel and Wiesel, 1962), their activity can also be modulated by broader contextual information outside of these receptive fields (so-called “extra-classical” receptive field effects) (Bolz and Gilbert, 1986). These response modulations, along with sensory occlusions or mismatches, have been cited as supporting evidence for some of the earliest theoretical studies on predictive coding (Rao and Ballard, 1999). For example, surround-suppression (and enhancement) experiments, where the neural response to a center grating is suppressed (or amplified) when surrounded by either the same or a different grating, can be viewed as examples of sensory mismatch stimulus studies (Jones et al., 2001; Bair et al., 2003; Keller et al., 2020b) (see Relevant Theoretical Models for details).

## Sensory-motor mismatches

Sensory-motor mismatches have been extensively used to test hypotheses related to predictive processing. For example, in visual experiments, the movement of images shown on a screen can be paired to the animal's own movement. This coupling can then be disrupted, allowing researchers to study how the brain responds to sensory-motor mismatches (see **Figure 1D**).

Keller et al. used a simplified virtual reality environment where the visual flow of a stimulus was controlled by an animal's running speed (Keller et al., 2012). Sensory-motor mismatch responses are induced when the coupling between the flow of the visual stimuli and the animal's running speed is transiently disrupted. This visual-flow feedback paradigm has significantly influenced subsequent research, enhancing our understanding of the visual tuning and receptive field properties of neurons that encode mismatch signals (Saleem et al., 2013; Zmarz and Keller, 2016; Muzzu and Saleem, 2021), and facilitating the identification of transcriptionally defined subpopulations of neurons encoding prediction error responses (O'Toole et al., 2023). Similar stimuli with feedback alterations have also been used in auditory and vocalization studies (Eliades and Wang, 2008; Keller and Hahnloser, 2009; Rummell et al., 2016; Audette et al., 2022; Audette and Schneider, 2023; Morandell et al., 2024), reinforcing the idea that predictive processing plays a role in sensory processing across sensory modalities.

## Sequential mismatches

Subjects can be exposed to temporal sequences of stimuli to establish contextual regularity that can then be systematically violated to evoke mismatch responses (see **Figure 1B-C**). The most common type of sequence violation in the literature is known as the sensory "oddball" (see **Figure 1C**). In this paradigm, a single stimulus ("standard" or "redundant") is repeatedly presented relatively rapidly (every 75ms to 2000 ms), but interspersed with rare "deviants" or "targets" which differ in specific properties from the redundant stimulus (e.g., visual stimulus orientation, auditory pitch, etc). These basic sequential oddball paradigms were first developed in human studies in the 1970s, where Electroencephalography (EEG) event related potentials such as the mismatch negativity (MMN) or the P300 have been identified as gross neurophysiological indices of prediction errors (see **Section VII**). More recently, oddball paradigms, including "visual oddballs", have been adapted to rodent studies to expand our understanding of the mechanisms that generate MMN and P300 responses at the circuit level (Gavornik and Bear, 2014; Hamm and Yuste, 2016; Garrett et al., 2020; Hamm et al., 2021a; Price et al., 2023; Wyrick et al., 2023; Gillon et al., 2024). Frequently, sequences of visual stimuli such as gratings or oriented patches are used (Hamm and Yuste, 2016; Homann et al., 2022; Gillon et al., 2024) as the orientation tuning of neurons in visual cortex is well characterized (de Vries et al., 2020; Muzzu and Saleem, 2021). Together with similar findings in auditory cortex (Parras et al., 2017; Auksztulewicz et al., 2023) and multimodal parietal cortex (Van Derveer et al., 2023), this body of work shows that mismatch and oddball signals can emerge in the firing of individual neurons in sensory cortices in an experience-dependent manner, even when mice are passively experiencing sensory stimulus sequences.

When using simple oddball paradigms, and especially when analyzing single neuron responses, careful selection of the comparison condition is critical to ensuring that altered responses to the deviant stimulus are correctly interpreted. Direct comparisons of deviant and redundant stimulus responses may not purely reflect bona-fide prediction errors, but may instead be confounded by "stimulus-specific adaptation" and other forms of feed-forward synaptic depression (Ross and Hamm, 2020; Shiramatsu and Takahashi, 2021). A popular way to avoid this pitfall – and to better isolate sensory prediction errors from simple adaptation – is the use of “many standards” control sequences. In these control sequences, the same oddball stimulus (i.e. an oriented grating) is presented in a seemingly random pattern (e.g., of gratings with various orientations) where it is neither redundant nor contextually deviant (Harms et al., 2014; Wyrick et al., 2023). Comparing responses to the same stimulus when it is rare and contextually deviant (i.e., in the oddball sequence) vs. rare but not contextually deviant (i.e., in the many standards control) helps to isolate prediction error-like responses from effects of alternative processes such as stimulus-specific adaptation.

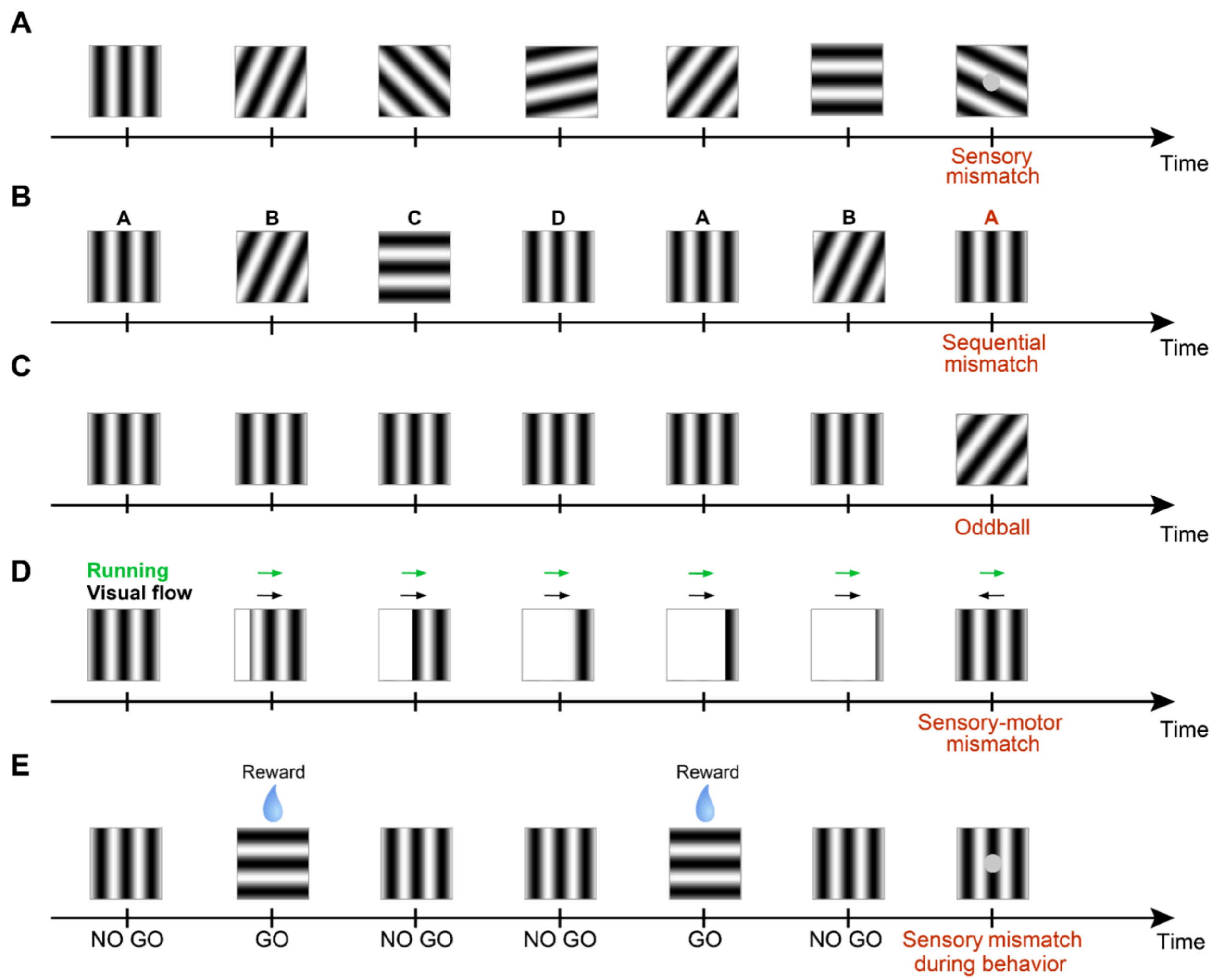

**Figure 1 - Example mismatch signals that have been experimentally evoked through a diversity of sensory stimuli.** For illustration purposes, we use visual gratings of various orientations. **A.** In sensory mismatch experiments, the animal is habituated to various stimuli that are presented one after another. A sensory mismatch, like the occlusion of part of a grating, is introduced for 1-10% of stimuli. **B.** In sequential mismatch experiments, the animal is habituated to a sequence of stimuli, for example a specific sequence of A, B, C, D oriented gratings. Sequential mismatches break this previously learned sequence, for example by replacing a C oriented grating with a A grating. **C.** In oddball experiments, a single stimulus is repeated most of the time, but at random times, a new stimulus is suddenly presented. **D.** In sensory-motor mismatch experiments, the motor behavior of the animal is coupled to a specific stimulus characteristic. In this example, the grating’s horizontal translation, or visual flow, follows the animal’s running speed. At random times, this relationship is broken, such that the stimulus does not follow the learned sensory-motor relationship. **E.** Sensory mismatch experiments can be performed passively or in a rewarded behavior task. For example, stimuli can be presented as part of a GO/NO GO experiment where the GO stimulus is rewarded with water. Behavioral experiments can provide an estimate of attention effects which are not constrained in passive behavioral tasks.

Oddballs can also be of higher order, for example when the deviations from expectations are set up through the repetition of sequences of stimuli (rather than the repetition of a single stimulus). For example, a repeated stimulus sequence A-A-A-B can randomly and rarely be replaced with A-A-A-A. Higher-order oddballs of this kind are referred to as “global” oddballs occurring across sequences, whereas the term “local” oddballs refers to deviations occurring within a given sequence (i.e. the Global/Local Task), originally implemented by Bekinschtein et al. (Bekinschtein et al., 2009). Comparing local and global oddball responses allows dissociating the effects of short- and long-term stimulus expectation (Westerberg et al., 2025).

### Omission oddballs

An omission paradigm is an important type of oddball response that can be highly informative about the circuits involved in predictive processing. This type of mismatch signal occurs when an expected stimulus within a sequence is omitted (Wacongne et al., 2011). Neuronal signatures related to the omission are interpreted as either a prediction error to the absence of a predictable stimulus or as an unfulfilled prediction. In either case, the potential confound of stimulus processing and stimulus change is avoided, which can be problematic in other paradigms.
Several factors have to be taken into account in order to reliably establish the genuine presence of neuronal coding of omission responses. One is the difference between an actual omission-specific response and a response due to the lack of stimulus-dependent bottom-up inhibition. This is because some neurons in the sensory cortex can be inhibited by the presence of a stimulus (Keller et al., 2020a). Thus, the lack of a frequently presented stimulus could appear as an increase in neuronal response, interpreted as a sequence-based prediction error, although it might be generated due to the absence of any stimulus in the receptive field. For example, if the stimulus A suppresses the neural activity in a group of neurons, if a repeated stimulus sequence of A-A-A-A is rarely replaced with A-A-A-X (X represents omission) those neurons might be mistakenly interpreted as “error” or “omission” neurons. Thus, stimulus type and position of omission in the sequence should be controlled to resolve such confounds.

In sequence based tasks, omission responses can be stimulus-specific, position-specific or both. A stimulus specific omission response contains significant information about the identity of the omitted stimuli. A position-specific omission response carries significant information about the position of the omitted stimuli in the sequence. Therefore, stimulus-specific omission responses could predict “what” was omitted and position-specific responses could predict “when/where” an omission occurred. The main advantage of studying specific omission responses is that with the lack of bottom-up inputs (Chien et al., 2019), the neural activity could only be due to the prior history, and not just a response to a change in the current stimuli.

### Mismatches during behavioral tasks

Active tasks in which the expected context is altered are used to increase effect sizes and/or increase the behavioral relevance of unexpected stimuli, for example, in mice trained to navigate an environment while learning a reward association (Shuler and Bear, 2006; Green et al., 2023; Furutachi et al., 2024). Alterations of the now familiar environment can then be introduced to create expectation mismatches (Fiser et al., 2016; Garner and Keller, 2022; Furutachi et al., 2024). In a navigation task, temporally structured sequences associated with specific locations are used to probe oddball responses via the occasional replacement of an expected element by an unexpected one (Furutachi et al., 2024). In the detection of change tasks, the reward can be made to coincide with the sensory mismatch (Garrett et al., 2020).

Mismatch responses can be modulated by the contextual relevance and attention given to the sensory features causing prediction errors. The relation between attention and mismatch signals remains complex (for review, see (de Lange et al., 2018)), with some work suggesting that attention can even counteract predictive suppression (Kok et al., 2012). It is certainly clear that attention is not necessary for mismatch signals to emerge, and that neural mismatch signals can be observed under anesthesia (e.g., (Chao et al., 2018)), although the level of anesthesia affects the cortical spread of mismatch signals (Nourski et al., 2018).

Several studies manipulated attention and stimulus predictability independently of each other yielding diverse findings. Some of these find that attention boosts specifically the response to the predictable sensory input, rather than the unpredicted input. Other studies find either no interaction or gain modulation of both the predicted and unpredicted input (Kok et al., 2012; Chennu et al., 2013; Auksztulewicz and Friston, 2015; Smout et al., 2019). The effect of attention may also depend on the processing level, as some studies suggest an attention-enhanced response to unpredicted inputs at later processing levels (Bekinschtein et al., 2009; Chennu et al., 2013; Kompus et al., 2020).

## 2. Relevant mechanisms of predictive processing across mismatch types

The terminology and conceptualization of error signals often differ between experimental and theoretical approaches. Experimental studies are conducted with a variety of underlying assumptions while theoretical models are explicitly designed with specific mechanisms in mind. This leads to varied conceptualizations and diverging terminology for similar ideas across each domain. For example, terms like "*world model*" and "*sensory inputs*" used by experimental groups, may correspond closely to "*internal model*" and "*teaching signal*" terms used by computational groups and grounded in theories of supervised learning (Jordan and Rumelhart, 1992). To promote clearer communication between experimentalists and theorists, we encourage the reader to refer to our **Glossary.**

### Do different error types engage distinct cortical networks and mechanisms?

While some studies have examined more than one type of mismatch stimulus (Gillon et al., 2024), a systematic comparison of oddballs, sensory mismatches and sensory-motor mismatches is still lacking. However, we can nonetheless anticipate potential differences by analyzing different experimental paradigms. In visual experiments, local sensory occlusions or sensory mismatches occur within a single presented image. To detect these errors, the visual cortex must rely on information from different receptive fields. This information can be provided by local projections within V1 or by immediate downstream areas, such as area LM in the mouse visual cortex (Marques et al., 2018). Responses to oddballs in stimulus sequences likely depend on short-term memory, which may arise through (1) local synaptic adaptation (Aitken et al., 2024), (2) local recurrence within the visual cortex (Reinhold et al., 2015) or (3) modulatory feedback from downstream areas, such as prefrontal cortex (Fiser et al., 2016; Bastos et al., 2023) or higher-order thalamic nuclei (Furutachi et al., 2024). In contrast, sensory-motor mismatch responses may result from the integration of motor cortex activity with visual input from LGN (Leinweber et al., 2017) or with auditory information at the level of the auditory cortex (Schneider et al., 2018). However, sensory-motor task designs vary widely and differences between them may influence their neural implementation. Experiments often engage a combination of complex motor processes such as locomotion, specific body movements like forelimb lever presses, continuous sensory cues like visual flow, or discrete sensory events like pure tone presentations. Each combination requires the brain to integrate information from various sources, possibly through different mechanisms. Supporting this, active tasks have recently been shown to engage large, brain-wide networks (Stringer et al., 2019), highlighting the need for caution when assigning a specific role to a particular brain area. Moreover, error responses observed during behavioral tasks may be influenced by neuromodulatory inputs (Collins et al., 2023) (see **Supplementary Text 2**), especially when these responses are tied to specific behavioral events (Ramadan et al., 2022).

In the next six sections, we review potential neuronal mechanisms underlying neuronal responses to mismatch stimuli from the larger to the smaller scale. Each section was written to be independently accessible. At the end, we review potential opportunities for integration across mechanisms.

# II. Distributed error computation

Experiments and theories have explored predictive processing mechanisms at various spatial scales. In this section, we review experiments and models that examined the involvement of multiple brain areas and cortical layers in predictive processing computations. We reserve finer-resolution mechanisms, such as the roles of individual cell types and dendrites, for later sections.

## 1. Experimental evidence

### Top-down inputs

For early on, most theoretical proposals for predictive coding involved distributed computations (see Glossary) of error signals (Rao and Ballard, 1999). Various studies have explored how distributed computations might shape local calculation of error signals, particularly in sensory-motor tasks or tasks involving complex sensory stimuli, which likely rely on a hierarchical feature representation. In one sequence-based oddball study, when mice were repeatedly shown sequences of gratings, inhibiting the activity in the PFC decreased the oddball response in V1, suggesting a role for top-down inputs in local mismatch computation or in modulating mismatch signals (Hamm et al., 2021a). More recent work in behaving mice has similarly demonstrated a role of thalamo-cortical projections from the pulvinar in generating mismatch responses (Furutachi et al., 2024). Here, pulvinar inputs appeared to feed back a mismatch signal, which amplified feed forward processing in V1. Other studies found stronger encoding of expected image identities in higher-order areas such as the retrosplenial cortex (RSP) compared to V1 when oddballs replaced images in sequences (Wyrick et al., 2023). Together, these observations highlight the role of higher-order areas and feedback (top down) connections in shaping error signaling in mouse V1. In contrast, Leinweber et al. showed that, while V1 receives strong axonal inputs from motor cortex area M2, these inputs primarily carry motor running signals, rather than mismatch information (Leinweber et al., 2017). This suggests that some aspects of the sensory mismatch computation may occur locally within V1. Similarly, Fiser et al. showed that axons from ACC to V1 carry stimulus predictions (Fiser et al., 2016), a finding later confirmed by Bastos et al. and Ross and Hamm (Bastos et al., 2023; Ross and Hamm, 2024). Taken together, the experimental observations support the possibility that sensory-motor mismatch responses and sequential oddballs recruit different top-down computational networks.

Connectivity data across species consistently support the existence of segregated feedforward and feedback pathways, distinguished by their cell types of origin and laminar termination patterns (Berezovskii et al., 2011; Markov et al., 2014). These studies reported that feedforward neurons rarely have a feedback collateral, which is significant considering the ubiquity of co-lateralization of inter-areal cortico-cortical neurons (Kennedy and Bullier, 1985). Feedforward and feedback connections are therefore distinct and this distinction forms the basis of anatomical definitions of the cortical hierarchy (Felleman and Van Essen, 1991; Markov et al., 2013; Harris et al., 2019). Yet, it is important to address the nature of feedforward (or bottom-up) and feedback (or top-down) connections in greater detail.

First, a detailed analysis of the hierarchical organization of visual areas in the mouse supports a relatively shallow hierarchy with fewer levels as compared to primates (Felleman and Van Essen, 1991; Markov et al., 2014; Harris et al., 2019; Siegle et al., 2021; Gămănuţ and Shimaoka, 2022; Burkhalter et al., 2023; Glatigny et al., 2024). For example, analyses of laminar connectivity patterns show that the cortical hierarchy, starting from V1, covers 10 hierarchical levels in macaque (Felleman and Van Essen, 1991; Markov et al., 2014; Harris et al., 2019; Siegle et al., 2021; Gămănuţ and Shimaoka, 2022; Burkhalter et al., 2023) but only 1.5 levels in mouse (Felleman and Van Essen, 1991; Markov et al., 2014; Harris et al., 2019; Siegle et al., 2021; Gămănuţ and Shimaoka, 2022; Burkhalter et al., 2023).

Second, while it might be tempting to assert that one type of connection carries predictions and the other prediction errors, it is not a necessary conclusion. For example, there is evidence that prediction errors are fed back to V1 (Furutachi et al., 2024). Indeed, each interareal connection comprises multiple components originating from different cell-type- or layer-specific populations of neurons in the source area, with different laminar termination patterns in the target area. Having two components per interareal connection leads to the conceptualization of dual counterstream architectures (Markov et al., 2013; Vezoli et al., 2021a), suggesting that discriminative and generative predictive coding might coexist in the cortex. In addition, the recent anatomical mapping of axonal projections into A1 or V1 shows that axonal afferents from LGN are not restricted to layer IV but terminate across all layers, although with significantly less density (Douglas and Martin, 1991; Constantinople and Bruno, 2013; Crandall et al., 2017; Chang and Kawai, 2018; Zhuang et al., 2020). Nevertheless, the projection into L4 is a distinguishing and unique feature of feedforward projections compared to feedback (which does not project to L4) in the sensory cortex, thus providing a solid basis for constructing a cortical hierarchy (Felleman and Van Essen, 1991; Markov et al., 2014; Harris et al., 2019; Siegle et al., 2021; Gămănuţ and Shimaoka, 2022; Burkhalter et al., 2023).

Third, it is important to note that there is no strict serial hierarchical organization, even though such organization is commonly assumed by hierarchical predictive coding models (Rao and Ballard, 1999). Rather, there are many connections between distant hierarchical levels (Felleman and Van Essen, 1991; Markov et al., 2014; Harris et al., 2019; Siegle et al., 2021; Gămănuţ and Shimaoka, 2022; Burkhalter et al., 2023)

## Global vs local oddballs

Studies in primates and humans have shown inconsistent results of global oddball encoding (see **Section I**). A study in primates and humans failed to identify a population of neurons in V1 and V4 responding to temporal mismatches. In this study, early and late temporal oddballs were introduced into a longer sequence, potentially creating “global” oddball responses (Solomon et al., 2021) (see earlier paragraphs). Other studies have shown that while population average does not show a robust global oddball response, population decoding methods are able to decode the global oddball condition with significantly above chance accuracy in frontal areas including prefrontal cortex and frontal eye field (Bellet et al., 2024; Xiong et al., 2024). This suggests that mechanisms beyond simple rate or magnitude codes are used within populations of neurons to encode complex predictions and prediction errors. Two recent studies in mice suggest that responses in the early visual and auditory sensory cortex primarily reflect local oddballs (Jamali et al., 2024; Westerberg et al., 2025). Extending the same paradigm to primates, revealed that global oddball responses are more prominent in higher-level cortical areas, and most pronounced in the prefrontal cortex (Westerberg et al., 2025). The same study suggests that local errors can largely be explained by low level mechanisms such as short-term adaptation and stimulus history. Hence, actual predictions might be largely restricted to non-sensory, cognitive areas (Gabhart et al., 2025). However, an ECoG study in primates also showed global oddball responses across electrodes placed on the (mostly mid-level) sensory cortex in at least one subject (Chao et al., 2018). Moreover, a human EEG/MEG study from the same group also supports a more widespread cortical distribution of global oddball responses (Wacongne et al., 2011). These apparent contradictions regarding global oddball responses in the early visual cortex warrant further scrutiny. One possibility is that the divergence rests on the non-local nature of slow-varying extracellular field potentials such as ECoG and EEG/MEG. Another possibility is that the extent with which global oddball responses are prominent in the sensory cortex depends on the experimental context.

## Cortical layers

Several experimental and theoretical studies have tried to assign specific or canonical roles to each cortical layer (Plebe, 2018) largely based on the stereotyped anatomy found across different cortical brain areas (Douglas et al., 1989; Barbas and Rempel-Clower, 1997; Douglas and Martin, 2004, 2007; Barbas and García-Cabezas, 2015; Harris and Shepherd, 2015). In addition, inhibitory neurons typically have stereotyped distribution profiles that are largely conserved across cortical areas (Tasic et al., 2018; Gouwens et al., 2020; Lichtenfeld et al., 2024). In the mouse, sensory-evoked activity profiles across cortical layers show layer-dependent organization: Hamm et al. found a higher proportion of oddball response encoding neurons in superficial cortical layers while responses to repeated stimuli significantly decreased across all layers (Hamm et al., 2021a). This result was later confirmed using electrophysiological recordings which differentiated L1 from L2/3 (Gallimore et al., 2023). Notably they reported reduced gamma synchrony between L1 and L2/3 during deviant stimuli, indicating potential functional differentiation of L1. Similarly, Jordan & Keller used intracellular recordings in mouse V1 to study how visual and motor inputs contribute to the generation of mismatch responses (Jordan and Keller, 2020). While L5 responded to the mismatch stimulus with primarily hyperpolarizing effects, only L2/3 exhibited visuomotor integration properties consistent with computing a visuomotor error directly. Consistently, ultra-high-field fMRI results in humans showed that expected events could be decoded with similar accuracy across cortical laminae, while unexpected events could only be decoded in superficial layers (Thomas et al., 2024).

However, Audette et al. recorded across layers during an audio-motor expectation task and found abundant error-like signals in auditory L2/3 and L5 (Audette et al., 2022). A key question that remains unanswered is whether L5 is actively computing these errors or dynamically relaying information from other regions.

In all of these studies however, there were no sharp boundaries in activity profiles. Notably, cortical layer boundaries are more loosely defined in mice (Harris and Shepherd, 2015), where the largest amount of cell-type specific error responses have been recorded. Further specialization may be more evident in mammals with thicker cortical tissue such as primates, or through additional measures such as receptor densities such as 5-HT$_2$ receptors or GABA receptors (Rapan et al., 2021). As a final point of note, the concept of a six-layered cortex is largely based on conventions established for primary sensory areas (Billings-Gagliardi et al., 1974), and may not be as well substantiated in other cortical areas (Buel & Hilgetag, 2015).

## 2. Relevant theoretical models

Hierarchical models, including predictive coding models are largely based on features of non-human primate (NHP) brains. In particular, in NHP sensory information is transmitted through sensory areas in a somewhat sequential manner (Schmolesky et al., 1998), with receptive fields becoming larger and more complex as they progress through the hierarchy (Desimone et al., 1985; Kusmierek and Rauschecker, 2009). Hence, some models of visual processing suggest that discrete computational steps (layers of the model) correspond to sequential sensory processing areas, such as V1 and V2, followed by V4 and eventually IT (Yamins et al., 2014). In addition, feedforward and feedback projection neurons are segregated in both mice and macaques (Berezovskii et al., 2011; Markov et al., 2014), indicating distinct functional roles of these two pathways. However, top down signals from higher levels of the cortical hierarchy can impact early visual cortex during processing of visual activity (Bullier, 2001), thereby allowing these areas to evoke non-sensory activation in visual cortex that modifies and augments sensory activation (Mumford, 1992; Roelfsema and de Lange, 2016). In other words, feedback projections can play an active role in the early visual cortex rather than just modifying feedforward signaling, as traditionally assumed. However, there seems to be clear species differences in mammalian cortical hierarchization. In macaques, the interareal connectivity graph density is 66% (i.e. two thirds of the possible interareal connections do exist) and such an interaction by proxy can concern many areas (Markov et al., 2013). However, in mice, the graph density is nearly 100% and virtually every cortical area can interact with each other (Horvát et al., 2016).

Rao & Ballard's predictive coding is a hierarchical model of sensory processing (Rao and Ballard, 1999), where each level encodes increasingly complex features, with the final level extracting the most abstract features of the input (Boutin et al., 2021). In classical predictive coding, predictions sent backward through the hierarchy become more granular as they approach the input level. This hierarchy generally aligns with the sensory processing regions observed in the brain. Within hierarchical predictive coding models, we can distinguish between "discriminative" (Whittington and Bogacz, 2017, 2019; Sacramento et al., 2018) and "generative" (Rao and Ballard, 1999; Friston and Kiebel, 2009; Mikulasch et al., 2023; Sennesh et al., 2024) models. In generative predictive coding, predictions flow down the hierarchy and prediction errors flow up. This is the currently dominating model in cognitive science and neuroscience. In contrast, discriminative predictive coding uses a reversed flow where predictions flow up and prediction errors flow down. In other words, generative predictive coding aims to predict the bottom-up input, whereas discriminative predictive coding focuses on predicting a top-down “label”. Teufel and Fletcher recently argued that considering both top-down and bottom-up forms of predictions may be essential for advancing predictive processing in neuroscience (Teufel and Fletcher, 2020). Experimental studies might have to account for the possibility of such an alternative signal flow as it could significantly impact the interpretation of experimental results.

### Early hierarchical models of cortical layers

Douglas and Martin synthesized anatomical, physiological, and computational observations to propose a unified model of “canonical” cortical processing, highlighting the importance of recurrent connectivity and the hierarchical organization of cortical layers (Douglas and Martin, 2004). According to this model, sensory or feedforward input primarily arrives in L4 from the thalamus or lower cortical areas, which then strongly projects to L2 and L3. These upper layers provide feedforward input to L4 of downstream cortical areas and also send projections to L5, which projects to other cortical areas and L6.

L6, in turn, sends projections back to L2/3, completing the so-called Canonical Cortical Microcircuit.

Additionally, local recurrent interactions between neurons are crucial, with ascending input targeting both pyramidal neurons and interneurons (Douglas and Martin, 1991). Recurrent connections amplify this input, generating an initial wave of excitation followed by a longer period of suppression (Douglas et al., 1995; Cossell et al., 2015). This notion of amplification via recurrent excitation is important because the synaptic projection strengths between hierarchical levels are notoriously weak, first at the level of thalamic input to the cortex and then between successive levels (Markov et al., 2011). In this manner, the local recurrent connectivity makes up over 80-90% of the total connectivity. Connections linking different levels make up 1 or 2%, similar to the LGN input to area V1. Of note, quantitative data regarding inter-laminar synaptic connectivity in mammalian cortex is limited (Binzegger et al., 2004).

Douglas and Martin also proposed a functional interpretation of the circuit, suggesting that L4 preprocesses the input, neurons in L2/3 collaborate to explore all possible interpretations and select one consistent with their subcortical inputs, while L5 uses these interpretations to produce an output to guide actions. Similar proposals have also been made by others, based on evolutionary considerations (Shepherd and Rowe, 2017). While these recent proposals prioritize pyramidal neuron types over cortical layers, insights from the layer-based perspective remain relevant (Adesnik and Naka, 2018).

This account just leaves L1, David Hubel's "crowning mystery" (Hubel, 1982). Traditionally L1 has been viewed as a major target of top-down projections from across the cortex (reviewed in (Markov and Kennedy, 2013)). This view point is largely supported by electrophysiological and anatomical findings in NHP and humans, suggesting that L1 constitutes a major convergence site for signals descending the cortical hierarchy (Cauller, 1995). A recent structural and functional study suggests that mouse L1 connectivity is more mixed than earlier work in primates suggests (Ledderose et al., 2023), perhaps echoing the ultra-dense mouse cortical matrix compared to macaque (Gămănuţ et al., 2018).

### Updated hierarchical models of cortical layers

Later anatomical findings led to a major revision of the Douglas and Martin model, stemming from the discovery that top-down feedback pathways have a dual origin (Markov et al., 2014). In addition to the classical feedback pathway originating from L6 in the infragranular layers, a second feedback pathway stems from the upper part of L2/3 in the supragranular layers. The two pathways differ in their topological aspects: the L2 feedback pathway projects over relatively short distances and tends to be point-to-point, this contrasts with the L6 pathway which is more long-distance and has a relatively diffuse topology and corresponds in this sense to the classically described feedback pathways (Rockland and Pandya, 1979).

### Predictive coding models of cortical layers

Predictive coding has been used as a framework to hypothesize the computations of each cortical layer (Rao and Ballard, 1999; Bastos et al., 2012; Mikulasch et al., 2023; Wang et al., 2024; Nejad et al., 2025). Earlier work (Bastos et al., 2012) proposed that prediction signaling is mediated by descending connections from deep layers (L5/6) of higher-order regions to superficial layers (L1) of lower-order regions. These predictions were compared against incoming ascending signals (thalamic inputs or prediction errors from lower cortical regions) arriving at Layer 4. Signals that were not effectively suppressed by descending predictions resulted in prediction errors, computed in superficial (L2/3) layers, and sent back up the hierarchy. In addition to laminar specificity, predictions and prediction errors were also proposed to be mediated by oscillatory activity in different frequency bands, with alpha/beta associated with predictions and gamma linked to prediction errors (Bastos et al., 2020; Vinck et al., 2024).

Because feedback pathways strictly avoid L4 in upstream areas (Markov et al., 2014), the question of where exactly feedback and feedforward pathways interact needs to be considered. One clue may be that the projection from L6 to L4 is one of the strongest inter-laminar pathways in the visual cortex (Binzegger et al., 2004). Hence, it may be the L6->L4 local inter-laminar connections which allows relaying feedback signals arriving in L6 to reach L4. The notion that the major site of feedback and feedforward convergence may actually be in L4 suggests a computational role for this layer which has been largely ignored and in this respect it is worth mentioning the high-specificity of glutamatergic cell-types found in primate L4 of area V1 (Jorstad et al., 2023), Interestingly, L6 to L4 is the pathway that was originally proposed for feedback-feedback convergence by Rao and Ballard (Rao and Ballard, 1999).

Inspired by self-supervised learning algorithms, Nejad et al. proposed another model in which L5 receives direct thalamic input, while L2/3 generates predictions of this input (Nejad et al., 2025). The temporal offset between L5 and L2/3 processing is crucial for L2/3's predictive function. Specifically, L2/3 processes information that is slightly delayed due to synaptic transmission from L4, creating a phase lag relative to the more immediate thalamic input to L5. These predictions from L2/3 are then sent down to L5 via L2/3 -> L5 synapses, where they are compared with the actual input to compute prediction errors in L5. This model aligns with observations on neuronal sparsity (Sakata and Harris, 2009) and error responses in sensory prediction tasks (Jordan and Keller, 2020).

Mikulasch et al.'s model, called the "dendritic hypothesis", follows the ideas of Douglas and Martin (Mikulasch et al., 2023). In this model, layer 2/3 (IT neurons) implements a sparse, predictive representation, while deeper layers likely process sensory input (L4) and compute the output of the microcircuit for long-range connections and motor control (L5/PT). Similar ideas were presented by Hawkins et al., in which a hierarchical network learns to predict longer temporal chunks of input data. In this framework, predictions are thought to be computed in layers 2/3, which implement a sparse, predictive and context-sensitive code, while deeper layers perform other functions such as belief calculation (Hawkins et al., 2009). This model was later extended (George and Hawkins, 2009; Bennett, 2020; Wang et al., 2024).

One important aspect of the cortical hierarchy that is largely ignored in predictive coding models is the highly parallel nature of hierarchical pathways, where each area projects in a distance dependent manner to many if not all upper and lower stream areas (Angelucci et al., 2023). The influence of interareal distance on the topology of feedforward and feedback projections has led to the idea of a dual counterstream architecture carrying distinct signals in upper and lower layers of the cortex (Vezoli et al., 2021a; Angelucci et al., 2023) and supported by recent human imaging studies suggesting a distinct role of feedback in L2 and L6 (Bergmann et al., 2024).

## 3. Divergence and convergence between experiments and theories

To a first approximation, experimental evidence supports the idea that sensory processing involves hierarchical feature representation, aligning with predictive coding theories. Studies have demonstrated that higher-order brain regions are more involved in processing complex sensory stimuli and errors, consistent with hierarchical models of sensory processing. For instance, the RSP shows responses to oddball sequences that differ from those in V1, aligning with the hierarchical processing theory proposed by Rao and Ballard (Wyrick et al., 2023).

Experimental findings indicate that top-down inputs play a significant role in error computation and modulation, both for oddball (Hamm et al., 2021a) and sensory-motor mismatches (Jordan and Keller, 2020; Audette et al., 2022). Inhibiting PFC activity during a visual oddball paradigm reduces responses to the mismatch (Hamm et al., 2021a), and this may work because PFC sends the necessary predictive information to V1 (Fiser et al., 2016; Bastos et al., 2023), rather than prediction errors. On the other hand, during a navigation task, higher order thalamo-cortical projections appear to send prediction errors directly to V1 (Furutachi et al., 2024), in contrast to the more passive feedback role of PFC (Ross and Hamm, 2024).

The functional differentiation of cortical layers observed in experiments corresponds with some aspects of theoretical predictions. For example, oddball responses are enriched in superficial layers across rodent and human studies (Jordan and Keller, 2020; Gallimore et al., 2023; Thomas et al., 2024) supporting models that propose that different layers compute distinct aspects of sensory processing and prediction errors (Rao and Ballard, 1999; Bastos et al., 2012; Rao, 2024; Wang et al., 2024; Nejad et al., 2025). These findings, however consistent, largely apply to the visual cortex. On the other hand, experiments in mouse auditory cortex have also shown auditory mismatch responses in both L2/3 and L5 (Audette et al., 2022; Audette and Schneider, 2023). Likewise, somatosensory mismatch (i.e., whisker stimulation) may be enriched in L4 and L6 of the barrel cortex in mice (Musall et al., 2017; English et al., 2023). Thus, while specific prediction errors may, indeed, show laminar specificity, the exact pattern may differ across modalities.

While hierarchical models like predictive coding assume a complex, multi-level hierarchy in sensory processing, experimental evidence suggests a shallower hierarchy in rodents compared to primates (Felleman and Van Essen, 1991; Harris et al., 2019; Siegle et al., 2021). This discrepancy indicates that highly hierarchical models may not fully capture the sensory processing dynamics in rodents, highlighting a potential limitation of these theories when applied across species. Comparative studies across multiple species, such as by Westerberg et al., thus seem particularly valuable for contrasting results in rodents and primate data, providing deeper insights into species-specific aspects of sensory processing (Westerberg et al., 2025).

Anatomical studies in both cats and rats reveal that thalamic inputs can bypass traditional hierarchical pathways, directly activating deep cortical layers (Douglas and Martin, 1991; Constantinople and Bruno, 2013). This finding challenges the classical bottom-up and top-down pathway definitions. Recently, it has even been proposed that the cortex is composed of two processing sheets with complementary roles (Keller and Sterzer, 2024; George et al., 2025).

Our ability to monitor neuronal activity across all cortical layers is nascent but expanding quickly and we anticipate significant progress in the coming years. Future experiments should leverage recent technical advancements to refine our understanding of computations across cortical layers during mismatch experiments (Weisenburger et al., 2019).

The role of corollary discharge remains less explored in predictive processing models compared to experimental research. Recent experimental evidence supports brain-wide behavioral modulation of cortical networks (Steinmetz et al., 2019), which challenges some aspects of hierarchical predictive coding and the concept of “explaining away”. However, it also suggests that distinct computations may occur across successive areas. It is likely that multiple computations occur within a single cortical column, involving a mix of local and global activity modes. For instance, recent experiments revealed varying decoding accuracy across layers for three distinct properties within a stimulus set (Tovar et al., 2020). Modeling work should integrate this possibility to align with cortical recordings.

# III. A diversity of predictive responses in single excitatory neurons

Predictive processing theories involve diverse responses across a large network of individual neurons. Specifically, predictive responses at the single neuron level involve anticipatory activity to forthcoming stimuli, suppressed or augmented responses to predicted stimuli, and augmented, suppressed, or otherwise altered responses to unpredicted stimuli. For example, in a visuomotor task, a moving visual stimulus can either match or differ from an animal's current locomotion. Neuronal responses in the cortex are modulated positively or negatively based on whether the stimulus speed exceeds or falls short of the expected motion (e.g., when the stimulus moves faster or slower than the animal's movement). Additionally, unsigned prediction errors can signal surprise or uncertainty, regardless of the stimulus valence or content. In this section, we first review experiments that have documented predictive modulation of excitatory neurons in the cortex. We consider the diversity of responses that have been observed in excitatory neurons in the cerebral cortex and the predictive processing theories explaining these observations. We then examine models proposing mechanisms by which these diversity of responses emerge.

In the next section (**Section IV**), we consider the diversity of responses and roles that have been identified for specific subtypes of inhibitory interneurons in the cortex and then relate these properties to their potential roles in the circuit overall in relation to predictive processing.

## 1. Experimental evidence

### The abiding relevance of the neuron doctrine

A core principle of modern neuroscience is the "neuron doctrine," which states that neurons are the basic structural and functional units of the nervous system. The neuron doctrine has its roots in Cajal's original discovery of neurons as individual cells (Jones, 1994, 1999), and the subsequent characterization of their neurophysiological properties. Early studies began with the *in situ* characterization of the biophysics underlying action potentials, and later expanded to the characterization of neuronal responses *in vivo*, culminating in a series of discoveries concerning the visual system, specifically how individual neurons in the visual cortex process information and respond to specific features like edges, orientation, and movement (Hubel and Wiesel, 1962, 1965). It is worth noting that, due to technical limitations at the time, these studies were primarily conducted one neuron at a time.

Early findings showed that individual cortical neurons are highly selective, or "tuned," to specific stimuli.. However, theoreticians even at the time recognized that the brain likely functions through the activity of large populations of neurons rather than relying solely on single neurons (e.g., (Bullock, 1959; Shepherd, 1972; Singer et al., 1997; Bullock et al., 2005; Guillery, 2005, 2007)). Nevertheless, the inability to study large-scale population activity at the time meant that much of the theoretical focus remained on the role of single neurons. Accordingly, the neuron doctrine evolved into the concept of neurons as “feature detectors” (Barlow, 1972; Parker and Newsome, 1998), also known as the “Barlow doctrine”. This concept suggests that individual neurons serve a fixed, singular function, similar to individual workers at a factory line.

The introduction of electrode arrays capable of measuring the simultaneous activity of hundreds, or even thousands of neurons *in vivo*, has shifted the theoretical focus towards understanding the function of neuronal collectives, also referred to as population activity. This trend has led to an increased scrutiny of the neuron doctrine (Yuste, 2015; Eichenbaum, 2018; Ebitz and Hayden, 2021). Discoveries such as context-dependent changes in neuronal tuning (Gilbert and Wiesel, 1990; Maier et al., 2007; Rigotti et al., 2013; Franke et al., 2022; Goldin et al., 2022; McFadyen et al., 2022; Popovkina and Pasupathy, 2022; Russell et al., 2024), representational drift (Deitch et al., 2021; Marks and Goard, 2021; Schoonover et al., 2021) and multiplexing (Jun et al., 2022; She et al., 2024) challenge the traditional notion that neurons have a fixed, hardwired role.

### Major subtypes of excitatory neurons

This flexibility notwithstanding, that neurons retain some category-specific role – e.g., roles for local interneurons or IT projecting pyramidal cells – seems plausible. Of course, excitatory and inhibitory neurons almost certainly play specific roles in brain circuits and computation: a point that is obvious from not only their postsynaptic actions but also by their morphologies and axonal projection patterns. But whether subtypes within these larger categories have circumscribed functions in relation to broader cognitive and perceptual outputs of the brain is less obvious. The emergence of transgenic and viral approaches to target and precision optical or novel pharmacological approaches record or manipulate molecularly-defined cell types techniques of the past three decades have confirmed that specific neuronal subtypes are linked to particular functions, even within a given sensory cortical brain area (e.g., (Fu et al., 2014; Kepecs and Fishell, 2014; Pakan et al., 2016; Dipoppa et al., 2018; Ferguson et al., 2023)).

Excitatory neurons, primarily glutamatergic neurons, can be classified into four major subtypes based on their laminar distribution, projections, and morphologies (or more specifically, their inputs, outputs, and internal structure) (Adesnik and Naka, 2018). Spiny stellate cells are predominantly located in layer 4, where they receive bottom-up input, typically from the thalamus, and project locally, mostly to neurons in layer 2/3. Intratelencephalic (IT) neurons are distributed across layers 2/3, 5, and 6 and project to other regions of the neocortex and the broader telencephalon. Evidence suggests that subpopulations of IT neurons may project either upward or downward within the cortical hierarchy– but rarely both – a distinction that carries significant implications for theories of predictive processing. Pyramidal tract (PT) neurons, primarily found in layer 5, project largely to the brainstem and spinal cord, while also sending collaterals to the neocortex, thalamus, and striatum. These collaterals may serve as substrates for corollary discharge or efference copies of motor output. Both IT and PT neurons are classified as pyramidal cells (PYRs) due to their triangular soma and prominent apical dendrites. Cortico-thalamic neurons, primarily located in layer 6, project directly to the thalamus, where they exert modulatory effects via mGluRs (Murray Sherman and Guillery, 2001; Liu et al., 2015). They often send collaterals to the thalamic reticular formation and layer 4 cells within the same cortical column. In intracortical recording studies, the explicit identification of excitatory neuron subtypes is typically absent. However, the laminar location of recorded neurons often provides indirect insights into their subtype, as well as their principal inputs and outputs.

These findings have influenced the study of predictive processing. Specifically, it raises the question: are distinct subpopulations of neurons responsible for specific computational steps in predictive processing, such as predicting sensory inputs, or calculating the discrepancy between sensory data and expectations?

### Neuronal responses to sequential mismatches

In standard visual oddball paradigms using oriented grating stimuli, excitatory neurons in the visual cortex exhibit the most pronounced prediction error-like responses compared to inhibitory interneurons. Hamm et al. conservatively estimated that approximately 11% of excitatory neurons in layer 2/3, among those responsive to a given stimulus orientation, show clear mismatch responses (Hamm et al., 2021a). In contrast, the percentage of mismatch-response neurons falls below chance levels in layers 4 and 5 - a finding corroborated using other types of mismatch features such as spatial frequency in V1 (Pak et al., 2021). Bastos et al. demonstrated that these prediction error responses are strongest in neurons specifically tuned to the orientation of the oddball stimulus but are also observed in neurons with limited or off target orientation selectivity, albeit to a much smaller degree (Bastos et al., 2023). This suggests that prediction error responses in the oddball paradigm involve a gain modulation mechanism. This is further supported by observations of gain modulation in stimulus selective PYRs in layer 2/3 of V1 during a navigation oddball paradigm, in which stimuli were sequentially presented along a linear corridor (Furutachi et al, 2024). Garrett et al. also found that excitatory neurons in superficial layers V1 demonstrated the largest responses to novel stimuli (Garrett et al., 2025) (see also (Westerberg et al., 2025)), but absent responses to stimulus omissions, in mice trained in a visual oddball detection task. These observations provide further support for a gain enhancement mechanism through the interaction of expectation with feed-forward input, as the absence of feed-forward input on an omission trial did not signal a prediction error in excitatory neurons. Accordingly, it is not clear whether excitatory neurons in V1 exhibit negative prediction errors during sequential oddball paradigms. In higher visual areas in parietal cortex of the ferret, a sequential oddball paradigm elicited prediction error responses consisting of increased responses in excitatory neurons that were selective for the deviant features and decreased – or silenced – responses in excitatory neurons that were selective for opposite features (Zhou et al., 2020). This is again consistent with a gain modulation mechanism, but also suggests divisive normalization and/or lateral inhibition are present as well.

In the auditory domain, systematic comparisons of prediction error responses across layers and subtypes of PYRs to sequential oddballs are absent, but i) direct recordings of layer 2/3 excitatory neurons in the mouse confirm the presence of such responses (Chen et al., 2015), and ii) recordings in A1 of non-human primates also suggest a subgranular enrichment based on current source density profiles (Lakatos et al., 2020). While this indicates some level of error selectivity within the excitatory neuronal population, detailed investigations into the specific excitatory neuron types (PT vs IT, for example) involved in these error responses have not yet been conducted. Recent work in rat auditory cortex identified clear prediction error response to omitted auditory stimuli (Lao-Rodríguez et al., 2023). These responses were robust across multiple auditory cortical regions and present in awake and anesthetized states. However, such omission responses were only present when very short (125ms) interstimulus intervals were used, and omission-responsive neurons responded also to frequency deviants as well. Thus, whether pure negative prediction error neurons are present in auditory cortex is unclear.

Divergence from this laminar pattern of enhanced prediction error responses in layer 2/3 has been found in other modalities. In a sequential oddball paradigm using whisker stimulation and recording in the barrel cortex of mice, Musall et al. found that prediction error-like responses were only present in granular input layers, emerging later (200ms) after the initial stimulus evoked responses (Musall et al., 2017). Later, the same group identified similar Bayesian surprise responses in layer 6 (English et al., 2023). Although cell-type (excitatory vs inhibitory) was not analysed, this pattern suggests that somatosensory mismatch responses diverge from visual responses.

One possibility for this observation is that layer 2/3 computes the prediction error when the competing features are represented locally, while other layers may exhibit the response when it is computed elsewhere and fed-back, forward, or laterally in the cortical hierarchy. In the visual mismatch studies cited above, the standard (predictable) stimulus features and the oddball (deviant) stimulus features were encoded within the same column of cortex (orientation or spatial frequency, in the mouse), while in the somatosensory mismatch studies, the standard and the deviant stimuli were distinct whiskers, which are encoded by distinct barrels in spatially separated columns of the cortex. Such differences may suggest that the layer-specific computation of prediction errors may depend on whether sensory features are encoded within a single cortical column or across spatially distributed columns, and invites further investigation.

## Sensory-motor mismatches neuronal responses

Locomotion, vocalizations, and other forms of motor outputs lead to predictable sensory inputs which the brain processes differently than sensory inputs arising from other sources – an observation well explained in a predictive coding framework. In systematic study of these phenomena, such as in sensory-motor mismatch experiments, both positive and negative prediction errors can be triggered by dissociating sensory feedback from the animal's actions or by omitting sensory feedback during an animal’s action. For example, various studies have used the locomotion-based sensory-motor mismatch approach to probe the presence of both positive and negative mismatch responses in neurons in V1 of mice (Attinger et al., 2017; Jordan and Keller, 2020). Intracellular recordings revealed distinct groups of excitatory neurons that responded oppositely to the same mismatch event, with some neurons depolarizing and others hyperpolarizing in response to sensory-motor mismatches. These responses formed a continuous distribution and were linked to the neurons' intrinsic electrophysiological properties, suggesting that neuron type characteristics may underlie positive and negative prediction error signals. Additionally, Fiser et al. demonstrated that, with experience, some layer 2/3 neurons become predictive by responding in anticipation of an expected visual stimulus (Fiser et al., 2016). When these expected stimuli were omitted, neuronal activity increased, particularly during trials that had previously exhibited strong predictive activity. This highlights the dynamic plasticity of neuronal populations in encoding predictive signals.

Audette et al. used an alternative sensory-motor mismatch paradigm where forelimb lever-pressing triggered a brief sound event at a fixed position in the movement (Audette et al., 2022). The consistent timing of sensory feedback during a single limb movement allowed for the analysis of extracellular neural activity in anticipation of the expected sound. A population of auditory cortex neurons across all layers signaled both the timing and identity of the anticipated sensory event (Audette et al., 2022; Zhou and Schneider, 2024). Altering the frequency of the sound revealed mismatch neurons, including a large population of neurons that did not respond to sounds in any other condition, including passive listening. In the auditory cortex, movement, prediction and mismatch neurons were largely carried by distinct neural populations, suggesting that predictive computations may be performed by specific, identifiable cell types (Audette et al., 2022; Audette and Schneider, 2023).

A separate group studying predictive suppression of self-generated responses in the auditory cortex of mice found that excitatory neurons across all layers of A1 display this attenuation, with deep layers 5/6 exhibiting the strongest suppression. Interestingly, this corollary discharge (see **Glossary**) was largely disrupted in a mouse model of schizophrenia (df16A+/- mice), consistent with some models of altered predictive processing in psychosis (Rummell et al., 2023) (see **Supplementary Text 1**).

Advances in single-cell transcriptomics have enabled a comprehensive classification of cortical cell types by analyzing gene expression profiles at the cellular level (Jorstad et al., 2023; Yao et al., 2023). These distinct cell types can now be experimentally targeted and appear to support specific subcircuits and dynamics (e.g., (Mohan et al., 2023; Musall et al., 2023)). O'Toole et al. combined photoconvertible markers with transcriptional profiling to identify and functionally characterize the cell types responsible for encoding positive and negative prediction errors during sensory-motor mismatches (O'Toole et al., 2023). Their results suggest that sub-populations of layer 2/3 excitatory neurons may encode positive (Rrad expressing) vs negative prediction errors (Adamts2 expressing) in a locomotive task. Whether this distinction holds true for other cortical regions and other mismatch types remains to be determined.

### Spatial integration and mismatch

In the visual system, the "classical receptive field" of a neuron is defined as a limited portion of the visual field that must be stimulated in order to modulate a neuron's response. However, top-down connections from higher levels of the cortical hierarchy, originating from neurons with much larger receptive fields, can influence the responses of target neurons at lower levels (see **Section II**, (Salin et al., 1992)). Under certain conditions, these top-down signals can cause neurons to respond to stimuli outside their classical receptive fields (Angelucci et al., 2023).

This phenomenon led to the concept of an "extra-classical receptive field", which extends beyond the classical receptive field. One well-studied example of an extra-classical receptive field effect is end-stopping, or surround suppression, in the visual cortex. This occurs when a stimulus extends beyond the boundaries of the classical receptive field, often resulting in a reduced neuronal response (Hubel and Wiesel, 1965; Xing and Heeger, 2001; Fu et al., 2024). Another example is the emergence of responses to illusory contours through local recurrence (Shin et al., 2025). This type of modulation highlights the influence of contextual information on visual processing.

A recent occlusion study found that layer 2/3 pyramidal neurons selective for an occluded region of an image encoded image-specific information in their responses. This suggests that these neurons signal the absence of predicted visual stimuli, corresponding to positive prediction errors (Seignette et al., 2024). The same study identified another population of layer 2/3 pyramidal neurons that responded to the presence of unpredicted visual stimuli, signaling negative prediction errors. Interestingly, the study also showed that layer 5 pyramidal neurons could be divided into subpopulations that preferred either contextual (predictive) input or sensory input. This division suggests that these neurons may also encode prediction errors. However, their responses were more complex and varied depending on the task engagement of the animal, highlighting a dynamic aspect of their role in processing prediction errors. Thus, consistent with studies employing sequential oddballs and sensorimotor mismatches, it appears that layer 2/3 excitatory neurons may exhibit the most reliable positive and negative prediction errors – at least in visual cortex.

### Precision signals

In hierarchical predictive coding , representations of sensory causes are updated through the precision-weighting of bottom-up prediction errors and top-down predictions (Rao and Ballard, 1999; Bastos et al., 2012). Precision, defined as the inverse of the variance, determines the relative reliance on sensory bottom-up inputs versus top-down predictions. When sensory input is noisier or more unreliable, top-down predictions, based on prior knowledge, dominate (Huang and Rao, 2011). For instance, when walking through a dark room, prior knowledge and predictions from other sensory modalities guide perceptual inferences about object shapes and surfaces influencing representations in the visual cortex.

Precision is not only influenced by the signal-to-noise ratio of the sensory input but also by internal and behavioral states. For example, attention can enhance sensory weighting through mechanisms such as gain modulation, neural synchronization, or reduced neural variability (Mitchell et al., 2007, 2009; Cohen and Maunsell, 2009; Harris and Thiele, 2011; Denfield et al., 2018; Thiele and Bellgrove, 2018). Similarly, arousal can increase the reliability and signal-to-noise ratio of sensory responses (Harris and Thiele, 2011; McGinley et al., 2015).

Although experimental data is not typically interpreted or reported in relation to precision signals in the literature, there is some evidence that prediction errors lead to increased precision in sensory regions when considered through this lens. For instance, Zhou et al., recording in higher visual areas in parietal cortex of the ferret, demonstrated that visual oddballs evoke increased responses in excitatory neurons coding for the deviant features and decreased responses – or suppressed activity – in excitatory neurons coding for opposite features from the deviant (relative to a many standards control) (Zhou et al., 2020). As discussed above, others have shown that prediction errors in mouse visual cortex primarily involve augmented responses across neural populations that depend on the feature selectivity each neuron in relation to the deviant stimulus– an effect which would amount to increased precision (Audette and Schneider, 2023; Furutachi et al., 2024).

On the prediction end, Bastos et al. examined the spatiotemporal pattern of top-down inputs to visual cortex from prefrontal cortex and showed that during highly predictable sequences (a visual oddball), the spatial standard deviation of activity across the population of axons in was higher than in a less predictable sequence (many standards control) (Bastos et al., 2023). Specifically, during the less predictable sequence, the distribution of activity across the population of PFC axons was more gaussian, while, during the predictable sequence, the distribution showed more high and more low activity axons. Further, the stimulus could be decoded from the spatiotemporal pattern of axonal activity better during the predictable oddball sequence than during the less-predictable control. Although authors did not interpret this in terms of precision, it is consistent with this aspect of the predictive coding model.

## 2. Relevant theoretical models

### Emergence of error neurons

From a normative perspective, the classical predictive coding framework (Rao and Ballard, 1999) treats prediction error as an instantaneous signal used to update the states of predictor neurons in order to minimize prediction error. Friston and colleagues (Friston, 2005; Friston and Kiebel, 2009), starting from free energy minimization, recovered the update equations for neural firing rates and synaptic weight. In both frameworks, the synaptic update rules can be interpreted as Hebbian learning.

Earlier predictive coding frameworks allowed the activity of error neurons to be positive or negative, which is biologically implausible. Synaptic weights could also take both positive and negative values, effectively allowing units to switch between excitatory and inhibitory roles (Ackley et al., 1985; Rumelhart et al., 1986) This type of flexibility is generally considered biologically implausible under Dale's principle (Eccles et al., 1954). To resolve these issues, Keller and colleagues (Keller and Mrsic-Flogel, 2018) proposed a circuit with separate positive and negative prediction error neurons with inhibition provided through local interneurons (see **Figure 2**). This is consistent with experimental work showing the emergence of positive and negative mismatch responses during learning (Attinger et al., 2017; Jordan and Keller, 2020).

## Models of pure sensory errors

Given the early availability of single-cell recordings in the visual system, some of the first models focused on the so-called "extra-classical" receptive field effects as a testing ground for model validation (Allman et al., 1985). Predictive coding was initially used to explain surround suppression via predictive inhibition in the retina (Srinivasan et al., 1982) and later extended to cortex (Rao and Ballard, 1999). Rao and Ballard (see **Figure 2A**) proposed that cortex performs hierarchical Bayesian inference by integrating bottom-up sensory information with top-down priors through local feedback loops between areas, a concept also suggested by others (Lee and Mumford, 2003). In their model, "end-stopping" occurs in specialized error neurons that compare inputs with top-down prior expectations, describing this phenomenon as a hierarchical computation.

In addition, predictive coding may operate in a non-hierarchical manner (**Figure 2B**). For example, when a sound consistently follows repeated visual flashes, the visual signal can act as a predictor for the upcoming auditory input. Conversely, when a flash follows repeated auditory tones, the auditory signal can serve as the predictor. In this scenario, prediction and error signals can propagate bidirectionally between V1 and A1 (Keller and Mrsic-Flogel, 2018). This non-hierarchical implementation is consistent with the shallow brain hypothesis, which proposes that the cortex, particularly in rodents, lacks a strictly hierarchical organization (Gămănuţ et al., 2018; Suzuki et al., 2023). Whether predictive coding is implemented hierarchically or non-hierarchically implies distinct patterns of functional connectivity between dedicated local error neurons and prediction neurons, thereby generating experimentally testable predictions (**Figure 2C-D**).

The functional connectivity is again different when error signals are encoded within specific dendritic compartments rather than by distinct neuronal populations **(Figure 2E)**, putting a stronger emphasis on local interneuron circuits implementing lateral inhibition (Mikulasch et al., 2023). Therefore, specific choices on the algorithmic level (e.g., hierarchical versus non-hierarchical), as well as different implementation strategies for predictive coding make distinct predictions on the circuit level that need to be tested in the future.

Alternatively, normalization mechanisms and figure-ground segmentation propose that mismatch responses organize and prioritize sensory information according to context. Predictive and sparse coding models, along with conceptually closely related normalization models (Lian and Burkitt, 2025), have highlighted the role of lateral competition as a potential origin of extra-classical receptive field effects in cortex (Lee et al., 2006; Carandini and Heeger, 2011; Spratling, 2011; Zhu and Rozell, 2013; Boutin et al., 2021). This type of computation referred to as a "explaining away" (see **Glossary**) through lateral inhibition[1] in cortical circuits. As a bar stimulus becomes longer, neurons with receptive fields near the end of the bar take over and "explain away" the stimulus. This process inhibits neurons with more central receptive fields whose activity is considered redundant.

Unlike models that explicitly represent prediction errors through dedicated neurons, Nejad et al. take a distinct approach rooted in deep learning frameworks. In their neocortical circuit model, L2/3 output learns to predict current sensory input in L5 by integrating past sensory information, relayed through L4, with contextual top-down input (Nejad et al., 2025). The network training relies on a self-supervised cost function, optimizing synaptic weights through backpropagation and gradient descent. In this setup, mismatch errors in L2/3 and L5 were defined as the derivatives of the self-supervised loss function with respect to neuronal activity in each layer. Thus, their model does not rely on specific "prediction error neurons”; instead, error signals are represented in the gradients, indicating how each neuron’s activity should adjust to reduce prediction errors. Although this gradient-based method differs from conventional models that simulate prediction errors directly in neuron activity, the authors propose that these gradient-based error signals could be implemented biologically through a multiplexing framework, where error signals from L5 are kept separate from inference signals and propagate to L2/3 as burst-like events (Payeur et al., 2021; Greedy et al., 2022; Friedenberger et al., 2023).

[1] Note that explaining away has been connected to two different network motifs: One, top-down inhibition, as in the cellular hypothesis, in which bottom-up input that has been explained away is canceled out (Clark, 2013). Or two, lateral inhibition between neurons (or cortical areas) that provide competing explanations for the same inputs (Moreno-Bote and Drugowitsch, 2015), as in the dendritic hypothesis (Mikulasch et al., 2023). The latter motif is also used in sparse coding models, where it improves the coding efficiency of the neural network.

### Models of temporal sensory errors

Early theoretical models of predictive coding did not incorporate a temporal component (Rao and Ballard, 1999; Bogacz, 2017). Models extending the predictive coding framework to the temporal domain emerged later (Friston et al., 2008). The effects reported in earlier experimental designs using the oddball paradigm may be explained by simpler mechanisms such as adaptation, without requiring top-down inputs (May, 2021). Later experiments demonstrated that more complex predictive models are necessary to account for sequential mismatch responses (Hamm and Yuste, 2016). Indeed, when Lieder et al. used EEG responses to MMN-inducing stimuli to compare the plausibility of mathematical models based on Bayesian inference with traditional adaptation-based models, they found that models incorporating explicit prediction errors better explained their recordings (Lieder et al., 2013).

Many other models have been proposed to explain sequence-based oddball responses (Wacongne et al., 2012; Chien et al., 2019; Auksztulewicz et al., 2023; Awwad et al., 2023; Lao-Rodríguez et al., 2023). For instance Wacongne et al. modeled sequential mismatch responses using a network of prediction and prediction error neurons. By being trained to predict stimuli over time using a spike-timing-dependent plasticity rule and input from a memory module, prediction neurons were able to internalize stimulus statistics allowing them to anticipate future inputs based on past patterns (Wacongne et al., 2012).

For tasks with temporal dynamics, models with leaky integrate and fire neurons have shown that network properties such as excitatory/inhibitory (E/I) balance and predictive coding responses in individual neurons emerge when efficient coding constraints are applied (Boerlin et al., 2013; Denève and Machens, 2016; Brendel et al., 2020). Some models implementing E/I balance generate mismatch responses when shifts occur in input stimulus distributions (Hertäg and Clopath, 2022). Millidge et al. demonstrated how local Hebbian plasticity can enable predictive coding networks to learn temporal relationships (Millidge et al., 2024). Similarly, Jiang and Rao proposed a network that learns hierarchical temporal features, with deeper layers reflecting relationships with increasingly complex and longer timescales (Jiang and Rao, 2024).

Predictive models that process sensory information in real-time must also overcome processing delays to accurately predict sensory inputs based on past events. Such models are taught to prospectively estimate errors, thereby allowing the network to anticipatively adjust its activity and correct for potential future discrepancies (Hogendoorn and Burkitt, 2019; Ellenberger et al., 2025; Senn et al., 2025), Nejad et al. proposed a computational model of a cortical column that processes both sequential and sensory-motor mismatches using a similar delay mechanism. Their model suggests that cortical layer L2/3 neurons learn to generate predictions of incoming sensory stimuli by comparing past sensory inputs, relayed via L4, with current thalamic inputs arriving at L5 (Nejad et al., 2025).

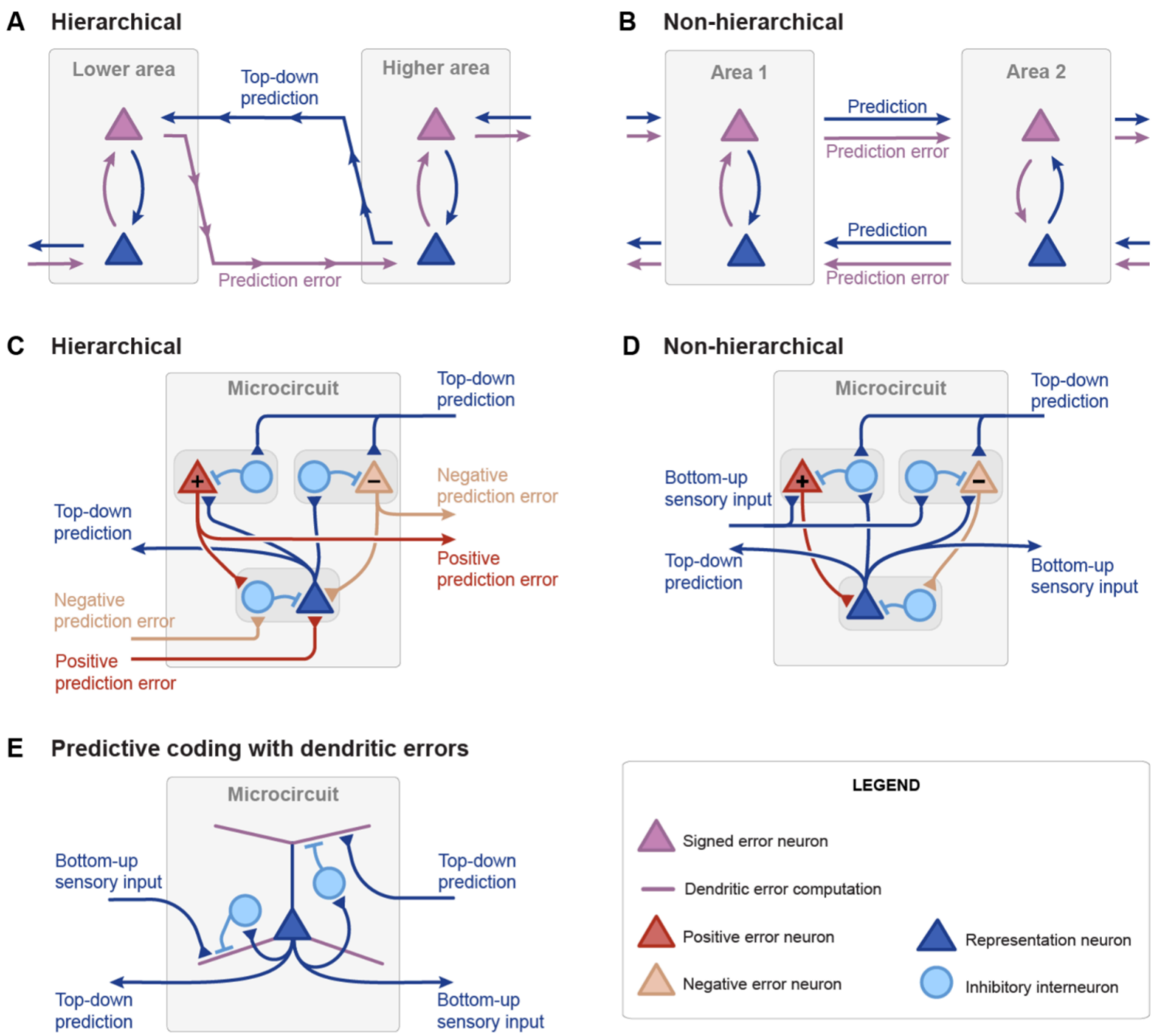

**Figure 2 - Diverse microcircuits proposed to implement predictive coding. A.** In the original predictive coding proposal, each brain area contains dedicated representation and error neurons that interact in a hierarchical architecture: error signals are passed up the hierarchy while predictions are sent down the hierarchy. The circuitry depicted here is based on Walsh et al. 2020. **B.** Recently, predictive coding has been proposed to operate in a non-hierarchical or partially parallel architecture, in which both prediction and error signals are transmitted to functionally relevant areas rather than strictly along a layered hierarchy (Keller and Mrsic-Flogel, 2018; Suzuki et al., 2023). **C, D.** Since neurons can only exhibit positive firing rates, not negative ones, many models assume dedicated positive and negative error neurons. This further requires distinct local circuits in both hierarchical and non-hierarchical architectures. In these models, positive and negative error neurons receive opposite input patterns from local representation neurons (excitatory versus inhibitory) and in turn exert opposite influences on those neurons. C shows a modified version of the hierarchical circuitry in A, whereas D depicts a non-hierarchical circuit proposed by Keller and Mrsic-Flogel, 2018. **E.** In most dendritic models of predictive coding, error signals are computed on dendritic branches rather than on dedicated error neurons. These models can be implemented hierarchically or non-hierarchically, where only predictions are exchanged between areas (Mikulasch et al, 2022, 2023).

Chien et al. hypothesized that cortical deviance-related activities are primarily generated locally through reciprocally connected neural circuits (Chien et al., 2019). To explore this, Chien et al. proposed a network model based on reciprocally coupled neural masses, with a focus on excitatory-inhibitory wiring patterns within the cortex. This model successfully reproduced properties of cortical deviance-related responses, including On/Off responses, omission responses and MMNs. However, since this network model consisted of rate coded neuronal populations with synaptic plasticity, Chien et al. did not assume a dedicated subset of neurons to error processing. Instead, their simulations suggest that the recurrent wiring patterns in the cortex provide a suitable environment for mismatch oriented calculations.

Other models that do not rely on dedicated error neurons have also been proposed. For example, it has been argued that sequential mismatch responses could be explained by changes in posterior variance due to neural sampling (Lee and Mumford, 2003; Aitchison and Lengyel, 2017), bottom-up attention to surprising stimuli (Aitchison and Lengyel, 2017; Westerberg et al., 2023), adaptive neural codes (Młynarski and Hermundstad, 2018), or updates to internal representations (Hawkins and Ahmad, 2016; van Driel et al., 2023). It is worth noting that many of these ideas also depend on computing prediction errors, which influence how sensory inputs are processed. For example, estimating the variance of the posterior for neural sampling requires computing squared prediction errors. However, unlike cellular predictive coding models, these computations could be performed in a less specific manner, such as by cortical interneurons (Garrett et al., 2020; Bastos et al., 2023; Furutachi et al., 2024; Ross and Hamm, 2024) and could differ depending on the complexity of the predictive sequence (Westerberg et al., 2025). This is covered more extensively in **section IV**.

### Model of sensory-motor errors

Models of sensory-motor mismatch responses typically rely on error-computing neurons. In these models, the sensory effects of self-generated movements are computed as corollary discharges which cancel out the effects of these movements on sensory representations (Jordan and Rumelhart, 1992; Wolpert and Miall, 1996). This mechanism is believed to enhance the system's ability to isolate external factors that are not the result of its own actions. In Bayesian theory, this can be understood as a form of explaining away through the use of the internal model (Moreno-Bote and Drugowitsch, 2015; Mikulasch et al., 2022). If multiple cortical areas jointly represent or “explain” sensory input (e.g., externally generated input in visual areas and self-generated input in motor areas), predictive coding suggests that these areas should actively subtract corollary discharge from the input they send to other areas. This would result in visual neurons encoding only to external motion, consistent with the “dendritic hypothesis” of predictive coding (Mikulasch et al., 2023).

A microcircuit implementation of explaining away that involves distinct types of inhibitory neurons has been proposed by (Hertäg and Sprekeler, 2020)**.** These models, involving the balance between excitatory and inhibitory neurons are discussed in more details in **Section IV**.

### Models of omission errors

Omitted-stimulus responses in the sensory cortex might arise due to the lack of suppression from the bottom-up input, causing disinhibited neuronal activity in the sensory cortex. Thus, omission of a frequent bottom-up input results in an absence of inhibition (disinhibition) elevating the omission-time response compared to the neural activity immediately prior to the omission. In the network model of (Chien et al., 2019), simulations showed that disinhibition played an important role in the generation of On/Off responses. On responses arose from a transient disinhibition prior to the network achieving a steady state, while Off responses were linked to release from prolonged disinhibition. Their simulations also suggest that cortical omitted-stimulus responses and MMN are fundamentally similar, as both reflect expectation violations, such as the timing, location, or identity of stimuli.

It should be noted that the potential role of stimulus omission is often overlooked in models of other types of sensory errors. For example, in an oddball task (AAAB), the positive error elicited by an unexpected stimulus B is often considered, but not the negative error elicited by the omission of the expected stimulus A. The emergence of global error signals, measurable using local field potentials, may therefore arise from a combination of positive and negative error signals collaborating to amplify deviant detection (Meng et al., 2025).

## Models of precision signals

In hierarchical predictive coding, precision is often modeled as a multiplicative modulation of prediction error signals, although different implementations exist: (1) **Gain modulation**: precision may be implemented as gain-modulation of the firing rate of prediction error responses (Ferguson and Cardin, 2020; Granier et al., 2024; Wilmes et al., 2025). In this case, the magnitude of the prediction error responses becomes ambiguous, as it reflects both the precision and the magnitude of the prediction error. (2) **Variability modulation**: alternatively, precision may affect the variability of neural responses or membrane potential, rather than altering the magnitude of responses (von Hünerbein et al., 2025). Precision could be multiplexed within the feedback itself, encoded in the variability of prediction error unit firing rates, making every prediction error inherently precision-weighted (Orbán et al., 2016). (3) **Synaptic modulation**: precision could also be implemented by modulating synaptic weights onto downstream neurons, originating either from prediction-error units or the representation neurons themselves (Hertäg et al., 2025). Flexible modulation of these weights, depending on behavioral context, may involve neuromodulatory pathways to the cortex such as dopaminergic and cholinergic connections (Yu and Dayan, 2005; Thiele and Bellgrove, 2018; Shine et al., 2021; Mei et al., 2022; Collins et al., 2023; Jordan and Keller, 2023; Pérez-González et al., 2024). (4) **Neural synchronization**: another potential mechanism involves changes in the synchronization of neural responses (Fries et al., 2001; Bastos et al., 2015b), influencing information transmission independently of firing rates.

Specific neuronal populations or pathways may implement a precision-weighting through mechanisms (1-4). If precision signals are conveyed through cortical feedback, a multiplexing mechanism may be required to separate precision signals from prediction signals. It has been proposed that precision-related feedback travels via short-range L2 pathways, while prediction feedback uses L6 pathways (Vezoli et al., 2021a). Specific populations of GABAergic interneurons may also play a role in encoding precision (Granier et al., 2024; Hertäg et al., 2025; Wilmes et al., 2025). Notably, Granier et al. propose that higher-level areas send precision estimates (or “confidence” signals) alongside the more classical predictions, and propose a role for disinhibitory circuits in mediating the entailed top-down gain modulation (Granier et al., 2024). Their theory predicts the existence of cortical second-order errors, comparing precision estimates with actual performance. Precision information may also be routed through apical dendrites, as discussed later in **Section V**, shaping the gain of pyramidal neuron responses (Shipp, 2016).

### Precision vs attention

(Friston, 2009) argues that "attention is simply the process of optimizing precision during hierarchical inference”. However, as reported by (Bowman et al., 2013), event-related potentials for repeated stimuli are enhanced when subjects are asked to attend to them. If attention is precision, these repeated stimuli should be fully predicted and thus produce a null precision-weighted prediction error. Furthermore, if precision-weighting serves to weight errors to adjust further predictions, then these large event-related potentials from fully predicted input should not update an internal model. This attention-based mechanism, which increases event-related potentials but does not provide a precision-weighted update to future predictions, makes the relationship between attention and precision less straightforward.

## 3. Divergence and convergence between experiments and theories

One of the central challenges in predictive coding is identifying the dynamic interaction between sensory afferents, error neurons, and internal predictions. While prominent predictive processing models include an explicit role for positive and negative error neurons to differentially signal whether inputs are larger or smaller than expected, there is no consensus that this is a theoretical requirement and experimental evidence clarifying this issue remains complexly ambiguous. Studies of sensory-motor mismatch responses demonstrate a clear distinction between positive and negative error neurons (Jordan and Keller, 2020), signalling when the motion information is more than or less than the animal’s current velocity. In experiments with omissions (e.g., a portion of a screen turning gray or the absence of a recurring stimulus), negative errors are typically defined as predicting a stimulus while it is actually absent, while positive errors involve predicting its absence when it is actually present. In behavioral paradigms involving predictions across multiple variables or dimensions, defining a 'positive' or 'negative' mismatch response can be more challenging.

In sequential oddball paradigms, experiments suggest that prediction errors to oddball stimuli reflect positive gain modulation in neurons selective for the unexpected stimulus (Bastos et al., 2023; Furutachi et al., 2024) – a finding which seems difficult to reconcile with the notion of dedicated positive prediction error neurons, although some work suggests that sensory cortical neurons can exhibit selectivity both for errors and stimulus features (Hamm et al., 2021a; Audette and Schneider, 2023). An omission oddball paradigm enables the assessment of negative error neurons, but studies in auditory and visual cortex evince limited or absent responses to stimulus omissions among excitatory neurons (Lao-Rodríguez et al., 2023; Garrett et al., 2025); but see **Section IV** for a discussion of omission responses in inhibitory interneurons.

The idea that stimulus size, actualized by the spike rate of individual neurons, results in neural activity that can be precisely compared against predicted values does not easily apply to primary sensory areas where normalization (a “canonical cortical function”) and localized feature extraction mechanisms quickly discard information about absolute stimulus intensity levels. Also prediction errors in visual cortex have been shown in multiple studies to boost representation of the unexpected stimulus features (Bastos et al., 2023; Furutachi et al., 2024; Ross and Hamm, 2024) or to signal broader contextual information (Hamm et al., 2021a; Audette and Schneider, 2023; Najafi et al., 2025), rather than a difference between current and expected inputs. These and other issues (e.g., defining 'size' in the auditory system, applying intuitions from rate-based models to spiking networks, etc.) make it challenging to relate external sensory inputs to the activity of individual neurons in experiments and models. Future work should aim to quantify intermediate representations from the earliest stages of sensory pathways (e.g., from the retina to the visual cortex).

The advances in the targeting of specific neuron types experimentally offer a valuable basis for refining predictive processing theories. These models can now start to reflect the circuit implementation more closely by incorporating the latest experimental insights. For example, building on recent work (O'Toole et al., 2023), one could propose that predictive processing is predominantly carried out within specific cortical layers, such as Layer 2/3 where different subtypes of pyramidal neurons (e.g., Rrad+, Adamts2+, and Agmat+ neurons) may play distinct roles in encoding positive errors, negative errors, and predictions, respectively. To bridge experiment and modeling work, it will be necessary to extend the classical predictive coding framework to include more complex interactions between different neuronal populations. This could involve incorporating additional factors such as neuromodulatory influences, the role of cortical feedback pathways, and the impact of behavioral state on the precision of error signals. For instance, precision signals, reflecting the certainty of predictions, might be routed through specific layers and circuits. For example, the concept of "second-order errors," as suggested by (Granier et al., 2024), introduces the idea that cortical areas not only predict sensory inputs but also estimate the precision or confidence of these predictions.

# IV. Role of Excitatory/Inhibitory balance and interneurons

The interplay between excitatory and inhibitory neurons has been proposed as a key mechanism through which predictive processing emerges. In this section, we review experimental evidence and related models that explore how excitatory and inhibitory subpopulations might collaborate in predictive processing. Here, we focus on cellular-level mechanisms, reserving the contributions of these subpopulations to dendritic processing for the next section.

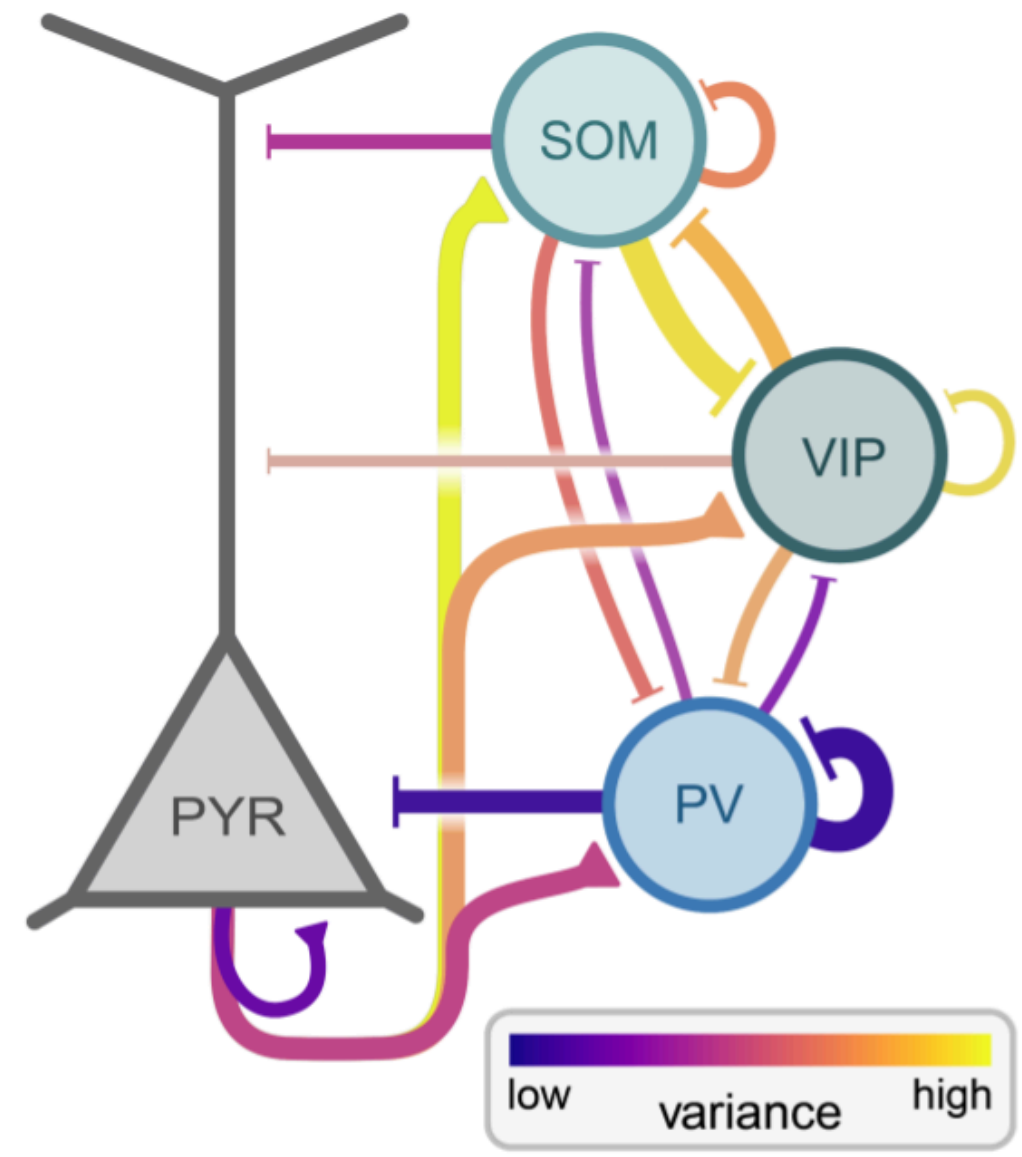

**Figure 3 - Intralaminar circuit diagram showing connections between major excitatory (PYR) and inhibitory (PV, SOM, VIP) cell subclasses, based on all cortical layers in mouse V1.** Line thickness depicts the relative weight (strength and probability of connection) of connections between cell subclasses. Line color shows the spike-to-spike variance in amplitude of synaptic signaling, which is strongly cell subclass-dependent. Excitatory synapse variance depends on the postsynaptic subclass. PV cells project low variance connections, whereas SOM and VIP project high variance connections. Adapted from Campagnola et al., *Science* 375: eabj5861, 2022, with permission from AAAS; modifications made with author permission.

## 1. Experimental evidence

In the framework of predictive processing, understanding the roles of different interneuron types is essential to delineating how bottom-up and top-down inputs are processed and integrated within cortical circuits. Importantly, the connectivity between pyramidal neurons, PV, VIP, SOM, and other interneuron subtypes is sufficiently stereotyped to facilitate investigation (**Figure 3**). The dense local connectivity and responsiveness of PV neurons to sensory-driven input position them well for facilitating the rapid processing of feed-forward, bottom-up signals. In contrast, VIP and NDNF interneurons are targeted heavily by projections from other cortical regions and deeper cortical layers, especially in higher-order areas (Wall et al., 2016; Huang et al., 2024b), making them well-suited to modulating cortical circuits based on contextual and predictive information. This connectivity supports their role in adjusting cortical processing in accordance with predictive signals, facilitating the brain's integration of expectations with sensory input. VIP interneurons preferentially suppress SOM neurons, thereby disinhibiting pyramidal neurons and enhancing selectivity for expected stimuli (Pfeffer et al., 2013; Pi et al., 2013; Karnani et al., 2014; Wall et al., 2016; Huang et al., 2024b). SOM interneurons, on the other hand, are distinct in their connectivity and are well-positioned to integrate lateral inputs, such as those required for surround-suppression effects in visual stimuli (Adesnik et al., 2012; Urban-Ciecko and Barth, 2016). They are also dendrite-targeting, further supporting a role in modulating excitatory neuron activity.

### PV neurons

Parvalbumin-expressing (PV) interneurons in the cortex may play a key role in predictive processing through their regulation of specific aspects of neural processing. By scaling the response amplitude of pyramidal neurons, PV neurons can amplify or reduce the activity of nearby pyramidal cells (Atallah et al., 2012). They have been shown to be central in shaping the precision of stimulus tuning in pyramidal neurons (Lee et al., 2012). PV-positive, fast-spiking basket cells, specifically, are involved in controlling cortical E/I balance via fast inhibition of cell bodies and basal dendrites (Ferguson and Gao, 2018). A prominent feature of this inhibition is that it is feature-specific, i.e., basket cells implement inhibition between neurons that receive similar feed-forward inputs (Chettih and Harvey, 2019; Najafi et al., 2020; Znamenskiy et al., 2024). This suggests that PV neurons could not only be used to maintain cortical E/I balance on fast timescales (Moore et al., 2018), but also to precisely cancel inputs predictable from ongoing neural activity, as suggested by predictive coding models (Uran et al., 2022). However, in a study by (Westerberg et al., 2025)), PV neurons were more responsive to oddball stimuli than to expected stimuli. Alternatively, PV neurons could contribute to the weighting of prediction errors based on their salience and reliability, or to amplifying relevant sensory inputs while suppressing distractors. Through this modulation, PV neurons may also enable differentiation of positive and negative prediction errors, adjusting the response amplitudes of excitatory neurons based on the reliability of predictions (Womelsdorf et al., 2014).

### VIP and SOM neurons

Many studies support potential roles for VIP and SOM neurons in predictive processing. These roles range from generating prediction errors during unexpected stimuli (i.e., via disinhibition), mediating predictive suppression leading up to expected stimuli, supporting paradigm-relevant feature selectivity (e.g., orientation preference) within local populations, or even conveying non-stimulus-specific attention signals linked to changes an animal's internal state, like its level of arousal. Given that VIP neurons are known to disinhibit pyramidal neurons via SOM neurons, these two interneuron populations have often been thought to play complementary roles in predictive processing. However, the research points to a more nuanced picture, in which the roles of these inhibitory neurons may be both context, task and subpopulation-specific.

#### VIP and SOM responses to surround suppression stimuli

As described in Section III, surround suppression is a phenomenon in visual processing whereby neurons show a reduced response to stimuli extending beyond their receptive field. Interneurons are thought to play a crucial role in surround suppression in V1 by modulating E/I balance. Accordingly, SOM interneurons in the superficial layers of the mouse V1, excited by horizontal cortical axons, have been observed to contribute to surround suppression, increasing their response as a stimulus grows in size (Adesnik et al., 2012). This finding is broadly consistent with a role for SOM neurons in inhibiting pyramidal neuron responses to expected stimuli. Learning has also been shown to increase selectivity for specific stimuli in subsets of PV and SOM neurons. Notably, this is not the case for VIP neurons, further supporting potentially distinct roles for these cell types in learning and memory processes (Khan et al., 2018).

### VIP and SOM responses to repeated stimulus sequences

SOM neurons have also been shown to decrease their activity in response to stimulus repetition, while VIP neurons increase their activity, see **Figure 4** (Heintz et al., 2022; Bastos et al., 2023). This finding appears to contradict a simple role for SOM neurons in inhibiting responses to expected stimuli, but corroborate the idea that VIP and SOM neurons play complementary roles. However, adding more complexity to the picture, the responses of SOM and VIP neurons are not opposite for novel stimuli compared to familiar ones. SOM neurons instead display control-level (Heintz et al., 2022; Bastos et al., 2023) and even decreased responses (Kato et al., 2015; Natan et al., 2015, 2017; Hayden et al., 2021) to novel stimuli. In contrast, Westerberg et al. found that SOM neurons showed enhanced responses to oddball stimuli (Westerberg et al., 2025). VIP neurons present a similarly complex picture. In a passive oddball paradigm, they appear to show decreased responses to novel stimuli compared to familiar stimuli (Heintz et al., 2022; Bastos et al., 2023). In contrast, in a rewarded task with image sequences, VIP neurons respond more strongly to novel than to familiar stimuli. In fact, VIP neurons develop ramping responses in anticipation of expected stimuli which continue to increase when the stimulus is omitted, but are dampened if the stimulus does appear as expected (Garrett et al., 2020, 2025). Notably, Najafi et al. showed that during these omissions, VIP neurons encode not stimulus predictions, but task-independent information shared with other brain areas (Najafi et al., 2025). Follow-up work by Huang et al. (2025) further demonstrated that neural responses to unexpected image omissions during passive behavior—in VIP, SOM, and excitatory neurons of the visual cortex—are invariant to temporal predictability. This study provides evidence that intrinsic stimulus-evoked dynamics, rather than temporal predictive processing, underlie the ramping and omission-evoked activity in the visual cortex during passive behavior (Huang et al., 2025).

Optogenetically and chemogenetically suppressing SOM responses (Hamm and Yuste, 2016; Heintz et al., 2022; Bastos et al., 2023) in visual cortex has also been shown to reduce visual oddball responses, instead of enhancing them. Together with the evidence cited above, this strongly suggests that the role of SOM neurons is not simply to cancel out expected inputs to excitatory neurons (Heintz et al., 2022; Bastos et al., 2023; Gabhart et al., 2025; Westerberg et al., 2025). In VIP neurons, the same effect is also observed whether their activity is enhanced or suppressed, pointing to a potential role in tuning local V1 circuit excitability for optimal detection of oddballs (Heintz et al., 2022; Bastos et al., 2023).

### VIP and SOM responses to sensory-motor mismatches

Attinger et al. combined two-photon imaging and optogenetic manipulation to examine how VIP neurons regulate SOM inhibition of pyramidal neurons during sensory-motor mismatches in primary visual cortex (Attinger et al., 2017). Mice were trained in either a closed-loop condition, where visual feedback was tied to an animal’s movements (i.e., visual flow match running speed) or an open-loop condition where the two were decoupled. Halting visual flow in these conditions created both visuomotor and purely visual mismatches in both conditions. Excitatory neurons in closed-loop reared mice showed enhanced responses only to visuomotor mismatches, and not motor-driven halts, while those in open-loop reared mice responded to both visuomotor and purely visual mismatches. These mismatch responses appeared to be inherited from concurrent decreases in SOM input. However, VIP neurons, which generally inhibit SOM neurons, showed enhanced responses only in response to visuomotor mismatches, and thus could not explain enhanced responses to purely visual mismatches. This suggests that VIP neurons in primary sensory areas are specifically involved in integrating predictive motor information and is consistent with findings from Najafi et al. showing that VIP neurons integrate information from other brain areas during familiar stimulus sequence presentations (Najafi et al., 2025).

### VIP and SOM neurons during navigation tasks

Lastly, it should be noted that the involvement of these interneuron subtypes extends also to more complex tasks. VIP neurons, along with pulvinar inputs, are involved in generating mismatch responses in visual navigation tasks (Furutachi et al., 2024). In a foraging task, SOM neurons have been found to fire in synchrony during course corrections, pointing to a role in adaptive motor control (Green et al., 2023).

Altogether, although it is clear that VIP and SOM neurons are involved in generating prediction errors, their exact roles remain unclear. It is possible that their roles are very sensitive to the exact parameters of the task and stimuli being tested. This is corroborated, for example, by the fact that their response patterns are so different for novel vs oddball stimuli, and for familiar vs repeated. The apparent contradictions in their responses might also be explained by the cortical network entering different context or task-specific regimes if these differ in how they recruit the VIP and SOM neuron populations. Previous research has shown how this type of canonical circuit can move between such regimes, swinging between a disinhibitory and an inhibitory mode (Tsodyks et al., 1997; Garcia Del Molino et al., 2017; Miller and Palmigiano, 2020; Beerendonk et al., 2024). Lastly, individual neuron responses are far more variable than the average patterns observed for each cell type population (Heintz et al., 2022; Bastos et al., 2023), suggesting that the appropriate level of analysis may be more granular than major inhibitory cell classes. Overall, more research is required to clarify the roles of SOM and VIP neurons in predictive processing, and the contextual and circuit factors that shape these roles.

### Layer 1 interneurons

Layer 1 is distinct from other cortical layers as it lacks excitatory cell bodies, and is primarily composed of L1 interneurons, apical dendrites from pyramidal neurons located outside L1, and dendrites from other inhibitory interneurons. L1 interneurons are therefore well positioned to have a broad influence on feedback inputs as they primarily target the superficial dendrites of pyramidal neurons (Huang et al., 2024a). Interneurons in L1, which provide prolonged inhibition to local dendrites, can be labeled using either LAMP5 or NDNF (Neuron-Derived Neurotrophic Factor) promoter genes (Tasic et al., 2018; Huang et al., 2024a). Huang et al. characterized running modulation and responses to visual gratings in LAMP5 interneurons of L1 (Huang et al., 2024a). In addition to being modulated by behavioral state, activity in LAMP5 interneurons state increased at lower contrast and following grating omissions, consistent with a role in top-down regulation.

NDNF interneurons, for their part, receive a wide array of cortical and subcortical inputs, suggesting a role in broadly integrating top-down inputs. Evidence to date also points to a delicate interplay between NDNF and SOM neurons. Whole-cell patch clamp recordings from NDNF interneurons in mouse auditory cortex slices has established that they can influence cortical processing by modulating incoming SOM synapses through GABAergic volume transmission (Naumann et al., 2025). In return, during exposure to auditory stimuli, NDNF interneuron activity is inhibited by SOM inputs in proportion to stimulus intensity.

Although both SOM AND NDNF interneurons target apical dendrites, during fear conditioning, the inhibition provided by NDNF interneurons in the auditory cortex lasts 4-5 times longer than that of SOM neurons (Abs et al., 2018). NDNF responses also increase after fear conditioning, unlike those of SOM neurons. This suggests that NDNF neurons, receiving top-down inputs, may compete with SOM neurons to regulate apical dendrites. Cohen-Kashi Malina et al. showed that L1 NDNF neurons not only inhibit apical dendrites (Cohen-Kashi Malina et al., 2021), but also disinhibit L2/3 pyramidal cell bodies by selectively inhibiting a subpopulation of PV neurons, thus controlling the ability of bottom-up inputs to shape pyramidal neuron activity. Together, these findings point to a role for L1 interneurons in shaping pyramidal neuron activity, through selective inhibition or disinhibition of localized synaptic inputs. The interplay between L1 feedback, SOM neurons, and bottom-up signals could be an important control point to shape learning (Doron et al., 2020).

## 2. Relevant theoretical models

Models often do not include distinct excitatory and inhibitory populations. This is particularly the case for deep learning models in which the sign of each synaptic weight is typically determined through learning. Such networks are incongruent with Dale's law according to which a neuron can have inhibitory (negative) or excitatory (positive) output synapses, but not both. In order to better study the role of inhibitory neurons in the brain and improve the biological plausibility of networks, models have been developed that do conform to Dale's law, with explicitly defined excitatory and inhibitory neurons.

### (E/I) balance

Predictive processing models that adhere to Dale's Law generally incorporate inhibitory inputs, at a minimum, to maintain excitation/inhibition (E/I) balance. Much of the early theoretical literature on E/I balance focused on why neural activity in cortex resides in the asynchronous irregular state (Renart et al., 2010), where neural firing, even of nearby neurons, often shows very low correlations. These network models showed that an asynchronous state is reached when the strengths of recurrent excitatory and inhibitory connections between neurons are of similar magnitude and are loosely balanced (van Vreeswijk and Sompolinsky, 1998; Brunel, 2000). Research into balance mechanisms revealed even further that if inhibitory connections can learn to precisely balance feed-forward inputs at the single neuron level, this enhances the efficiency of encoding in the feedforward input stream (Denève and Machens, 2016). The underlying principle is that, when a neuron encodes a specific part of the input stream, lateral inhibition removes this information from the inputs to all other neurons in the population. As a result, only the unencoded input (i.e., unpredicted) remains to drive network activity. This, in turn, decorrelates neural spiking and promotes efficient coding (Vinck and Bosman, 2016). This decorrelating balance is typically thought to be mediated by PV interneurons, which provide fast, strong lateral inhibition, primarily targeting basal dendrites and cell bodies.

PV interneurons are central for cortical gamma oscillations (Cardin et al., 2009), often modeled by Pyramidal Interneuron Network Gamma (PING) networks. In these models, when a network is balanced, it shows efficient coding, reduced firing rates and stable gamma oscillations (Traub et al., 1997; Jadi and Sejnowski, 2014). Theoretical work further suggests that gamma oscillations may reflect optimal sensory processing and arise naturally in E/I balanced networks built with transmission delays (Chalk et al., 2016; Echeveste et al., 2020). However, an alternative theoretical perspective is that gamma-band oscillations increase with predictions errors, e.g., due to increased firing rates and metabolic demands (Bastos et al., 2012). Experimental results are inconsistent when it comes to the relationship between stimulus predictability and the power and synchronization of gamma-band oscillations (see **Section VII**).

## From (E/I) balance to error neurons: role of VIP, SOM and PV neurons

Some models explicitly incorporate the *in vivo* connectivity patterns observed in the sensory cortex between PV, SOM, VIP and pyramidal neurons (see e.g., **Figure 3**). Hertäg et al. investigated how this canonical interneuron motif can give rise to excitatory neurons that exhibit response patterns characteristic of negative prediction error or positive prediction error neurons. In their model, excitatory neurons are modeled with an apical dendrite compartment, specifically targeted by SOM interneurons. They demonstrate that excitatory neurons develop positive and negative error responses when compartment-specific E/I balance is established in their inputs (Hertäg and Sprekeler, 2020; Hertäg and Clopath, 2022). This E/I balance is achieved through a combination of excitatory, inhibitory, disinhibitory, and dis-disinhibitory pathways with balanced pathway strengths (see **Figure 5**).

In these circuits, few constraints on the interneuron inputs are required to ensure that prediction error neurons can emerge. However, the distribution of sensory inputs and their predictions does bias the ratio of negative to positive prediction error neurons that develop during learning. Specifically, when PV and SOM neurons are predominantly driven by feedforward sensory input, excitatory neurons are more likely to develop into negative prediction error neurons. Conversely, when VIP neurons are predominantly targeted by feedforward input, excitatory neurons are more likely to exhibit response patterns aligned with those of positive prediction error neurons.

E/I balance can also be achieved through inhibitory plasticity (Vogels et al., 2011). In this type of network, the emergence of negative prediction error and positive prediction error neurons also results naturally from the network's efforts to establish an E/I balance that generalizes to all regularly encountered inputs. The type of stimuli encountered during learning, whether predicted or unpredicted, can also influence and bias the ratio of negative prediction error and positive prediction error neurons (Hertäg and Sprekeler, 2020; Hertäg and Clopath, 2022). Notably, positive and negative prediction error neurons can also emerge in a recurrent network from brief disruptions of E/I balance (Asabuki et al., 2025) without the need to incorporate a dendritic compartment.

E/I balance may also be maintained across longer time windows, with deviations from the balance generating fluctuating dynamics and rhythmic activity. Recent work by Lee and colleagues shows that balance-based interactions between separate subnetworks of excitatory and inhibitory neurons can lead to the emergence of rhythmic fluctuations, and the preservation of balanced and stable neural representations over longer time scales (Lee et al., 2025). Such rhythmic fluctuations may mechanistically underlie gamma oscillations observed in cortex (Spyropoulos et al., 2022), and could be involved in frequency-dependent hierarchical communications conveying bottom-up prediction errors and top-down signals, as described in **Section VII** (Bastos et al., 2015a; Mejias et al., 2016).

Separately, to reconcile the role of predictions in learning and recall, Barron et al. proposed a conceptual model in which different inhibitory subtypes enable top-down predictions to either inhibit or modulate the responses of pyramidal neurons, depending on their precision (Barron et al., 2020). Specifically, the model suggests that SOM interneurons channel the direct suppressive effect of predictions required to generate prediction error signals. In contrast, VIP-mediated disinhibition enables predictions to precisely modulate neural responses for memory recall. Relatedly, divisive inhibition by PV neurons has been suggested to enable the modulation of prediction errors based on prediction uncertainty (Wilmes et al., 2025).

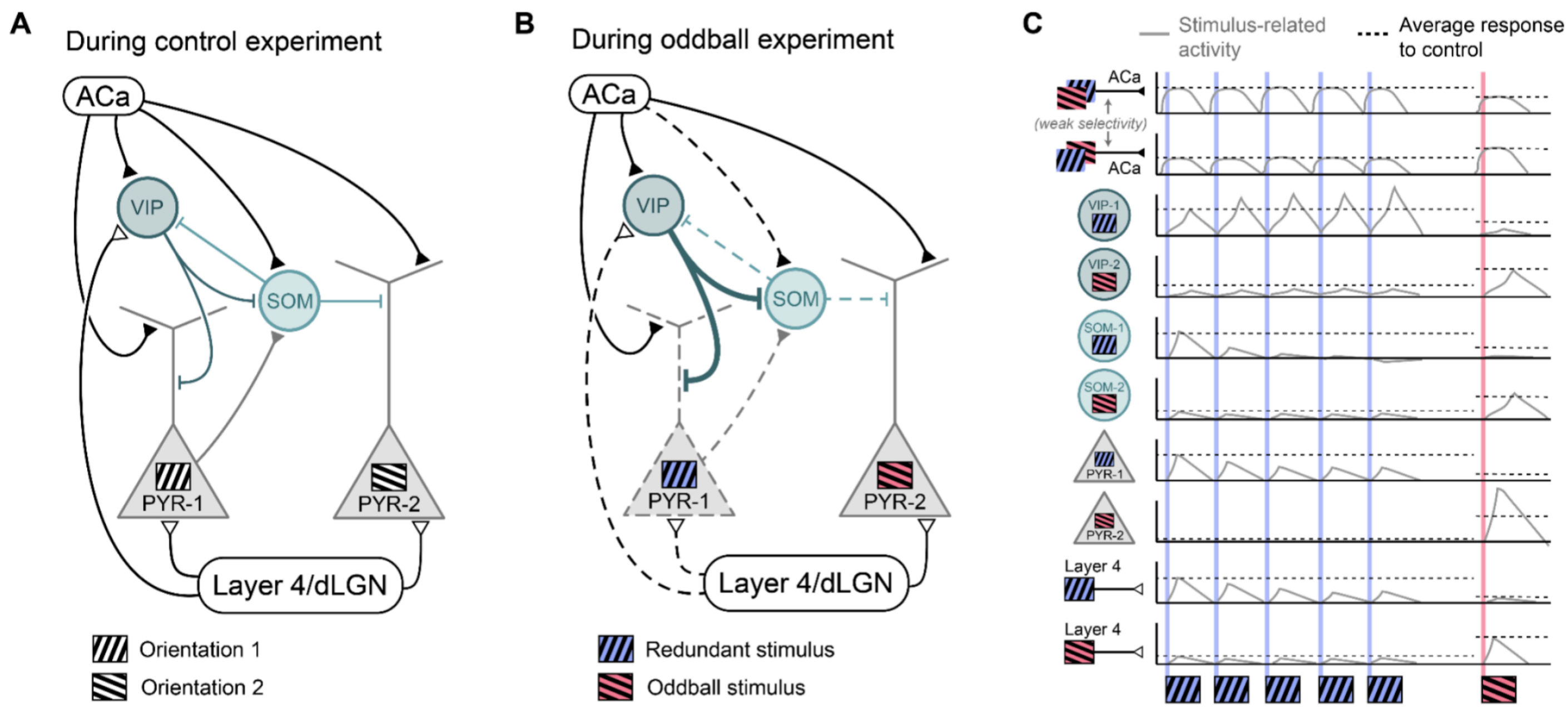

**Figure 4 - Summary figure highlighting potential key pathways underlying sequential oddball responses.** **A**. Simplified circuitry in layer 2/3 of V1. Uniform line thicknesses are used to represent the baseline (or unadapted) relative excitability of all neuron types when a many-standards control sequence is presented, in which all orientations of a visual stimulus are equally likely. Certain connections (e.g. between VIPs and PYR-2) and cell populations (e.g. orientation selective SOMs) are left out for clarity. **B**. State of the circuit during a basic visual oddball sequence. Neurons with increased output line thicknesses are more excitable, while neurons with dotted output lines exhibit stimulus specific adaptation (decreased excitability for their preferred orientation). The circuit state depicted here best corresponds to the state immediately before an oddball stimulus is presented. **C**. Activity profiles during the oddball sequence. Lines represent activity traces of different cell types with different orientation preferences (i.e., preferring the redundant or the oddball stimulus). Horizontal dotted lines depict the maximum response level of a cell population to the same stimulus orientation, but in the control context (panel **A**). In summary, top-down modulation bolsters VIP neuron response to the redundant stimulus, leading up to the oddball. This serves to strongly inhibit SOM neuron activity, releasing PYRs selective to non-redundant orientations from tonic inhibition and increasing their excitability when the (preferred) oddball orientation arrives. Adapted from Bastos et al., *Cell Rep* 42:113133, 2023. Originally licensed CC BY-NC-ND 4.0; modifications made with author permission.

Recent modeling work has further shown that experimentally observed cell-type-specific novelty responses (Garrett et al. 2025) emerge naturally from optimizing a combination of computational objectives, including prediction error minimization, reward prediction error minimization, and energy efficiency (Sharafeldin and Choi, 2025). In the proposed circuit model, a subpopulation of SOM neurons receives feedback signals and encodes expectations, subtractively inhibiting excitatory neurons so they can compute prediction errors. VIP neurons encode the uncertainty of predictions, modulating prediction errors both directly and indirectly via the VIP-SOM disinhibitory circuit to shape input representations in pyramidal neurons. Finally, PV neurons encode the inverse uncertainty in incoming sensory signals and divisively inhibit pyramidal neurons that receive feedforward sensory signals. When the circuit model is trained on naturalistic visual change detection tasks similar to those used experimentally (Garrett et al., 2020), distinct responses to local, global, and omission novelty emerge across the circuit. This suggests that local inhibitory circuit configurations alone can explain the emergence of novelty responses observed in sensory cortices without the need for hard-coded novelty detection mechanisms.

### Contribution of NDNF inhibitory neurons

In a recent model (Naumann et al., 2025), extending the interneuron circuit described above to include NDNF interneurons, revealing the effects of competition between SOM- and NDNF-mediated dendritic inhibition on pyramidal neuron activity. By operating over longer timescales than SOM interneurons (Abs et al., 2018), NDNF interneurons can shift dendritic inhibition from fast to slow timescales, modulating information flow in pyramidal neurons. In this model, NDNF neurons are hypothesized to release ambient GABA into L1. Being confined to the superficial layers while sparing deeper cortical layers, this inhibition primarily activates slow GABA(B) receptors at SOM output synapses in L1. As a result, NDNF interneurons are able to counterbalance the inhibition they receive from SOM interneurons, resulting in a form of mutual inhibition that amplifies weak signals to NDNF interneurons. This model suggests that NDNF interneurons, with their unique properties, could be ideally positioned to influence the relative balance of bottom-up and top-down inputs to excitatory pyramidal neurons.

## 3. Divergence and convergence between experiments and theories

Experimental findings and models converge on the idea that different inhibitory interneurons (e.g., PV, SOM, VIP, and NDNF) have distinct roles in processing feed-forward and feedback inputs. PV neurons are consistently associated with maintaining E/I balance and encoding feed-forward inputs, while VIP and NDNF neurons are more involved in integrating top-down feedback.

Experimental evidence also consistently supports the involvement of VIP and SOM neurons in encoding prediction errors. Numerous studies have shown that these neurons distinguish between predictable or familiar and novel stimuli, and that suppressing their activity affects prediction error signaling. This aligns with theoretical models suggesting roles for them in predictive processing and error signaling. The disinhibitory effect of VIP neurons on pyramidal neurons, which enhances sensory processing during mismatches, also supports models proposing that inhibitory subtypes modulate predictive processing through complex inhibitory circuits. In addition, recent work established that activity in VIP interneurons is essential for the computation of cognitive prediction errors in the ACC in a task-switching paradigm(Cole et al., 2024).

However, there are conflicting reports on the activity patterns of SOM neurons in response to predicted versus novel stimuli. Some studies show decreased activity in response to predicted stimuli, while others report reduced spiking in response to novelty. This discrepancy suggests the existence of different SOM neuron subtypes with distinct roles, which current models may not fully capture. Similar discrepancies in reports exist for VIP neurons, with a potential link to their role in integrating motor inputs. Future models could investigate potentially distinct roles for subpopulations of SOM and VIP neurons to account for their divergent roles in encoding novelty and prediction errors. It is also possible that the predictability of a stimulus based on local features plays a significant role in shaping SOM neuron responses. For instance, for homogeneous stimuli like grating patterns, local visual receptive fields can reliably be used to predict distant receptive fields. This predictability is much lower for natural images due to their complex structure. This difference in stimulus predictability may help explain the distinct roles of SOM neurons observed, with grating stimuli providing higher baseline predictability and potentially eliciting different network responses compared to natural images (Uran et al., 2022).

The temporal dynamics of E/I balance and the specific roles of different inhibitory neurons in balancing feed-forward versus recurrent inputs also remain areas of divergence. Experimental studies show variability in how rapidly different interneurons (e.g., PV versus SOM neurons) achieve balance. Incorporating the variability in the temporal dynamics of E/I balance maintained by different interneurons could significantly affect the dynamics observed in models of neural processing.

The role of NDNF interneurons in predictive processing is still poorly understood, with experimental evidence suggesting an involvement in long-lasting inhibition and top-down input integration. In terms of their potential role in predictive coding, theoretical models are ahead of experimental findings. Experiments measuring NDNF responses to mismatch stimuli are needed to better refine and constrain these models.

Each interneuron type likely cannot be studied in isolation as they are part of a highly integrated network. Given how heavily interconnected PV, VIP and SOM neurons are, it is unlikely that they perform neatly independent functions. Instead, they likely implement their computations cooperatively. To allow our understanding of this intricate network to converge, a close collaboration between experiments and theory is needed.

# V. Dendritic computations with apical dendrites

Dendrites are complex neuronal compartments that greatly expand the computational repertoire of individual neurons. Some predictive processing theories have postulated specific roles to dendrites. In this section, we review experimental evidence and related models that examine how dendrites, and in particular the apical dendrites of pyramidal neurons, contribute to predictive processing.

## 1. Experimental evidence

### Properties of apical dendrites

How a neuron integrates the inputs it receives is heavily influenced by its dendritic structure. Pyramidal neurons are notable for having two distinct sets of dendrites: basal and apical dendrites. Whereas the basal dendrites of pyramidal neurons extend out from the cell body, the apical dendrites are connected via the apical trunk, a thicker dendrite which extends toward the pia and branches into distal dendrites to form the apical tuft (Larkman, 1991). In the cortex, the apical tuft of L2/3 and L5 neurons extends into L1 where it is innervated by feedback connections from other cortical regions (Schuman et al., 2021; Young et al., 2021).

Research into the electrophysiological properties of pyramidal neurons has shown high compartmentalization of activity in the apical tuft dendrites, particularly in L5 pyramidal neurons (Larkum et al., 2022). However, the presence of voltage-gated ion channels in the apical trunk enables strong depolarizing events such as dendritic spikes to be triggered. Experiments have also shown that strong depolarization events localized to the cell body, like action potentials, can backpropagate up the apical trunk. When these backpropagating action potentials are coordinated with activity in the apical tuft, they can generate long-lasting depolarization events known as dendritic plateau potentials (Larkum et al., 1999; Antic et al., 2010; Hay et al., 2016). These nonlinear events involving the apical tuft have been proposed to allow pyramidal neurons to perform complex computations, such as detection of coinciding inputs, multiplexing (Hay et al., 2016; Naud and Sprekeler, 2018) and functioning as XOR gates (Gidon et al., 2020). Notably, dendritic plateau potentials can induce long-term synaptic changes, supporting a role in learning-related plasticity (Holthoff et al., 2004; Sjöström and Häusser, 2006; Hardie and Spruston, 2009; Gambino et al., 2014; Cichon and Gan, 2015; Mateos-Aparicio and Rodríguez-Moreno, 2019). In fact, as discussed in **Section VI**, the bursts of action potentials at the cell body that accompany dendritic plateau potentials have been shown to produce large synaptic changes at the basal dendrites, potentially enabling rapid new learning (Gordon et al., 2006; Bittner et al., 2017; Schiller et al., 2018; Caya-Bissonnette et al., 2023).

There is some evidence that apical dendrite activity is highly coupled to cell body activity, even outside of large plateau events (Beaulieu-Laroche et al., 2019; Francioni et al., 2019). However, the level of coupling observed drops off heavily as distance and branching complexity increase, supporting the conception of apical dendrites as a computationally distinct compartment in pyramidal neurons (Hill et al., 2013; Francioni et al., 2019; Kerlin et al., 2019; Landau et al., 2022). In addition, distal dendrites generate about ten times as many sodium spikes as the soma, further indicating that local computations within the dendrite are likely largely decoupled from the somatic spiking (Moore et al., 2017). Overall, it is important to note that dendritic events are diverse and that their characteristics vary across cell types, likely conferring distinct processing properties to the dendrites of different neurons (see (Larkum et al., 2022), for a review).

### Predictive inputs to apical dendrites

Supporting the role of apical dendrites in error computation, axons carrying motor and predictive visual signals to V1 specifically target distal apical dendrites in L1 (Leinweber et al., 2017). Similarly, axons from the prefrontal cortex to L1 of V1 carry stimulus-specific predictions in mice trained to expect specific image sequences (Fiser et al., 2016). In a multisensory task where an auditory cue predicted a visual stimulus, it was shown that axons arriving in L1 of V1 from the auditory cortex increasingly encoded information about the cued visual stimulus across learning. Conversely, V1 responses measured at the cell bodies of L2/3 neurons showed increased stimulus-specific suppression across learning (Garner and Keller, 2022). Importantly, optogenetic silencing of these incoming axons reinstated responses to the cued visual stimuli, indicating a role for top-down projections to dendrites in predictive suppression. In contrast, a study in parietal cortex found that movement was encoded anticipatorily in cell bodies, but not in putative apical dendrites (Moore et al., 2017).

### Stimulus responses in apical dendrites

Task-relevant stimulus selectivity increases in the L5 tuft dendrites of the barrel cortex when mice are trained on a discrimination task, but not when they are merely exposed to the stimuli (Benezra et al., 2025). This increase in selectivity persists after training, pointing to a role for apical dendrites in learning to encode task-relevant stimuli in primary sensory cortices. Gillon et al. studied how apical dendrites responded to unexpected stimuli by presenting mice with repeating image sequences, featuring occasional oddballs (Gillon et al., 2024). They measured calcium activity over several days in the cell bodies or apical dendrites of L2/3 and L5 pyramidal neurons. Consistent with previous studies, they found L2/3 and L5 neurons that responded selectively to the oddball stimulus (around 20-30%, in the first session). With experience, oddball selectivity decreased across cell bodies but increased across dendrites. Notably, the least selective dendrites in the first session tended to show the greatest increase in selectivity by the second session. Gillon et al. also studied responses to a passively viewed visual flow stimulus and found that a different oddball response pattern emerged. Oddball responses were primarily found in L2/3 neurons, consistent with previous findings that neurons sensitive to visuomotor disruptions are scarcer in L5 than in L2/3 (Jordan and Keller, 2020). Unlike for the image sequence oddballs, visual flow oddball responses increased across sessions not only across L2/3 cell bodies but also across L2/3 apical dendrites. Together, these results indicate that apical dendrites can develop oddball responses with experience either in coordination with or independently of the cell body population, depending on the type of stimulus being presented.

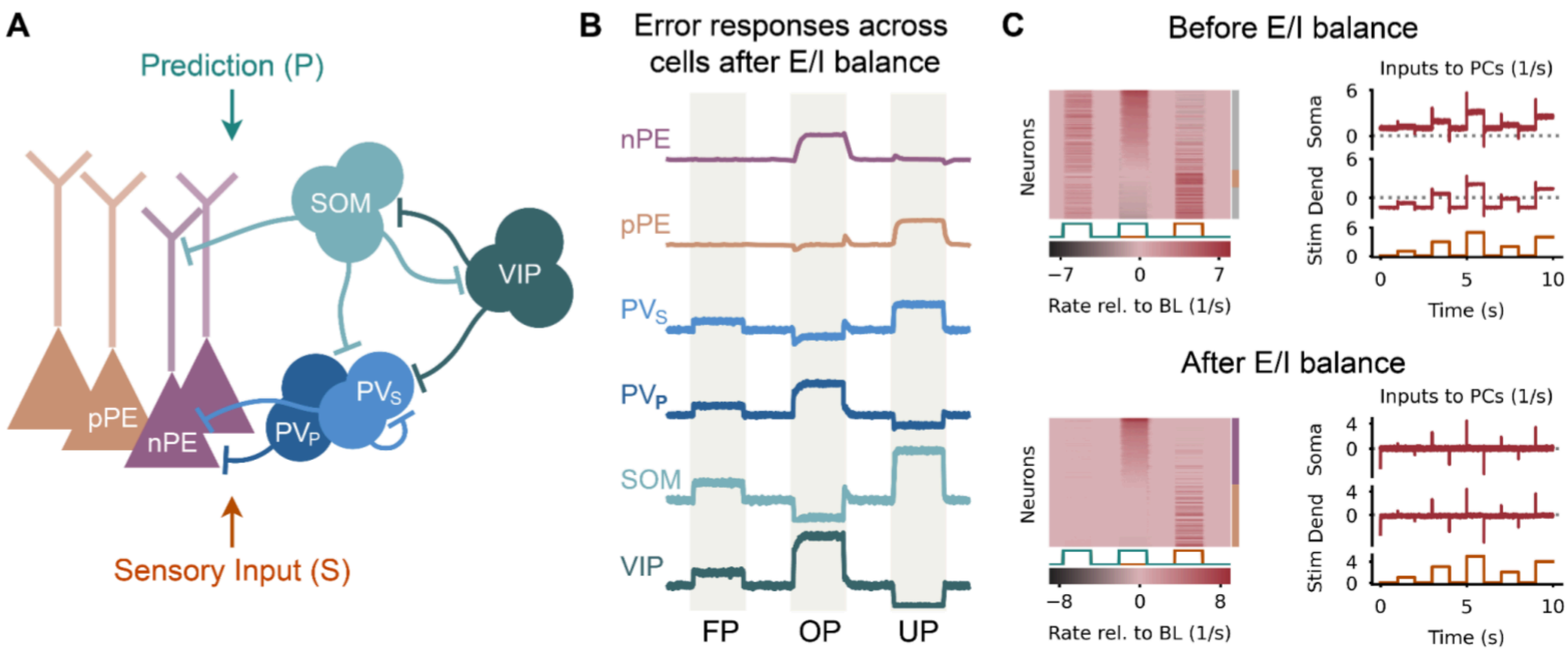

**Figure 5 - A canonical microcircuit with negative and positive prediction error neurons. A.** Negative Prediction Error neurons (nPE) and positive Prediction Error neurons (pPE) are modelled as two-compartment pyramidal cells, with predictions arriving top-down and sensory input arriving bottom-up. The microcircuit includes two types of soma-projecting PV interneurons: $PV_S$ driven by sensory input, and $PV_P$ driven by prediction, along with dendrite-projecting SOM interneurons and VIP interneurons (driven by both sensory input and prediction). **B.** Mean-field rate-responses of the different cell populations during fully predicted (FP) inputs (where prediction equals sensory input), overpredicted (OP) inputs (where the prediction is stronger than the sensory input) and for underpredicted (UP) inputs (where the prediction is weaker than the input) **C. Left:** Rate responses of the excitatory neurons before and after learning during the three phases (see B). Green and red traces above the color bars show prediction and sensory inputs, respectively. The vertical colored bar on the right shows the classification of each excitatory neuron as either an nPE neuron (purple), a pPE neuron (orange), or unclassified (gray). **Right:** Total inputs to the somata and dendrites of pyramidal cells during the presentation of FP stimuli (used in learning). Before learning (top), the total input is unbalanced (i.e., excitatory and inhibitory inputs do not cancel). After learning (bottom), inputs are balanced (excitatory and inhibitory inputs cancel), except for brief onset and offset responses. Adapted from Hertäg & Clopath, *Proc Natl Acad Sci USA* 119:e2115699119, 2022. Originally licensed CC BY-NC-ND 4.0; modifications made with author permission.

## Error encoding in apical dendrites

Francioni et al. used a brain-computer interface (BCI) paradigm to probe the role of apical dendrites in error signaling (Francioni et al., 2025). In their experiment, the orientation of a visually displayed grating was controlled by the activity of eight neurons in the retrosplenial cortex. With training, mice learned to shape the activity of these retrosplenial cortex neurons to rotate the grating toward a rewarded target orientation. Simultaneous calcium recordings near the apical tuft revealed the presence of neuron-specific error-like responses in the apical dendrites. Specifically, apical dendrites responded differently to failed versus successful trials, and their responses also differed based on whether a neuron's BCI role was to push the grating clockwise or counterclockwise.

## 2. Relevant theoretical models

In computational neuroscience modeling, including deep learning, neurons are often approximated as single integration site units without dendrites, in which inputs from all synapses are typically integrated linearly, as if all arriving at a single location. In contrast, in studies looking at the role of apical dendrites in predictive coding, the apical tuft is typically modeled as a separate compartment that is nonlinearly connected to a basal compartment. Notably, the basal compartment often comprises both the cell body and the basal dendrites, but it can also be divided into two separate compartments. In these models, and based on experimental evidence (Harris and Mrsic-Flogel, 2013), the basal compartment typically receives feedforward inputs, while the apical compartment is targeted by feedback inputs from higher-order areas.

### Initial models integrating apical dendritic compartments

Early models of apical dendrites explored various computations they might perform (Poirazi and Papoutsi, 2020). For example, apical dendrites could enable neurons to solve nonlinear classification tasks like the XOR operation (Körding and König, 2001; Payeur et al., 2021) or solve nonlinear classification tasks typically performed by deep neural networks (Bicknell and Häusser, 2021). Pyramidal neurons with apical dendrites have also been proposed to multiplex and demultiplex incoming information with bursts and single action potentials potentially carrying different types of information (Naud and Sprekeler, 2018). Additionally, apical dendrites may play a role in higher cognitive functions such as attention (LaBerge, 2005) and conscious information processing (Spratling, 2002). Notably, however, the apical dendrites of L2/3 and L5 neurons differ morphologically, endowing them with different properties (Larkum et al., 2007). Designing models that incorporate cell-type specific properties might reveal different functions for apical dendrites across layers.

### Predictive coding in apical dendrites

In the context of predictive coding, the observation that apical dendrites are primarily targeted by top-down connections of higher-level areas (Harris and Mrsic-Flogel, 2013) suggests they might receive top-down predictions or related signals. Although models of predictive processing can readily be built without dendrite-like compartments (Rao and Ballard, 1999), introducing these compartments can help address important biological plausibility limitations. For example, apical dendrites could regulate the gain of pyramidal neuron activity based on the precision of feedback predictions, as described in Section III (Shipp, 2016). General support for this idea comes from the involvement of L1 in such diverse processes as attention and arousal (Schuman et al., 2021). Alternatively, in Hertäg et al.'s circuitry model described in **Section IV**, which incorporates PV, VIP and SOM neurons, and an apical dendrite compartment, the latter receives predictive inputs, while stimulus input is targeted to the cell body and basal dendrite compartment (Hertäg and Clopath, 2022). As described above, the network is trained, through inhibitory plasticity, to achieve E/I balance in each compartment, and excess activity emerges in the dendrites or cell body during prediction errors (Figure 5). In this model, the apical dendrite compartments become sites of prediction error computation, comparing (excitatory) top-down predictions with (inhibitory) sensory signals, transformed by the interneuron network. Notably, the basal dendrite and cell body compartment is also a locus of prediction error calculation, but for the (excitatory) bottom-up sensory signals (**Figure 5**).

In contrast to the idea of error computation, some models focus on the ability of basal (bottom-up) and apical (top-down) input to cooperatively induce large, extended neural events, as described above (Anon, 2024). Since these large events only occur when both input streams match, they may be used for coincidence detection (Hay et al., 2016). Another hypothesis is that, with experience, apical dendrites learn to predict spiking at the cell body. This type of model, a dendritic predictive coding model, has been shown to learn simple supervised tasks using only local voltage-dependent plasticity rules (Urbanczik and Senn, 2014) (**Figure 6**). This type of model has also shown promise in associative memory formation, reinforcement learning, and temporal prediction tasks (Brea et al., 2016). In the context of hierarchical predictive coding, dendritic predictive coding may explain how neurons are able to align the predictive inputs received at their apical dendrites with the stimulus selective inputs received at their basal dendrites, such that the predictive signals can effectively be used as priors over the sensory signals (Mikulasch et al., 2023). Notably, if lateral inhibition is added between neurons, this type of model can also develop a form of biased competition (Spratling, 2008) which ensures that different neurons end up encode different stimulus characteristics.

Other work has shown that learning with predictive apical dendrites in a deep network can reproduce key features of predictive processing, like the emergence of neurons that preferentially respond to expected or unexpected stimuli (Zhang et al., 2025). This model differs from the previous examples as it is trained using energy optimization on a classification task, instead of an explicit predictive coding objective. A secondary somatodendritic mismatch loss is included, but unlike the previously discussed models, it is not the primary goal. Nonetheless, this work further illustrates the potential role of apical dendrites in predictive learning.

As the next section shows, efforts to scale biologically plausible learning to deeper networks with apical dendrites have largely focused on a different class of models, which we call dendritic error backpropagation models. With these models, the broad aim is to reproduce backpropagation-like learning with local learning rules (**Figure 6**).

### Apical dendrites and backpropagation

The backpropagation algorithm (i.e., in which errors are propagated backward through a network via the chain rule) used in deep learning has proven to be one of the most effective ways to train a neural network to perform complex tasks, like matching complex natural images to specific object classes (supervised learning), seeking out rewards (reinforcement learning) or uncovering hidden structure in data (unsupervised learning). However, as a model for learning in the brain it faces numerous bioplausibility challenges. In particular, the backpropagation algorithm requires information that is spatially nonlocal (e.g., encoded by synaptically distant neurons) and collected at different timepoints to be accurately distributed to all neurons in the network (Lillicrap et al., 2020).

The idea that apical dendrites might be key to addressing this problem has gained a lot of traction in the past decade. As mentioned above, Urbanczik and Senn demonstrated how learning in apical dendrites could help align prediction and stimulus streams during network training (Urbanczik and Senn, 2014). Building on this work, a class of dendritic error backpropagation models has emerged in which the express aim is to use apical dendrite compartments, and other biologically-inspired structures, to approximate backpropagation learning in the hopes of explaining how the brain is able to learn to perform highly complex tasks.

Like those described in the previous section, these models leverage predictive coding-like dynamics combined with Hebbian plasticity, and the idea that top-down connections primarily target apical dendrites (Millidge et al., 2022; Song et al., 2024). However, in contrast to dendritic predictive coding models, errors between sensory inputs and predictions are computed in the apical compartment, but then transmitted back to the cell body compartment where they drive learning (**Figure 6**).

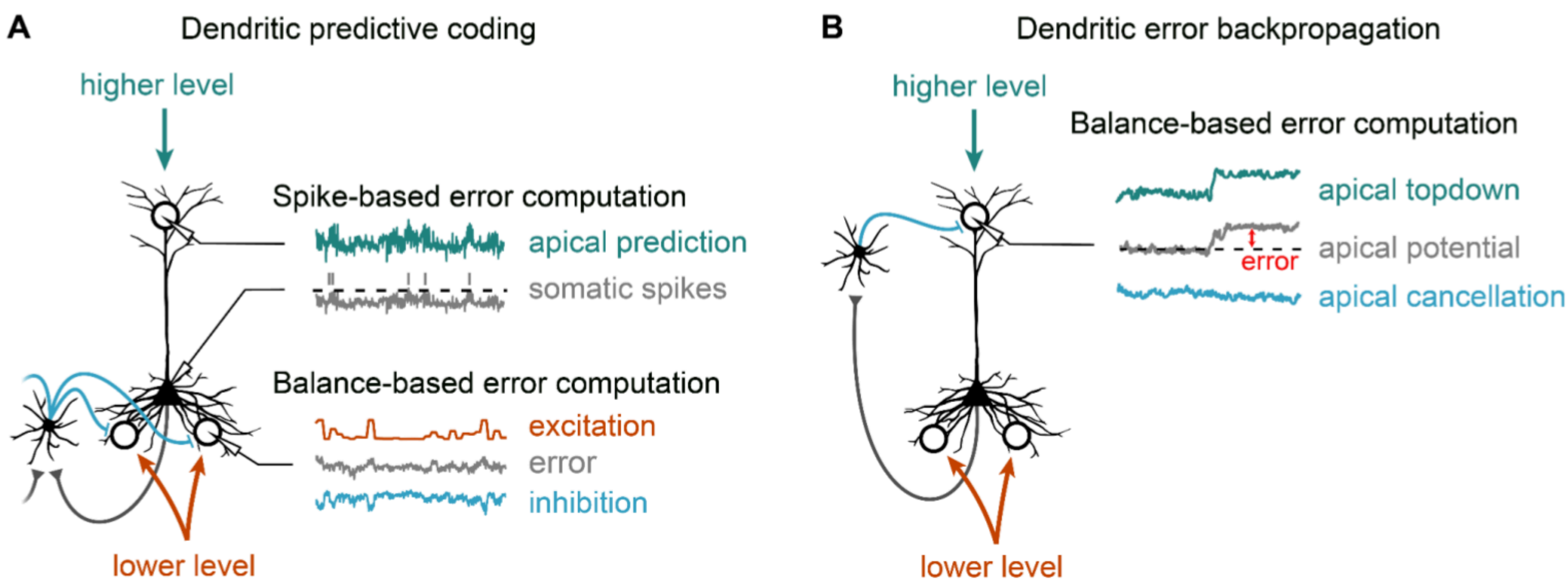

**Figure 6 - Different models of dendritic error computation. A.** Models of dendritic predictive coding (Mikulasch et al., 2023) propose distinct mechanisms of error computation in basal and apical dendrites. Apical dendrites compute errors as a mismatch between apical potential and somatic spiking, while basal dendrites compute errors as a mismatch between excitation and inhibition. **B.** Models of dendritic error backpropagation (Sacramento et al., 2018; Greedy et al., 2022) employ an inverted model of predictive coding, and thus engage in balance-based error computation in apical dendrites. An important distinction between these two models is whether learning balances apical or proximal dendritic potentials. Adapted from Mikulasch et al., *Trends Neurosci* 46:45-59, 2023, licensed under CC BY 4.0 and from Sacramento et al., *Advances in Neural Information Processing Systems*, 2018, with author permission.

The first dendritic error backpropagation models showed how a network of two-compartment spiking neurons trained with a local learning rule could learn to approximate backpropagation in a handwritten digit classification task (Guerguiev et al., 2017; Sacramento et al., 2018). The Burstprop model (Payeur et al., 2021) extends this model by incorporating additional biological observations: the distinct role of single spike and bursting events, the role of distal apical dendrites in generating bursts (Larkum et al., 2009) and the ability of connection-specific short-term synaptic plasticity (STP) to multiplex these signals through burst-dependent plasticity (Friedenberger et al., 2023). This allows neurons to separately transmit stimulus and error-related information, and learn continuously. However, as with Guerguiev et al.'s model, during training, the response to every input must go through two consecutive phases, impairing the model's biologically plausible. The first phase is needed to calculate baseline burst rates induced by the input, while the second phase reveals how the burst rate changes based on errors received from the top brain area. The difference between these two burst rates is fed to the learning rule, which approximates rate-dependent long-term synaptic plasticity as observed by (Sjöström et al., 2001).

Independently of Burstprop, Sacramento et al. showed how inhibition targeted to apical dendrites not only allows a network to learn continuously in time, but also removed the need for two phases (Sacramento et al., 2018). Lastly, Greedy et al.'s Bursting cortico-cortical Networks (BurstCCN) brings together features of both the Sacramento et al. model and the Burstprop model (Sacramento et al. and Burstprop), improving both biological plausibility and performance (Greedy et al., 2022). As in Sacramento et al.'s model, the inclusion of dendrite-targeting interneurons circumvents the need for multiple phases, and, as with Burstprop, connection-specific STP and burst-dependent plasticity enable continuous learning through multiplexing. Notably, BurstCCN predicts a key role for inhibitory interneurons in learning: specifically, that short-term synaptic facilitation at synapses of SOM interneurons is critical to reliably decoding burst events and thereby propagating accurate prediction errors across the cortex.

When designing biologically plausible algorithms to approximate backpropagation, it is challenging to ensure they can generalize to multi-layer networks, which are needed for more complex tasks (Richards and Lillicrap, 2019). Greedy et al. show in image classification tasks that, compared to previous models, learning with BurstCCN across multiple layers is better aligned to backpropagation. Nonetheless, more work is needed to enable dendritic error backpropagation models to approach the performance of backpropagation on complex learning tasks.

In summary, a variety of models have been developed to leverage the potential computational power of apical dendrites. These can be broadly categorized into two categories: dendritic predictive coding models and dendritic error backpropagation models (see **Figure 6**). In addition to requiring different synaptic plasticity rules, a core difference between these classes of models is where the error is computed. In the former, errors are computed at the cell body, whereas the latter presents a sort of 'inverted' predictive coding in which errors are computed at the apical dendrite (Whittington and Bogacz, 2019).

## 3, Divergence and convergence between experiments and theories

Experimental findings and theoretical models increasingly support the idea that apical dendrites play a crucial role in complex neural computations and learning processes. Both suggest that apical dendrites integrate feedback inputs and contribute to the generation of nonlinear events such as dendritic spikes and plateau potentials, which may be essential for functions like predictive coding and error computation. For instance, the experimental evidence shows that task-relevant selectivity can emerge in apical dendrites during learning tasks (Benezra et al., 2025; Francioni et al., 2025) or through passive viewing (Gillon et al., 2024), and suggests that apical dendrites may encode neuron-specific error signals (Francioni et al., 2025). Similarly, models incorporating apical dendrites have been shown to enable biologically plausible networks to perform more complex learning tasks (Guerguiev et al., 2017; Sacramento et al., 2018; Greedy et al., 2022).

Despite these convergences, significant gaps remain between experimental observations and theoretical models particularly regarding the site of error computation. Specifically, although this information could significantly help narrow down the space of biologically plausible dendritic networks, it remains unknown whether or not prediction errors are encoded in apical dendrites. In addition, while models like Burstprop (Payeur et al., 2021) and BurstCCN (Greedy et al., 2022) assume that burst activity and apical dendritic potentials constitute key learning signals, direct experimental evidence linking these predictions to observable learning processes remains limited.

Lastly, dendritic arbors are highly complex and compartmentalized (Larkum et al., 2022). Theoretical models built using only two or three-compartment neurons are thus very likely to heavily oversimplify and underestimate the complex computations dendrites can engage in. A better understanding of activity at the sub-dendritic level during predictive processing is needed to further illuminate the computational advantages these complex structures offer.

# VI. Synaptic plasticity and learning dynamics

Predictive responses result from dynamic learning processes occurring between neurons. In this section, we review experimental and theoretical literature exploring various learning rules underlying predictive processing. Rather than covering the extensive literature on synaptic learning, we focus specifically on studies relevant to predictive processing.

## 1. Experimental evidence

In general, predictive coding models are too detailed and context-dependent to be genetically encoded. Therefore, these models must be acquired through experience-dependent processes. Although many mechanisms can alter cellular excitability, changes in synaptic weights have traditionally been the main focus for learning. The literature on synaptic plasticity is vast, revealing a wide variety of synaptic plasticity rules across neuronal cell types and sub-neuronal compartments.

### Results from *in vivo* chronic recordings across days

Various experiments have shown that predictive spatiotemporal representations can form over days. In V1, passive multi day exposure to a sequence of gratings causes sequence-evoked LFP potentiation that decreases when the same images are shown with an unexpected order (Gavornik and Bear, 2014). The potentiation was highly specific for the timing of stimulus presentation (see **Figure 7**), with small changes in element duration causing evoked response magnitudes to drop significantly, and evoked-like responses appearing when an expected, but omitted, element transition would have occurred. The same paradigm also revealed sequence-specific latency shifts occurring in ACC in parallel with changes in V1 (Sidorov et al., 2020). Subsequent work recording single unit activity electrophysiologically (Price et al., 2023) or via calcium imaging (Knudstrup et al., 2025) confirmed that cells at the layer 4/5 border and in superficial layers are significantly modulated by learned expectations about when a stimulus will occur. These multi-day experiments differ from in-session oddball responses which are insensitive to the relative timing of sequential expectations (Knudstrup et al., 2025).

Fiser and colleagues trained mice to run down a virtual corridor for a reward at the end while a sequence of two images ABAB was presented on the walls of the corridor (Fiser et al., 2016). Over 4 days of training, a subset of neurons developed responses that predicted the occurrence of an image and occurred at or before the image was presented, while most other neurons responded with a typical latency. Supporting the idea of predictions being fed back down the cortical hierarchy, predictive responses also emerged in axons from the anterior cingulate cortex (ACC) measured in V1. In addition, the omission of an expected grating evoked a strong response in a small subset of V1 neurons. Similarly, Leinweber et al. trained mice on a ball in a virtual reality environment to associate wide-field image motion, resulting from their own movement on the ball, that was in the opposite direction as would occur in real life (Leinweber et al., 2017). Training lasted until mice reached a criterion performance in running down the virtual corridor, taking up to 10 days. Before training, activity of ACC axons in V1 correlated more with ipsiversive turns (which cause higher velocity optic flow) than contraversive, but after training, the correlation reversed its direction, corresponding to a bias in the ipsiversive direction for the new trained visual experience.

As described in **Section V**, Gillon et al. passively exposed mice to oddball sequences (Gillon et al., 2024). Over three days of testing, they found that the oddball response to Gabor spatial images increased in L2/3 and L5 dendrites and decreased in L5 somas. For visual flow reversals, they found increased oddball responses in L2/3 somas and dendrites along with relatively little change in L5 neurons. Together, these results demonstrated significant changes in oddball responses across multiple days, although some cell types and compartments showed increases while others showed decreases.

While multi-day predictive coding seems to be based on consolidated synaptic plasticity, and have been tied to M2 muscarinic acetylcholine receptors (Sarkar et al., 2024), it has been proposed that visual oddball deviants characterized using calcium imaging are a consequence of relatively simple and ubiquitous adaptation mechanisms (Homann et al., 2022). Evoked potentials measured in the LFP, however, have recently been shown to scale inversely with predictability in a manner inconsistent with a simple adaptation model (see **Figure 7**). While these accord with some computational models that anticipate response scaling based on event probability, it is not clear how to resolve the apparent contradictions between calcium imaging and LFP based studies. It is important to note that the LFP reflects incoming synaptic currents rather than local neural spiking (Einevoll et al., 2013).

### Experimentally observed synaptic plasticity mechanisms.

Given the observed fact that predictive neural responses can be learned, what are some of the synaptic plasticity rules that may account for them? Spike-timing plasticity was first reported in excitatory synapses in 1989 (Markram et al., 1997; Bi and Poo, 1998). After an explosion of research in the 1990s and 2000s, the literature has stabilized with a set of major review articles that are still current (Abbott and Nelson, 2000; Caporale and Dan, 2008; Feldman, 2012) that describe a variety of timing windows for LTP and LTD, depending on cell type, brain area, and species. In general, these rules involve potentiation for pre-post spike pairs within a ~10 ms causal temporal order (pre before post), and depression for spike pairs in an acausal order (post before pre) in time windows varying from 10 ms up to 100 ms or more. This rule is ideal for learning temporal sequences of neural activity and thus contributing to predictive processing.

E-to-I synapses exhibit potentiation in almost all cases and mostly require spike pairing with a causal order (Bannon et al., 2020). These rules have been observed to differ across types of inhibitory neurons, including fast-spiking (FS), low-threshold spiking (LTS), and SOM neurons (Bannon et al., 2020). These forms of plasticity are also effectively anti-Hebbian, as they strengthen disynaptic inhibitory feedback.

The classic long-term plasticity of I-to-E synapses involves potentiation for pre- and postsynaptic spikes paired within ±20 ms with no requirement of a causal temporal order (Woodin et al., 2003). Spikes paired at longer time intervals (±50 ms) instead trigger depression. This potentiation of inhibitory synapses is effectively anti-Hebbian, creating a form of negative feedback loop (Kilman et al., 2002; Woodin et al., 2003; Hartmann et al., 2008). This plasticity rule is well-known for its role in establishing E/I balance in the neocortex (Hartmann et al., 2008; Vogels et al., 2011). This E/I balance may also implement the learning and cancellation of temporally precise predictions (Herstel and Wierenga, 2021).

Subsequent experiments show great diversity of timing rules for I-to-E plasticity (Abbott and Nelson, 2000; Hennequin et al., 2017; Zappacosta et al., 2018; Capogna et al., 2021), such that experiments do not currently provide strong constraints on the assumptions that go into models of predictive processing. However, one important detail that is not typically incorporated into models of E/I balance is that plasticity in I-to-E synapses appears to require activation of nearby E-to-E synapses. This potentially changes the stability properties of the plasticity rules.

### Behavioral Time Scale Synaptic Plasticity

A specific form of synaptic plasticity called Behavioral Time Scale Synaptic Plasticity (BTSP) could play a role in predictive processing. BTSP was first observed at CA3 inputs to the basal dendrites of pyramidal neurons of hippocampal area CA1 *in vivo* within a handful of trials (Gordon et al., 2006; Bittner et al., 2017; Schiller et al., 2018; Milstein et al., 2021; Grienberger and Magee, 2022; Caya-Bissonnette et al., 2023; Fan et al., 2023). BTSP appears to occur when coincident activation of the apical dendrite tuft and the cell body gives rise to plateau potentials in the dendrites and bursting at the cell body. Characterized as a short-term non-Hebbian plasticity mechanism, BTSP is distinct from LTP, LTD or STDP, as presynaptic activity alters synaptic efficacies over hundreds to thousands of milliseconds. Specifically, unlike STDP, this type of plasticity enables presynaptic inputs that may be behaviourally relevant, but are neither directly causal nor very close in time to postsynaptic activation to be potentiated. Strikingly, the magnitude of the synaptic potentiation is quite large, sufficient for one or few-shot learning (Gordon et al., 2006; Bittner et al., 2017; Schiller et al., 2018; Milstein et al., 2021; Grienberger and Magee, 2022; Caya-Bissonnette et al., 2023; Fan et al., 2023). A BTSP-based learning rule appears ideal therefore to enable rapid learning of more temporally distant associations, potentially enabling predictions over longer time spans (Hamid et al., 2021).

It should be noted that plasticity rules observed on distal or apical dendrites can differ significantly in their strength and time window compared to the proximal dendrites or cell body (Gordon et al., 2006; Bittner et al., 2017; Schiller et al., 2018). In fact, detailed biophysical computational models showed that the same stimulus frequency that induces synaptic potentiation at the proximal dendrite would induce synaptic depotentiation at the distal dendrites, largely due to the dendritic distance-dependent attenuation of the back-propagating action potential (Kumar and Mehta, 2011). These findings seem to reinforce the intertwined relationship between the temporally-relevant BTSP rules discussed above (Gordon et al., 2006; Bittner et al., 2017; Schiller et al., 2018) and the spatially-relevant dendritic specialization.

### Depolarization-induced suppression of inhibition

Another prominent form of plasticity that is not typically included in predictive processing models is depolarization-induced suppression of inhibition (DSI) (Kullmann et al., 2012; Barberis, 2020). DSI is a short-term plasticity mechanism that involves disinhibition driven by a reduction of GABA release. In DSI, endocannabinoids bind to presynaptic EBC receptors (Kano et al., 2009). This retrograde message gives rise to a net enhancement of synapses on a time scale of ~60 seconds. This subset of enhanced synapses can serve as a “trail of breadcrumbs” to bias a neural network towards reactivating previous patterns of activity (Pang and Fairhall, 2019). Such a mechanism may be important for within-session phenomena like the oddball response and may be involved in multi-day plasticity experiments.

### Investigations of prediction-error related plasticity using BCI paradigms

Brain-Computer Interface (BCI) paradigms have provided a unique avenue to study “covert" learning. In BCIs, action potentials from individual neurons (invasive), local field potentials (invasive) or EEG signals (non-invasive) are decoded in real-time to directly control a disembodied agent within the task space (Oweiss and Badreldin, 2015). Beyond control in the task space, the decoded control signal can also directly stimulate other neurons to achieve a desirable functional outcome. For example, they can induce Hebbian-like plasticity between ‘trigger’ and ‘target’ neurons (Packer et al., 2015; Zhang et al., 2018) or between spatially/functionally distinct brain areas (Jackson et al., 2006).

BCI learning tasks demonstrate rapid internal model formation within a few trials (Badreldin et al., 2013; Clancy et al., 2014; Oweiss and Badreldin, 2015; Balasubramanian et al., 2017; Vaidya et al., 2018). Additionally, BCI paradigms also allow for the systematic manipulation of the internal model by altering the decoder coefficients and observing how the neural population changes its dynamics in response.

These features allow the prediction error to be rapidly and precisely measured, and demonstrate the power of the paradigm for studying mechanisms of learning and adaptation. Recent work suggested that BCI learning could be supported by BTSP within a few trials (Chueh et al., 2025). Future work could use BCI to study how prediction error representation within dendritic compartments and functional connectivity between neurons co-evolve together as a function of task learning.

## 2. Relevant theoretical models

In most predictive coding theories, predictions must be learned through plasticity mechanisms in the brain. In the classical theory of predictive coding (Rao and Ballard, 1999), a local Hebbian learning rule updates synaptic weights. However, translating such a theory into biological mechanisms remains challenging. First, synaptic weights could move between positive and negative values, violating Dale's law. In addition, the Hebbian learning rule relies on error neurons that can have positive and negative activity, which, as discussed before (see **Section III**), is incompatible with spiking neural activity. One possibility is to encode errors in deviations with respect to a baseline firing rate (Alonso and Neftci, 2021), but this is impractical considering the low firing rates in cortex.

As discussed previously, certain models avoid the implausibility of negative and positive activity in single error neurons by using different neural populations to compute positive and negative errors. In a model by Hertäg and Clopath, negative and positive error signals emerge through updates to inhibitory connections that optimize an E/I balance on post-synaptic pyramidal neurons with Hebbian-like plasticity (Hertäg and Clopath, 2022). However, in their model, certain weight updates depend on activity beyond the presynaptic and postsynaptic neurons, and are therefore not purely local.

To address this limitation, another model explored how positive and negative error signals can emerge under fully local learning rules guided by a global neuromodulatory signal, such as noradrenaline (Meng and Wang, 2026). This type of learning rule is often referred to as neo-Hebbian or three-factor learning rules, which have been reviewed in detail elsewhere (Gerstner et al. 2018).

Alternatively, the dendritic predictive coding hypothesis (see **Section V**) proposed two different learning mechanisms (Mikulasch et al., 2023). Apical dendrites might compute the error by comparing predictive input to activity at the cell body. In this case, the apical potential should ideally be predictive of spiking at the cell body (Urbanczik and Senn, 2014). This idea leads to a voltage-dependent plasticity (VDP) rule of apical targeting synapses that combines postsynaptic spiking and the apical potential. On the other hand, basal dendrites might compute the error by balancing bottom-up (excitatory) input with lateral (inhibitory) connections, learned via VDP (Denève and Machens, 2016; Mikulasch et al., 2021). To learn feed-forward weights to basal dendrites, the balanced membrane potentials can be exploited by another VDP rule combining postsynaptic spiking and the dendritic potential (Mikulasch et al., 2021). Another possibility is to learn feed-forward weights with Hebbian-like plasticity (Brendel et al., 2020), which however might result in faulty learning with inhibitory delays (Mikulasch et al., 2021).

Still, there are open questions about how these models relate to plasticity in the cortex. For example, hierarchical predictive coding algorithms, whether cellular or dendritic, often assume that feedforward and feedback weights are symmetric, which is also called the “weight transport problem” (Lillicrap et al., 2020). While this pre-established perfect weight symmetry is implausible, it has been shown that through an additional weight decay term, weights can be aligned sufficiently to enable learning without symmetry (Alonso and Neftci, 2021).

The plasticity rules we discussed so far describe plasticity of bottom-up and top-down connections in the microcircuit (or lateral inhibitory connections supporting their computations) in the context of hierarchical predictive coding (Rao and Ballard, 1999). However, predictions could also be generated as sequences within neural populations. Proposed plasticity rules that could enable this are based on Hebbian-like plasticity (STDP) (Kappel et al., 2014; Bouhadjar et al., 2022), VDP as proposed for apical dendrites (Brea et al., 2016; Millidge et al., 2024), error neurons (Millidge et al., 2024), or more complex gradient-based learning rules (Bellec et al., 2020; Saponati and Vinck, 2023). Many of these concepts for within-level learning are thus similar to what has been proposed for between-level learning.

## 3. Divergence and convergence between experiments and theories

Given the diversity of cell types and synaptic plasticity rules, it is unlikely that the core computations underlying predictive processing are entirely determined by a single cell type or plasticity rule. However, it may still be the case that particular cell types and plasticity rules make stronger or conceptually more significant contributions. In this light, we here contrast two broad categories of mechanisms.

The first category consists of mechanisms driven primarily by plasticity in inhibitory pathways. This approach is consistent with the predominant predictive coding concept of subtracting predictable information from the raw bottom-up sensory data. For the cortex to learn these subtractions, it must strengthen specific I-to-E synapses that represent predictable information and/or modify other synapses to amplify the activity of specific inhibitory neurons. In both cases, this plasticity must selectively shape the subtraction of accurately predicted information from the pyramidal neurons encoding that information.

The second category involves mechanisms primarily driven by plasticity within excitatory pathways. This hypothesis is motivated by the biophysical evidence showing that the E-to-E synapses, particularly those on the spines of pyramidal neurons, have significant potential for learning and encoding specific prediction. In particular, many E-to-E synapses are known to exhibit spike-timing-dependent plasticity (STDP), which strengthens synapses between neurons activated in a causal temporal order (Markram et al., 1997, 2011; Caporale and Dan, 2008; Feldman, 2012). This causal plasticity has been shown in models to enhance predictable information (Masquelier et al., 2009; Saponati and Vinck, 2023) and has been implicated in experiments involving similar computations (Mehta et al., 2000b; Yao and Dan, 2001; Yao et al., 2004; Saponati and Vinck, 2023). Predictive processing implemented in a single neuron leads to STDP-like kernels and efficient encoding and anticipation of temporal sequences (Saponati and Vinck, 2023). Of course, if plasticity in excitatory synapses plays a primary role, then plasticity in inhibitory pathways must still be involved. This is because the neocortex must maintain a proper balance between excitation and inhibition. Thus, if specific excitatory pathways are potentiated due to temporal correlations in the input to a local circuit, then inhibitory pathways will tend to be strengthened to maintain E/I balance (Vogels et al., 2011; Yang and Sun, 2018; Zhou and Yu, 2018). While this homeostatic inhibitory plasticity may play a strong role in shaping neural activity, it could be conceptually “secondary” or “reactive” and may exhibit less specificity than plasticity in excitatory pathways. Inhibitory pathways may also be strengthened for other purposes, such as learning multiple temporal sequences in different pyramidal ensembles (Masquelier et al., 2009). Arguably, this kind of inhibitory plasticity might not be viewed as secondary to excitatory plasticity but instead as equally important in allowing for multiple temporal sequences to be learned.

As reviewed earlier, LTP, LTD, STDP, BTSP, and DSI plasticity rules are well-supported by substantial experimental literature. However, the detailed mathematical form and parameters of these learning rules vary widely in this literature. Thus, translating these learning rules into integrated predictive coding models at the level of individual neurons or small neuronal networks remains an area of active research. There is notably a lack of clear principles on how these learning rules interact with the diverse mechanisms discussed in **Sections II**, **III**, **IV**, and **V**.

A key barrier may be the insufficient data on the neuronal distribution of individual ion channels, as well as limited insights into simultaneous spine, dendritic, and somatic activity. Recent advances in electron microscopy connectivity mapping offer hope that the necessary anatomical data will provide critical information on connectivity. However, these datasets will not address ion channel distributions. Both in vitro and in vivo studies are urgently needed to better constrain existing models and bridge these knowledge gaps.

## VII: From single neuron activity to inter-areal signal flow and whole-brain activity patterns

The oddball sequence paradigm originated in human research, where it has been widely used for almost fifty years to study brain responses to unexpected or deviant stimuli (Squires et al., 1975). While experiments in mammalian model organisms allow for the systematic study of neural activity at a single-cell level, techniques like Electroencephalograph (EEG) and magnetoencephalography (MEG) in humans have enabled researchers to capture large-scale neural changes in response to deviant stimuli. The study of ERP phenomena like the MMN and P300 responses remains prominent in both basic and clinical neuroscience, providing a noninvasive way to investigate human cognition and a reliable marker of neurological conditions like schizophrenia (Avissar et al., 2018; Mazer et al., 2024). In addition to these EEG and MEG studies, a large body of work in humans has used either ECoG or fMRI signals to study predictive coding, often using visual or auditory oddball (e.g., (Thomas et al., 2024)). While these signals offer better spatial resolution than EEG and MEG, they still fall short of single neuron resolution. While ECoG and higher field fMRI may reflect the local spiking activity in a given region, they also integrate this activity with incoming synaptic inputs, creating a combined signal (Logothetis et al., 2001; Schneider et al., 2021). Relating findings at the level of single neurons in animal models to meso- and macroscopic signals in humans is therefore challenging and requires modeling approaches, such as dynamic causal modeling.

Studies using techniques with high temporal resolution (ECoG, EEG, MEG) often focus on either oscillatory activity or event-related potentials. Oscillatory activity is divided into multiple spectral bands, which are each associated with characteristic circuit motifs and spatial scales (Buzsáki and Draguhn, 2004; Womelsdorf et al., 2014). A central question is whether the emergence of distinct rhythms can be explained as consequences from the recurrent dynamics implementing predictive coding, and whether distinct rhythms play specific functional roles in predictive processing. As oscillations often alternate in time with aperiodic transients, which are reflected in evoked potentials, another important question is the distinct contribution of these transients vs. oscillations. Several theories, reviewed in more detail below, have proposed functions for specific dynamics, e.g., transient or oscillatory dynamics, in predictive processing (Bastos et al., 2012; Singer, 2021; Vinck et al., 2024).

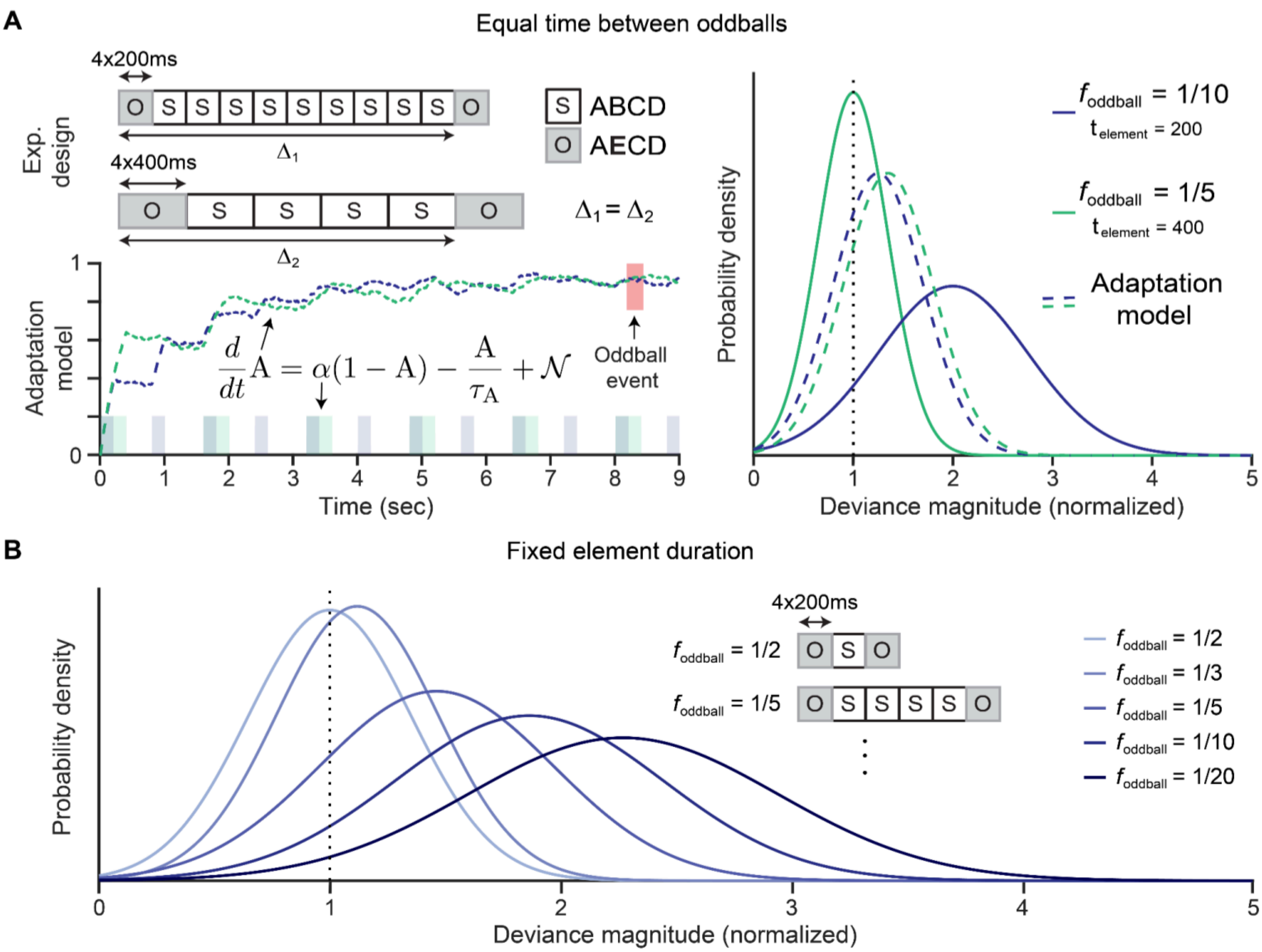

**Figure 7 - Evidence from the local field potentials in mouse V1 shows complex deviant responses that are inconsistent with a simple adaptation model. A.** Deviant responses can be evoked by occasionally replacing element B in a standard 4-element visual sequences (ABCD) with the oddball element E (AECD). By doubling the element duration ($t_{element}$) and halving the oddball frequency ($f_{oddball}$), it is possible to compare deviant response produced by oddballs occurring at the same temporal interval (Δ) but different frequencies. A simple illustrative saturating exponential model of neural adaptation (left, bottom) shows that the overall level of adaptation for cells responding to element A is approximately the same for both cases when the oddball is presented. In this simulated model, adaptation (A) increases with activity ($\alpha$, non-zero during time periods indicated by shaded regions), includes additive gaussian noise (*N*), and decays with a fixed time constant ($\tau_A$=15 sec, which is ~1/2 the decay constant estimated by Homann et al., 2022). If deviant responses scaled based on this adaptation value, their magnitudes would be approximately equivalent, as indicated by the dashed distributions (right). Instead, responses that occur less are often larger than those that occur more often, even though they share the same inter-oddball interval. **B.** Experiments varying the oddball frequency for a fixed element duration likewise show that deviance responses get larger as the oddball becomes less frequent. All distributions in this figure represent the average difference between standard and deviant LFP responses with parameters fit from data recorded in mice in Knudstrup et al. (2024).

## 1. Experimental Evidence

### Narrow-band gamma oscillations vs. broadband fluctuations

There are conflicting results and interpretations concerning gamma-band (30-80Hz) oscillations, which result from balanced interactions between excitatory and inhibitory neurons in the local microcircuit (Cardin et al., 2009). In LFP/EEG/MEG/ECoG studies it is crucial to distinguish between broadband fluctuations, which reflect enhanced spiking and/or synaptic activity, and narrow-band gamma oscillations which reflect synchronized activity in the 30-80Hz range. High-frequency gamma power is a common marker of firing rates used in human ECoG studies (Miller, 2019). In the early visual system, temporal (stimulus repetition vs. novel stimuli) and spatial stimulus predictability (spatially homogeneous stimuli) promote increased narrow-band gamma activity (e.g., (Vinck and Bosman, 2016; Peter et al., 2019, 2021; Shirhatti et al., 2022; Uran et al., 2022)) that shows an opposite correlation to broadband fluctuations and spiking rates. Uran et al. showed that gamma tracks predictability for natural images in a monotonic manner, and specifically reflects low-level visual predictability (Uran et al., 2022), consistent with the dendritic predictive coding hypothesis (Vinck et al., 2024) (see Section V, and Relevant Theoretical Models, Section VII). However, other studies have found enhanced gamma-band power for sensory (sequential) mismatches (e.g., (Arnal et al., 2011; Chao et al., 2018; Bastos et al., 2020; Gallimore et al., 2023; Xiong et al., 2024)).

One possible explanation for these contrasting findings on gamma-band power and oscillations is that there is a difference in the correlates of gamma-band activity across cortical areas or sensory modalities, or types of sensory mismatches (e.g., spatial vs. sequential, see Section I). It has also been proposed that these different findings on gamma-power reflect the distinction between narrow-band gamma oscillations vs. broadband fluctuations. Analyses indicate that in the auditory system, increases in gamma-power reflect broadband rather than narrow-band gamma fluctuations (Canales-Johnson et al., 2021). To understand these differences in the literature, decomposition techniques (Canales-Johnson et al., 2021; Gelens et al., 2024) on LFP signals as well as spike-spike and spike-LFP measures (Ray and Maunsell, 2011) are critical to distinguish narrow-band gamma oscillations from broadband fluctuations.

### Alpha/beta activity, and its relation with gamma

Several studies have observed that alpha/beta (10-30Hz) activity is suppressed with sensory mismatches (e.g., (Chao et al., 2018; Jiang et al., 2022), but see (Todorovic and de Lange, 2012; Nougaret et al., 2024)). One interpretation is that these findings support the theoretical model that alpha/beta oscillations are involved in the feedback transmission of sensory predictions (Arnal and Giraud, 2012; Bastos et al., 2012) (see Relevant Theoretical Models below). An alternative interpretation is that the suppression of alpha/beta sensory mismatches reflects a negative prediction error or results as a consequence of increases in firing rates (Chao et al., 2018; Vinck et al., 2024), noting that the suppression of alpha/beta with increased cortical activation is a phenomenon that occurs under many behavioral conditions (e.g., (Jensen and Mazaheri, 2010; Miller et al., 2012; Canales-Johnson et al., 2020a; Weisz et al., 2020), but see (Richter et al., 2019)). A general problem is that it is difficult to make inferences about prediction signals based on the observation of sensory mismatch signals, rather than sensory prediction signals directly (Chao et al., 2018; Vinck et al., 2024), which applies to firing rate correlates as well (see Section III-IV).

Several studies have observed an anti-correlation between alpha/beta and gamma power with sensory mismatches (Chao et al., 2018; Bastos et al., 2020; Lundqvist et al., 2020; Xiong et al., 2024) which may suggest a causal relation between these two variables: Either alpha/beta suppression causing enhanced gamma power in lower hierarchical levels (Bastos et al., 2020) (“predictive routing”), or enhanced gamma power in lower hierarchical levels leading to alpha/beta suppression in higher hierarchical levels (Chao et al., 2018). It is also possible changes in both frequency bands are driven by a third factor (e.g., attention). However some studies find positive correlations between narrow-band gamma oscillations and beta power (Richter et al., 2017, 2019). Furthermore, as reviewed above, several studies suggest narrow-band gamma oscillations are decreased for temporal and spatial sensory mismatches in the visual system. One factor that has been proposed to explain these discrepancies is the distinction between narrow-band gamma oscillations and broadband fluctuations (Vinck et al., 2024).

Causal studies are also crucial to test for relations between oscillatory phenomena in different frequency bands. A recent study used propofol to induce loss-of-consciousness in macaque monkeys. Propofol has been shown to reduce alpha/beta power, and spiking activity throughout cortex, with the reduction in spiking being more pronounced in frontal areas (Bastos et al., 2021). Counterintuitively, during an auditory oddball paradigm, sensory-mismatch-related gamma-frequency power and late-period (after 100ms post-oddball) spiking responses were enhanced (Xiong et al., 2024) The interpretation put forward by (Xiong et al., 2024) is that propofol induces a loss of inhibitory control mediated by top-down alpha/beta oscillations on sensory responses in lower hierarchical levels (Bastos et al., 2020) (consistent with predictive routing, see Relevant Theoretical Models below). In predictive coding terms, propofol would reduce predictive inhibition via alpha/beta oscillations, leading to increased prediction errors indexed by gamma power (i.e. assuming propofol does not affect gamma power via other pathways). Future work should perform more specific causal manipulations that act to increase or decrease alpha/beta and gamma to test their specific roles and their corresponding circuits to predictive processing.

### Functional connectivity studies.

Several studies have suggested stronger feedforward than feedback Granger-causal influences between LFP signals in the gamma-frequency range in visual cortex (Bosman et al., 2012; van Kerkoerle et al., 2014; Bastos et al., 2015a). Another study linked sensory mismatch signals with feedforward gamma-band synchronization (Bastos et al., 2020); see Relevant Theoretical Models below). The interpretation of these correlations at the level of neural signaling remains debated. One interpretation is that gamma-band synchronization provides an effective mechanism for feedforward communication, based on the idea that synchronization leads to effective summation of excitatory synaptic potentials (Fries, 2015). Another interpretation is that the feedforward gamma influences reflect a prevalence of high-frequency gamma power in early visual areas, consistent with a model in which the influences of gamma on postsynaptic target areas may remain largely subthreshold (Buzsáki and Schomburg, 2015; Schneider et al., 2021). Recent empirical studies suggest that gamma is particularly effective in driving fast-spiking interneurons rather than excitatory neurons in downstream areas (Buzsáki and Schomburg, 2015; Schneider et al., 2023; Spyropoulos et al., 2024), suggesting an alternative interpretation that gamma could dampen the transmission of predicted sensory information, rather than enhance the transmission of unpredicted sensory information (Vinck et al., 2024) (see Relevant Theoretical Models below).

Studies have also shown a stronger association of alpha/beta frequencies with Granger-causal feedback than feedforward influences in visual cortex (van Kerkoerle et al., 2014; Bastos et al., 2015a; Michalareas et al., 2016), auditory (Fontolan et al., 2014) and olfactory systems (Martin and Ravel, 2014; David et al., 2015), which may suggest that these frequencies subserve the communication of sensory predictions (see Relevant Theoretical Models below). A possible mechanism underlying these Granger causality findings is that alpha/beta signaling is particularly effective in driving e.g., apical dendrites or specific interneurons with slower kinetics (SOM interneurons, (Chen et al., 2017)). Another interpretation is that the relation between alpha/beta and feedback influences mainly reflects the distribution of power across cortical hierarchy, and that beta may also be associated with feedforward influences from intermediate hierarchical to higher hierarchical levels (Brovelli et al., 2004; Salazar et al., 2012; Vinck et al., 2024).

Causal approaches have begun to elucidate whether distinct frequency bands are specifically involved in feedforward or feedback transmission. Van Kerkoerle et al. show that electric stimulation (pulsed, at 200Hz) of V1 leads to enhanced approx. 60Hz gamma activity in area V4, while stimulation of V4 (pulsed, at 200 Hz) leads to increased alpha/beta activity in V1 (van Kerkoerle et al., 2014). This finding points to an association of gamma with feedforward transmission and alpha/beta with feedback transmission, although it remains to be established whether e.g., enhanced gamma in V4 due to V1 microstimulation results from local mechanisms in V4 or propagation of V1 gamma. Veniero et al. apply TMS in Frontal Eye Fields and find evidence for a beta reset in occipital areas, supporting an association of beta oscillations with feedback processing (Veniero et al., 2021). Another causal approach is to directly stimulate the feedforward or feedback pathway rhythmically and test for the differences in inter-areal propagation across different frequencies. For example (Schneider et al., 2023) (see also (Soula et al., 2023)) find that lower frequency inputs into V1 propagate effectively across layers and both excitatory and inhibitory neurons while high frequencies predominantly drive fast-spiking interneurons in the input layer 4, without propagation to layer 2/3.

In general, the interpretation of connectivity findings at the LFP level is complicated because of the influence of afferent synaptic inputs on LFP signals (Logothetis et al., 2001; Buzsáki and Schomburg, 2015; Pesaran et al., 2018; Schneider et al., 2021). and further studies are required linking LFP connectivity with measurements of cell-type-specific spiking activity. There is ongoing debate about the question whether inter-areal coherence is a mechanism for communication (CTC) or whether coherence is a consequence of communication (CTCOM) (Fries, 2015; Pesaran et al., 2018; Schneider et al., 2021), and the extent to which coherence can be explained by connectivity and power, or synchronization mechanisms (Schneider et al., 2021; Vezoli et al., 2021b). Regardless of the functional and mechanistic interpretations of connectivity measures like coherence and Granger-causality, they show very robust correlations with anatomical measures of connection strength and hierarchical distance (Bastos et al., 2015b; Vezoli et al., 2021a, 2021b).

### Distribution of rhythms and mismatch/prediction signals across layers

There is ongoing debate about the distribution of rhythms across cortical layers. This point is highly relevant for predictive coding because theories propose canonical functions for specific frequency bands in predictive processing, and have associated them with specific laminar compartments from which either feedforward (primarily superficial L3) or feedback projections (primarily deep L6) originate (Bastos et al., 2012).

A number of studies have concluded that gamma oscillations are stronger in superficial layers, while alpha/beta oscillations are stronger in deep layers (see (Bollimunta et al., 2008; Buffalo et al., 2011; Mendoza-Halliday et al., 2024). However other studies reported contradicting evidence with e.g., stronger alpha power in superficial layers and prominent gamma-band oscillations in deep layers (e.g., (Haegens et al., 2015; Halgren et al., 2019; Gieselmann and Thiele, 2022)) and there is ongoing debate about this point (Vinck et al., 2023; Mendoza-Halliday et al., 2024; Mackey et al., 2025) that revolves in part around various technical issues concerning LFP signals such as references, using unipolar vs. bipolar /current source density signals, and the link of LFP signals to unit activity. An additional level of uncertainty (which holds true for many spiking recordings as well) relates to the spatial density of the recordings made, with an inter-contact spacing often not reaching sub-layer resolution. There is a critical need for high-resolution laminar recordings due to the dual counterstream architecture showing that L2 is a feedback layer and L3 a feedforward layer (Markov et al., 2014; Vezoli et al., 2021a). Hence, the notion of a 'superficial' laminar compartment can easily mix-up feedforward and feedback layers (e.g., (Barzegaran and Plomp, 2022)). Computational modeling studies (e.g., (Lee et al., 2013)) will also contribute greatly to our understanding of how layer specific cell types (Glatigny et al., 2024; Lichtenfeld et al., 2024), along with their circuitry and connectivity (Campagnola et al., 2022) contributes to the formation of layer-specific oscillations.

The idea that different layers subserve distinct functional roles is further supported by recent human laminar fMRI studies suggesting that feedback to different laminar compartments contributes to distinct top-down generative processes (Muckli et al., 2015; Kok et al., 2016; Bergmann et al., 2024). Further translational efforts are needed to make the link between whole-brain laminar human studies exploiting perceptual reports and in-depth animal investigations across layers using similar paradigms (see Experimental proposals).

### Distribution of rhythms across areas

Hierarchical processing may be strongly influenced and constrained by the distribution of cortical rhythms across areas: Some studies and theories suggest that rhythms are canonical parts of cortical microcircuits across the hierarchy (Fries, 2009, 2015; Bastos et al., 2015a; Barzegaran and Plomp, 2022; Mendoza-Halliday et al., 2024). An alternative view is that the main axis of diversity in rhythms is not between cortical layers but rather between cortical areas (Vinck et al., 2023, 2024). Since there are major hierarchical gradients in excitatory recurrent connectivity and PV/SOM ratios, it is plausible that the characteristic frequencies of networks vary across the hierarchy (Wang, 2020). For example (Vezoli et al., 2021a) and (Hoffman et al., 2024) find evidence for frequency-specific networks and a high degree of diversity in rhythmic dynamics across the cortical sheet, where distinct rhythms are mainly expressed in specific networks (e.g., gamma oscillations in early visual cortex (Hoffman et al., 2024)), although there are significant functional connectivity links between these modules (Vezoli et al., 2021b). It remains to be further tested to what extent cortical gradients or a cortical hierarchy of timescales explains this diversity (see Relevant Theoretical Models below). It has also been proposed that the smaller receptive field sizes in early visual areas lead to increased redundancy and predictability of sensory inputs across space, thereby promoting gamma oscillations associated with E/I balance (Vinck and Bosman, 2016) (see Relevant Theoretical Models below)

Nonetheless, the diversity in dynamics needs to be addressed by theories that propose general functions for rhythms in cortical computation. While communication between areas sharing similar rhythms may depend on coupling between oscillations at the same frequency (Fries, 2015), evidence also points toward cross-frequency coupling (Bonnefond et al., 2017; Bastos et al., 2018; Márton et al., 2019; Esghaei et al., 2022). Other perspectives emphasize the non-linear nature and high-dimensional nature of neural communication (Singer, 2021; Vinck et al., 2023) which may require broadband communication rather than narrow-band phenomena (see Relevant Theoretical Models below). Such non-linear, broadband communication causes interactions across frequency bands rather than interactions within the same frequency band, as is the case with linear signal transfer (Vinck et al., 2023). For instance, recent work suggests that non-linear rather than linear (e.g., WPLI, Granger causality) connectivity measures capture differences in sensory inference during ambiguous perception (Canales-Johnson et al., 2020b, 2023), and that relations between sensory mismatch signals across the cortical hierarchy are information-synergistic (Gelens et al., 2024). Computational models suggest such synergistic relations between distributed error signals can result from non-linear recurrent dynamics between nodes (Gelens et al., 2024).

### Transients vs. oscillations

Event-related potentials are an effective way to measure transient, non-oscillatory activity. Increases in ERP amplitude for sensory mismatches have been shown in a large number of studies across species (e.g., (Todorovic and de Lange, 2012; Parras et al., 2017; Blenkmann et al., 2019; Gelens et al., 2024)). These increases ERP amplitudes are usually distinguished into different components, including the leading to the discovery of several error-like responses, including the mismatch negativity (MMN; peaking 150-250 ms post- stimulus onset) (Ford et al., 1976; Näätänen et al., 1978); the P300a, (250-350 ms) and the P300b (350-500 ms) (Chennu and Bekinschtein, 2012). Notably, these potentials, while influenced by attention, are often found and studied without participants being engaged in an active task (Chennu and Bekinschtein, 2012). The MMN was proposed to reflect an early “perceptual prediction error” (Friston, 2005), while the P300a is associated with attention orienting and the P300b, with context-updating and memory processing (Polich, 2007). Some links have been made between attention-dependent ERP signals noninvasively recorded in humans and their neurobiological sources. For example, the EEG event-related potential known as Selection Negativity, which is evoked by mismatch stimuli in both human and nonhuman primates, has been associated with increased activation in sensory cortical areas when attention is focused on the stimuli (Mehta et al., 2000a).

Empirical evidence suggests that sensory inference is rapid and occurs already at relatively early latencies (around 120-150ms, for review see (DiCarlo et al., 2012). Such rapid sensory inference is compatible with the emergence of mismatch signals at relatively early latencies (e.g., (Parras et al., 2017)) but poses challenges to the idea that oscillatory phenomena contribute to sensory inference via signaling predictions or prediction errors. The reason is that rhythms tend to be disrupted by transient activation of networks. Some studies suggest rhythms emerge at longer latencies after 100ms in the early visual cortex, well after the initial feedforward sweep (Gieselmann and Thiele, 2008). Furthermore, processing via rhythms requires integration of multiple cycles, which is especially a problem for proposals that e.g., alpha oscillations contribute to sensory inference (Vinck et al., 2024). However, an alternative perspective is that rhythmicity may already be expressed in stimulus-locked, evoked synchronization, or a phase-reset of ongoing oscillations. A more precise characterization of the temporal evolution of neural dynamics across cortical areas is required to elucidate these issues.

## 2. Relevant theoretical models

There have been several theoretical models proposed linking oscillations and transients to predictive processing.

### Gamma prediction error, alpha/beta prediction model

Arnal and Giraud, and Bastos et al. have proposed that the feedforward propagation of prediction errors depends on gamma-frequency oscillations in superficial layers of cortex while predictions rely on oscillations at alpha/beta frequencies (Arnal and Giraud, 2012; Bastos et al., 2012). This dual-frequency model relates to theoretical results for predictive coding models by Friston and Kiebel that predictions should be encoded at longer time scales than prediction errors (Friston and Kiebel, 2009). According to the Chao et al. model, updating of the internal model due to a sensory prediction error thus leads to a disruption of alpha/beta power in higher hierarchical levels (Chao et al., 2018). Bastos et al. propose that this disruption of alpha/beta beta power then leads to an increase of gamma-power in lower hierarchical levels (Bastos et al., 2020).

### Dual roles for transients and oscillations

Vinck et al. proposed that sensory prediction errors lead to sensory inference via transient activations that increase energy in a broad frequency range, and lead to signal propagation across cortical areas via non-linear, recurrent dynamics (Vinck et al., 2023, 2024). In this model, it is proposed that oscillations rather play a complementary role by stabilizing neural representations and to facilitate plasticity processes in the later phases of sensory processing.

### Gamma oscillations as a consequence of stimulus predictability and efficient coding

As reviewed in Section V, narrow-band gamma oscillations were proposed to systematically increase with the spatiotemporal predictability of sensory inputs in a local cortical region (Vinck and Bosman, 2016; Singer, 2021). A mechanistic explanation is that when sensory inputs match the predictions, there is an increase in E/I balance promoting the emergence of gamma oscillations, while leading to sparse coding (Mikulasch et al., 2023; Vinck et al., 2024). The recruitment of somatostatin interneurons by horizontal or top-down predictions may also play an important role (Börgers et al., 2008; Jadi and Sejnowski, 2014; Veit et al., 2017). Theoretical work further suggests that gamma oscillations reflect optimal sensory processing and arise in E/I balanced networks with transmission delays (Chalk et al., 2016; Echeveste et al., 2020).

### Oscillations and hierarchical time scales

A systematic increase in time-scales across the cortical hierarchy may reflect predictive coding in an hierarchical system, as the formation of predictions requires integration on longer time scales. Such increases in time-scales have been observed across macaque and human cortex (Murray et al., 2014; Gao et al., 2020) and may be linked to various hierarchical gradients as well as emergent dynamics due to inter-areal interactions (Chaudhuri et al., 2015; Gao et al., 2020). Such increased time-scales from tens of milliseconds to hundreds of milliseconds may be paralleled in differences in oscillatory behavior from gamma to beta to theta frequencies across the cortical hierarchy (Vinck et al., 2023), but see (Hoffman et al., 2024). In this perspective, beta oscillations may reflect processing at intermediate levels of the hierarchy, and could therefore reflect both bottom-up processing to higher hierarchical levels and top-down processing to lower hierarchical levels.

### Prospective coding via theta sequences

Predictive processing has also been linked to sequential firing at theta frequencies in the hippocampus. Here firing at different phases of the theta cycle reflects either encoding of the animal’s future or past spatial location (Dragoi and Buzsáki, 2006). As prospective neural coding may more generally rely on sequences that compress predictions of the future in the sequential activations of a neural ensemble, such sequences may be orchestrated by oscillations, either theta frequencies in ACC/hippocampus (Dragoi and Buzsáki, 2006; Womelsdorf et al., 2010) or gamma frequencies in visual cortex (Vinck et al., 2010). Such prospective coding of the future via temporal sequences however needs to be distinguished from hierarchical predictive coding models.

### Traveling waves in hierarchical predictive coding

A recent computational model suggests that oscillatory traveling waves, spanning multiple cortical nodes, at alpha-band frequencies naturally emerge in an hierarchically organized network performing predictive coding, due to the negative feedback loops that exist between nodes (Alamia and VanRullen, 2019). This theory predicts forward traveling waves when sensory evidence dominates inference, and backward traveling waves when priors dominate inference (Alamia and VanRullen, 2019). Such traveling wave phenomena may provide a mechanism for system bifurcations leading to effective feedforward or feedback propagation of information (Alamia and VanRullen, 2019).

## 3. Divergence and convergence between experiments and theories

The idea that aperiodic transients, which convey energy in a broad frequency range and can be accessed via event related potentials, play an important role in sensory inference by updating the internal model based on novel sensory evidence appears consistent with the empirical evidence above. Whether these transients convey information via firing rate codes or temporal sequences remains a topic of debate however, with recent work suggesting encoding of sensory information via temporal sequences during transients (Yiling et al., 2023; Xie et al., 2024; Sotomayor-Gómez et al., 2025) . In this context it is interesting to note that predictive coding is formulated in terms of rate coding, but does not offer a formalism for computation via sequences where information is carried by the relative timing of spikes between neurons. The theory of prospective coding via theta sequences poses a challenge to the rate coding dogma of predictive coding theory (Dragoi and Buzsáki, 2006). Furthermore it is unclear to what extent oscillations play a complementary role to transients in predictive coding. Future work carefully dissecting transient and rhythmic activity is necessary to answer this question.

The synergistic nature of transients across the cortical hierarchy may indicate a distributed, synergistic encoding of prediction errors rather than independent computation of prediction errors at each level (Gelens et al., 2024). This requires the consideration of non-linear, recurrent interactions in predictive coding models. In general, it is an open question whether sensory predictions and error signals are encoded in a localized or rather distributed manner.

There are contradicting theories concerning gamma-band oscillations and there is empirical evidence supporting different theoretical frameworks. As discussed above, a central issue is the distinction between broadband and narrow-band gamma oscillations, which requires better quantification in studies and more consistent terminology. In addition, the terminology concerning oscillations can be refined, as oscillations can refer to limit cycle behavior as well as quasi-oscillations / damped harmonic oscillations that have entirely different characteristics and computational consequences. For example, computational models suggest that efficient coding may be facilitated by stochastic quasi-oscillations, consistent with the stochastic nature of oscillations in vivo (Burns et al., 2011; Spyropoulos et al., 2022), but not by limit cycle oscillations (Chalk et al., 2016). Hence theories and empirical studies would profit from more precise terminology and quantification concerning quasi-oscillations, limit cycle oscillations and broadband fluctuations.

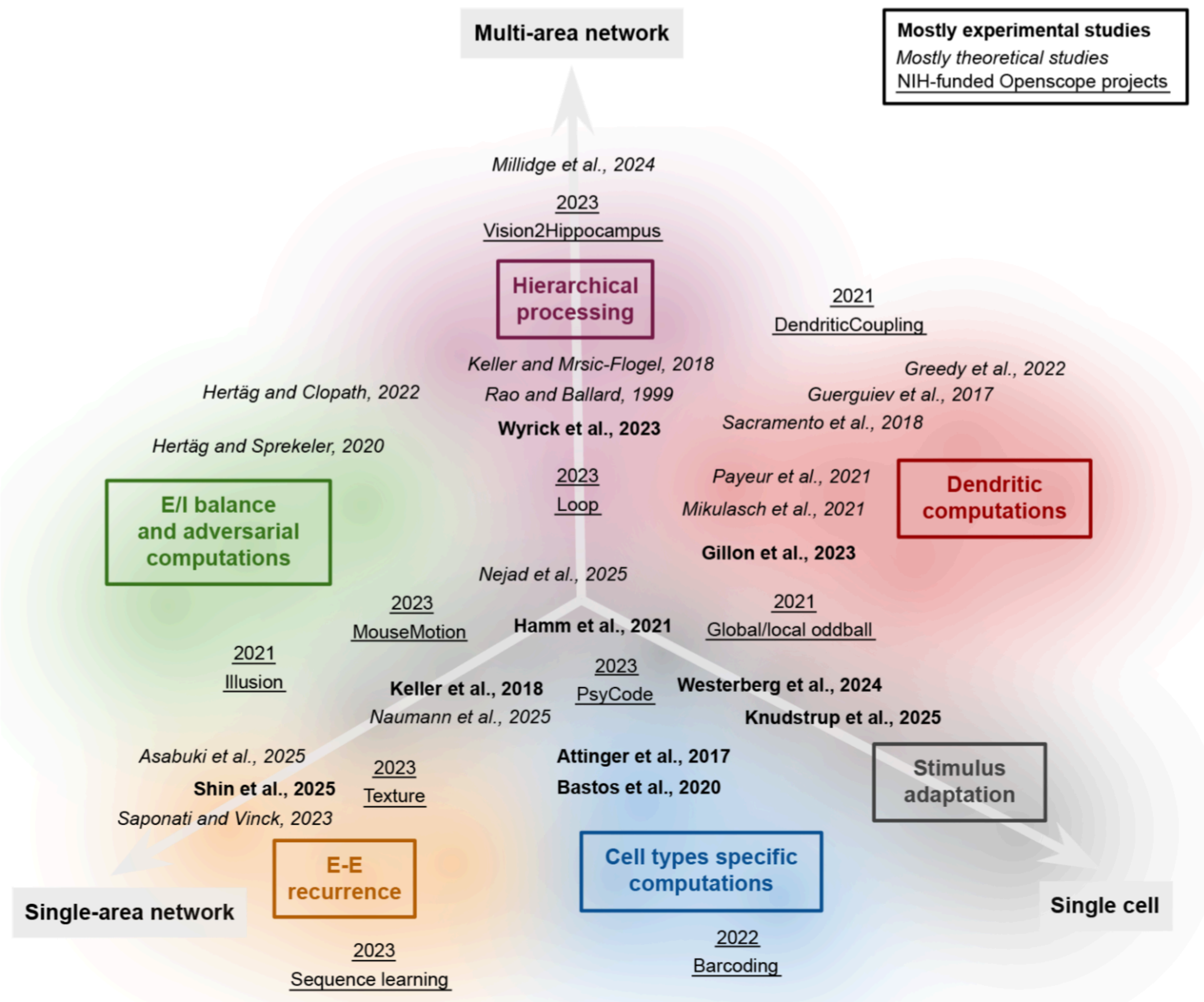

**Figure 8 - Thematic distribution of a subset of the predictive processing studies cited here.** Experimental (in **bold**) and theoretical (in *italics*) publications are positioned across three scale axes (single cell, single-area network, multi-area network) and six mechanistic themes. All NIH-funded OpenScope projects at the time of this perspective are also included (underlined).

In light of the empirical evidence reviewed above, it remains an open question whether alpha/beta oscillations are characteristic for specific hierarchical levels, or whether they play a more specific role in feedback processing. Large-scale recordings across multiple laminar compartments and areas are required to better characterize the distribution of rhythms across areas and the functional influences between areas in different behavioral states, as well as the relation of rhythms to cortical gradients. Because rhythms depend strongly on behavioral state and conditions (Steriade et al., 1993; McGinley et al., 2015), and because this dependence may be area-specific, it is difficult to generalize from a “static snapshot” of the distribution of rhythms across the cortex based on one behavioral condition. For example, during sleep or quiescence, low-frequency rhythms may be a characteristic feature of early sensory areas, while low-frequency rhythms may be a signature of active processing in higher cortical areas (e.g., during cognitive control) (Lacaux et al., 2024). Thus, the same frequency (e.g., theta or alpha) can be associated with higher or lower areas depending on task demands and alertness levels (Vinck et al., 2023; Lacaux et al., 2024).

Whether alpha/beta oscillations have a suppressive influence on downstream targets requires more work, by examining the specific consequences of these oscillations on the firing of different cell types and specific neural compartments. Furthermore, there are contradicting experimental findings and theoretical proposals concerning e.g., the role of beta (cf. (Richter et al., 2017; Bastos et al., 2020)) that remain to be understood. A challenge for computational models would be to understand if such a mechanism of suppression via alpha/beta network oscillations can account for the specific computations implied in predictive coding. Furthermore, the influence of top-down feedback on lower hierarchical levels is not strictly suppressive in predictive coding models. In fact, top-down feedback has excitatory effects on pyramidal neurons and apical error signals in dendritic predictive coding theories reviewed above (see Section V). Top-down feedback has mixed effects in classic, cellular predictive coding theories (see Section III-V). For instance, the effect of top-down feedback on positive error units in lower hierarchical units is subtractive while the effect on the representations and negative error units ends up being excitatory.

Theta oscillations may be a signature of processing in the highest hierarchical levels like hippocampus and ACC (Dragoi and Buzsáki, 2006; Womelsdorf et al., 2010; Murray et al., 2014; Gao et al., 2020; Vinck et al., 2023) and theta has been proposed to orchestrate top-down feedback across distributed cortical areas, e.g., via theta-specific resonances and theta-to-gamma cross-frequency coupling (for review see,(Sirota et al., 2008; Womelsdorf et al., 2010; Liebe et al., 2012; Vinck et al., 2023). This may suggest that predictions of the future may be broadcasted across the cortex via theta sequences. However a study on macaque visual cortex finds that theta frequencies are more strongly associated with feedforward than feedback influences (Bastos et al., 2015a), and theta frequencies may be associated with various forms of rhythmic sampling of the environment like eye movements, whisking and respiration that modulate sensory responses, complicating the relation between theta frequencies and top-down processing (Berg and Kleinfeld, 2003; Schroeder et al., 2010; Bosman et al., 2012).

To conclude, as is the case for spike rate coding, the empirical evidence is multi-interpretable and there are different and sometimes competing theoretical models interpreting the same empirical data. The understanding of the role of oscillations in predictive coding faces several specific challenges: (1) The lack of specific and frequency-band-limited causal manipulations of oscillatory phenomena. (2) The indirect, hybrid and meso/macroscopic nature of LFP/EEG/MEG signals that are often used to quantify oscillations. These require careful inference (Pesaran et al., 2018): e.g., decomposition of local and afferent synaptic inputs, mitigation of volume conduction, and separation of broadband vs. narrow-band phenomena. Ideally, inferences about oscillations are based on spike-spike and spike-field analyses. (3) The fact that cause and consequence are difficult to separate in recurrent systems: Changes in oscillations and synchronization may be both a consequence or a cause of changes in neural firing rates. (4) The fact that oscillations at specific frequencies, in contrast to spike rates, may be present only in some brain areas and only under some conditions (Hermes et al., 2015; Hoffman et al., 2024). Experimenters may optimize their experiments in order to boost oscillations (e.g., using artificial grating stimuli in the visual cortex), leading to generalization problems (Hermes et al., 2015, 2019). (5) While predictive coding models are typically based on spike rate coding, there is a limited computational understanding of the role that oscillatory phenomena could play in computational predictive processing models. A promising computational approach that may be specifically applied to predictive processing is to endow recurrent neural networks with (e.g., Kuramoto or damped harmonic) oscillatory neural units and compare their performance to recurrent networks with non-oscillatory units. Such an approach applied to sound and image classification suggests that adding oscillatory properties to recurrent networks can boost computational performance, by facilitating integration over time and mitigating the classic vanish/exploding gradient problem (Rusch et al., 2023; Piantadosi et al., 2024).

# Review summary

Overall, our review highlights the considerable interest that the field of predictive processing has garnered in recent decades. This underscores that predictive processing is a powerful framework for studying brain function, as it translates concepts from self-supervised learning and statistics into the tasks the brain must solve in real-life conditions.

We identified several major research areas. Predictive processing studies have typically examined mechanisms spanning from **single-cell computations** to **networks** within individual brain areas. Models and experiments have proposed numerous frameworks to organize these networks across **multiple brain regions** (See **Figure 8**).

In the context of making predictions, we noted six important computational primitives that have been uncovered:

1. **Stimulus adaptation** mechanisms enable individual neurons to reduce or alter their response to highly recurring stimuli.
2. **Dendritic computation** has gained interest in modeling studies but remains an emerging area experimentally.
3. **Cell-type specific computation**—particularly explored with transcriptomics—continues to be examined in great details experimentally, though theoretical work is nascent, except for some notable models involving inhibitory neurons.
4. **Recurrence** among pyramidal neurons has gained significant attention, both experimentally and in machine learning contexts.
5. **Excitatory/inhibitory (E/I) balance** offers a compelling framework for creating a competitive representational learning environment between excitatory and inhibitory neurons.
6. **Hierarchical processing** across brain areas was an early focus of predictive processing research, and most experimental studies to date reveal that the relationships between brain areas are more complex than initially proposed.

Projects submitted in recent years to the OpenScope program generally fall within these categories, reflecting the current distribution of interests in the predictive processing community (see **Figure 8**). However, we have identified several critical areas that remain under-explored, potentially slowing progress.

**Theoretical models of predictive processing often assume a uniform set of mechanisms across different sensory modalities** and cortical stages, which may not fully capture the diversity observed in experimental settings or distinct cortical properties. Experiments that shift and ideally quantify the relative contributions of adaptation, local, and global computations are needed. Realistic models with

access to simulated ground truth (Galván Fraile et al., 2024) could be used to validate metrics that quantify the contribution of cell-level adaptation versus network effects on predictive suppression and error signals. Together, these metrics and experiments would enable the community to directly compare the roles of these different computations in different error types, and provide a clear set of targets against which to test models of predictive processing.

It is important to note that many other computational goals, beyond predictive processing, such as normalization, figure-ground segregation, and salience detection, may provide alternative explanations of mismatch responses (Schwartz and Simoncelli, 2001; Li, 2002; Kirchberger et al., 2023; Cuevas et al., 2025). In this context, error and mismatch signals might be better understood simply as computational primitives allowing the cortex to perform these computations. For example, one goal of salience is to move the fovea to relevant locations, which could be partly supported by mismatch responses. Similarly, figure-ground segmentation, important for scene segmentation and object recognition, could also be initially supported by mismatch responses (Poort et al., 2016). Interestingly, the differential modulation of inhibitory cell types, with enhanced VIP (Vasoactive Intestinal Polypeptide) and suppressed SOM (somatostatin expressing) responses, has been shown to support figure-ground modulation (Kirchberger et al., 2021). Moreover, this is governed by top-down activation as optogenetic silencing of feedback connections eliminates the figure-ground modulation, similar to what is observed during sequential oddball paradigms (Hamm et al., 2021a; Bastos et al., 2023).
In addition, while predictive processing models have evolved to incorporate temporal predictions, integrating these models with experimental findings remains challenging. These challenges arise in part due to the highly complex temporal dynamics of neural signals, influenced by diverse neural integration properties, variable transmission delays, and perhaps most importantly the highly recurrent nature of cortical circuits. Experiments that present temporal sequences over varying time scales, from milliseconds to seconds, make it challenging to accurately capture temporal structure in predictive coding models.

Neurons and networks appear to involve diverse mechanisms to make predictions. Even at the level of primary sensory cortices, different modalities may engage divergent mechanisms for predictive processing. Integrative studies that examine multiple mechanisms, regions, and modalities simultaneously are scarce. Addressing this gap could be an important step for the field in the coming years. The explanatory power of the current data is limited due to the number of conditions tested, as well as the number of cell types and brain areas simultaneously recorded. Studies tend to rely on the existing literature to choose the most promising targets to record from in a given task. This, in turn, can reinforce biases in our understanding of the differences and similarities between the mechanisms that underlie neural responses to various mismatch types.

While predictive coding was proposed as a general framework, the community has yet to demonstrate that a single algorithm can account for all mismatch responses recorded, or to identify ways in which different types of mismatch stimuli engage different mechanisms in the brain. We propose that future experiments characterize error signals generated by a wide range of mismatch stimuli, with consistent temporal and spatial structures. This approach would enable both experimentalists and theorists to test whether predictive processing in the brain relies on canonical computations or identify how these computations vary depending on the stimulus properties.

# Experimental proposal

Based on the previously introduced literature background, we propose the following experiments. These aim to resolve existing divergences between the experimental and theoretical domains in the field and deepen our understanding of the mechanisms underlying predictive processing.

## Background

In **Section I**, we reviewed a broad range of mismatch stimuli used to investigate the possible mechanisms underlying predictive processing. In **Sections II**, **IV**, and **V**, we examined evidence supporting these proposed mechanisms, which include E/I balance, dendritic processing, and hierarchical computation. Our review suggests that different types of mismatch stimuli may recruit distinct sets of computational capabilities.

The cellular substrates of prediction errors differ across paradigms. In multiple visual oddball paradigms, prediction errors are mostly limited to layer 2/3 pyramidal neurons (Jordan and Keller, 2020; Hamm et al., 2021a; Pak et al., 2021; Gallimore et al., 2023). However, in auditory sensorimotor paradigms using pitch deviants, mismatch responses are present in layer 5 and layer 2/3 (Lakatos et al., 2020; Audette and Schneider, 2023; Obara et al., 2023; Xiong et al., 2024). In somatosensory oddball paradigms, prediction errors have been identified across multiple layers, including layers 4 and 6 of the barrel cortex (Musall et al., 2017) and layer 2/3 (Han and Helmchen, 2024). One possibility is that whether the conflicting features (of the deviant vs the predicted stimulus) are represented locally (e.g., orientations within mouse V1) or in distinct cortical areas (e.g., pitches or whisker stimulations) could determine the nature of the computation and thereby affect the spatiotemporal properties of the prediction error.

The neuronal pathways underlying prediction differ for visuomotor, sequential and spatial mismatches. For instance, a study on visuomotor mismatch responses found opposing influences of visual and motor inputs on the activity of individual L2/3 V1 neurons (Jordan and Keller, 2020). In contrast, a study of navigational mismatch responses found enhanced visual input driven by pulvinar inputs onto inhibitory populations (Furutachi et al., 2024). In addition, sensory occlusions in visual experiments may engage local projections within V1 or its immediate downstream areas (Cuevas et al., 2025), while oddball sequences could involve mechanisms like adaptation (Aitken et al., 2024), feedback from higher cortical areas (Hamm et al., 2021a; Obara et al., 2023) and interactions among local inhibitory interneurons (Hamm and Yuste, 2016; Bastos et al., 2023; Najafi et al., 2025).

It is possible that different mechanisms are engaged in passive vs active contexts. Experimental work on motor corollary discharges — notably from Georg Keller's laboratory (Keller et al., 2012) and David Schneider's laboratory (Audette et al., 2022) – has identified many neurons whose activity is suppressed in "expected" conditions. These studies point to a critical role for top-down inputs from frontal and motor areas in shaping predictions. In contrast, studies examining neural activity in response to passive exposure to temporal sequences of stimuli have shown more modest suppression (Hamm et al., 2021a; Homann et al., 2022; Price et al., 2023; Gillon et al., 2024; Westerberg et al., 2025), typically only affecting around 10% or less (Hamm et al., 2021a; Homann et al., 2022; Price et al., 2023; Gillon et al., 2024) of the recorded neurons.

Neurons have access to a wide range of biophysical mechanisms to form predictions and compute prediction errors. These mechanisms operate at different scales and likely interact in complex ways rather than acting in isolation. (1) Stimulus adaptation mechanisms allow individual neurons to adjust their responses to recurring stimuli, potentially enhancing their sensitivity to different or changing inputs. (2) Dendritic computation enables integration of multiple inputs within a neuron, potentially supporting the calculation of prediction errors in a shared post-synaptic compartment. (3) Cell-type specific computation, particularly involving inhibitory neurons, has been highlighted in models where distinct neuronal subtypes contribute uniquely to predictive tasks. (4) Recurrence among pyramidal neurons, combined with mechanisms such as spike-timing-dependent plasticity, allows groups of neurons to generate and refine specific temporal sequences of activity (Saponati and Vinck, 2023). (5) Excitatory/inhibitory (E/I) balance, often mediated by diverse inhibitory subtypes, provides a framework for making complex predictions, especially when integrated with top-down signals (Hertäg and Sprekeler, 2020). Finally, (6) hierarchical processing across brain areas can enable sophisticated, multi-modal predictive capabilities (Leinweber et al., 2017; Hamm et al., 2021b).

These mechanisms have been extensively studied both experimentally and theoretically in the context of predictive processing, but their interactions remain poorly understood. Specifically predictive mechanisms may not simply sum their effects but may also cooperate and compete depending on task demands and neural constraints. For example, temporal predictions during sequential oddball protocols involve not only local adaptation (Knudstrup et al., 2025), but also the integration of top-down inputs from higher cortical areas (Hamm et al., 2021a). These overlapping influences make it difficult to isolate the contributions of each mechanism, complicating data interpretation. As a result, many experimental mismatch protocols now include carefully controlled conditions to account for these combined effects to ensure that the observed responses can be more accurately attributed to specific underlying predictive processes. Real-world stimuli likely engage multiple, if not all, of these prediction mechanisms simultaneously, but whether and how they work together synergistically remains unclear. Gaining a deeper understanding of this integration could help resolve some of the discrepancies observed across experimental studies, as highlighted in our section outlining divergent results and interpretations.

We propose to design a set of different types of mismatch stimuli, and to record neural responses across the full set through several recording sessions in the same animals. Using this foundational dataset, we aim to train and validate predictive processing models that integrate and combine mechanisms operating at different scales. To maximize the applicability of this dataset, we propose to collect, in different animals, Neuropixels and two-photon imaging datasets. We hypothesize that adaptation, recurrence, top-down inputs, and E/I balance, although likely continuously engaged, show varying contributions depending on the nature and difficulty of the prediction task. By systematically varying mismatch protocols, we aim to investigate how neurons dynamically employ these predictive mechanisms. Here, we define the prediction set as the stimulus or set of stimuli expected in a given paradigm based on prior presentations, spatial continuity, or behavioral state and locomotion.

Recent work in both anatomy and neurophysiology points to important differences in the visual sensory hierarchies of mice and monkeys (Glatigny et al., 2024). What this work shows is that compared to mice, monkeys have a much steeper visual cortex gradient that scales the hierarchy and potentially allows for multiple levels of signal processing prior to sensory information (or putative prediction errors) reaching the prefrontal cortex. Therefore, it is likely that prediction error computations may be partially evolutionarily distinct and partially conserved in rodents compared to primates. As a result, it is critical to investigate similar tasks in both species to better appreciate the mapping from mouse to primate circuits involved in predictive processing. Importantly, it was recently demonstrated that some forms of predictive processing emerge very late in hierarchical processing, such as area AM in mice (near the top of the visual cortical hierarchy) and area FEF/PFC of monkeys (Westerberg et al., 2025). This underscores the need to record from higher-order cortical areas, in addition to the visual cortex, to determine whether certain predictions arise early or late in processing.

We anticipate that the modeling community will leverage these datasets to quantitatively compare models in terms of their ability to account for the different patterns of neuronal activity observed across different experiments.

# Specific Aims

The Specific Aim of this proposal is to elucidate the potential relationship between four most commonly studied mismatch stimuli and their associated error signals, as well as different neuronal implementations of predictive processing.

### Hypothesis-Driven Framework

We hypothesize that…

*H0*: mechanisms of predictive processing fundamentally differ depending on predictive set and prediction error types, and recruit different neuronal mechanisms.

Alternatively, we hypothesize that…

*H1*: a unified predictive processing mechanism drives all mismatch processing in the mammalian cortex.

To determine which of these two alternative hypotheses is correct, we propose to experimentally examine three major types of mismatch stimuli that have dominated the literature: temporal, motor, and omission. Further, we will examine two types of *prediction sets*: spatiotemporal (passive) and sensorimotor (active). Importantly, in all experiments, the region recorded (V1) will include spatially intermixed neurons selective for features of both the expected stimulus *and* the mismatch stimuli. For example, orientation and direction tuning are spatially mixed in V1, and the standard oddball paradigms (Zhou et al., 2020; Hamm et al., 2021a; Pak et al., 2021; Homann et al., 2022) and sensorimotor mismatch paradigms (Jordan and Keller, 2020) involve predicted stimuli and mismatch stimuli that differ in orientation or direction.

### Significance and Innovation

We will collect the first comprehensive set of neuronal data enabling direct comparison across different mismatch error signals. Additionally, our methodology will be designed to integrate two-photon imaging and electrophysiological recordings, leveraging the strengths of both techniques.

Resolving these alternative hypotheses will mark major progress for the field, unifying conflicting findings and clarifying how differences in experimental design shape interpretations of predictive processing.

Additionally, the resulting dataset will be a pivotal resource for validating mechanistic computational models across multiple mismatch types, advancing our understanding of predictive processing in the brain.

# Proposed Experiment

## 1. Stimulus set design

Our stimulus set will be designed to contain a few validated mismatch stimuli (see **Figure 9**). Particular attention will be given to the experimental design to allow fitting models of:

- Synaptic adaptation
- Positive and negative errors
- Short-term memory that could emerge through local recurrence.
- E/I balance

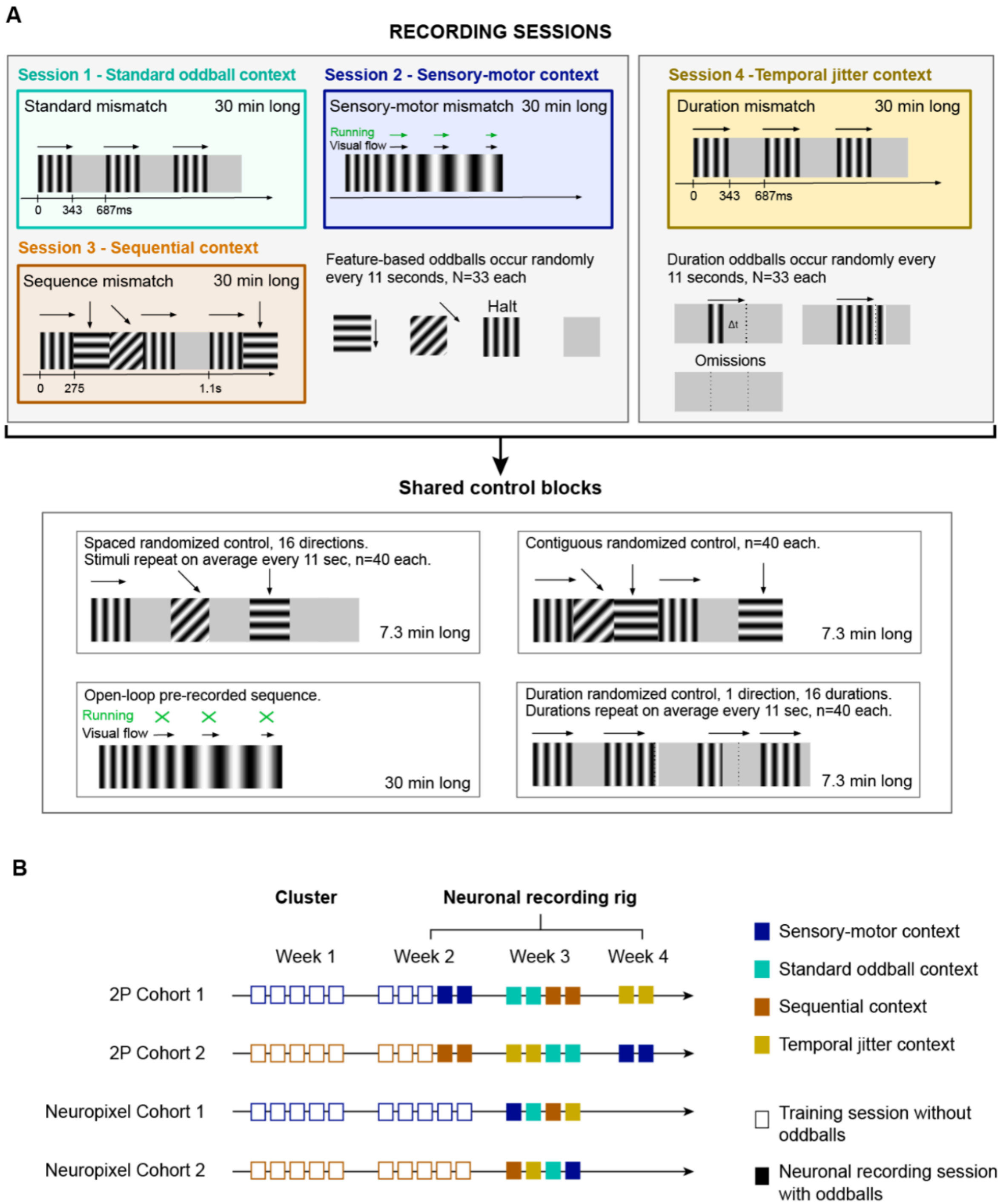

**Figure 9 - Proposed experiment design. A.** Session design for the four neuronal recording contexts. Each session consists of an oddball block followed by four shared control blocks. Session 1 (Standard oddball): Animals are habituated to repeated drifting gratings of one orientation. Oddballs are created by either presenting one of two alternative orientations, halting the grating drift, or omitting a grating. Session 2 (Sensorimotor mismatch): Animals are habituated to a closed-loop visuo-motor paradigm in which the running disk controls visual flow. The same types of oddballs as used for Session 1 are introduced on recording days. Session 3 (Sequence mismatch): Animals are habituated to rapidly repeating 4-stimulus sequences. The same types of oddballs as used for Session 1 are inserted in the third sequence position. Session 4 (Temporal jitter): Animals are habituated to repeated gratings presented for a specific duration (e.g., 343 ms). To probe duration-based prediction errors, oddballs are created by jittering the duration of the gratings or omitting them entirely and jittered blocks. In all sessions, oddballs occur on average every 11 seconds. All sessions also share four control blocks: a spaced randomized control (16 orientations separated by gray periods), a contiguous randomized control (different orientations without gaps), an open-loop replay of a pre-recorded sensorimotor sequence, and a duration randomized control. **B.** Cohort and timeline design. Four cohorts of mice are recorded over four weeks: two using two-photon calcium imaging (2P Cohort 1 and 2, pan-neuronal lines, 8 chronic sessions per mouse across V1 and LM at four cortical depths) and two using Neuropixels probes (Neuropixel Cohort 1 and 2, SOM-optotagging lines, 4 sessions per mouse). Each animal experiences all four session contexts, though in a different order depending on the cohort it is in, enabling direct comparison of mismatch responses recorded with complementary techniques.

Since, we aim to bridge various visual stimuli designs piloted, analyzed, and deployed over the last decades, we will use sequences of drifting gratings. We will present five types of oddballs: a drifting grating halt, two alternative drifting orientations,an omission and temporal jittered oddballs. All oddballs will be introduced in four different session contexts: standard mismatch, sensory-motor mismatch, sequential mismatch and temporal jitter mismatch. These contexts will be separated based on the session and habituation design. Individual animals will experience all 4 contexts in different orders. Two cohorts of separate animals will be recorded with Neuropixels probes and multi-area two-photon imaging.

*Session 1 – Standard mismatch:* Animals will be habituated to a series of drifting gratings of the same orientation. Various mismatch stimuli will be introduced randomly: differing orientations, omissions, and spatial oddballs.

*Session 2 – sensorimotor mismatch*: Animals will be habituated to a closed-loop visuo-motor running disk where the rotation of the disk will directly control visual flow on a screen in front of the mouse. On recording days, the same mismatch stimuli as in Session 1 will be introduced.

*Session 3 – sequence mismatch:* Animals will be habituated to rapid sequences of 4 stimuli. The same sequences will repeat, once per second, for 37 minutes. The same mismatch stimuli as in Session 1 will be introduced in the third position in the sequence order, once every 11 seconds, on average.

*Session 4 – Temporal Mismatch, Duration:* Animals will be exposed to a sequence of drifting gratings of the same orientation. Duration mismatches will be created by introducing gratings with durations that are either shorter or longer than expected. To examine temporal prediction and error signaling, we will use experimental manipulations similar to those used in previous work (Huang et al., 2025). This involves sessions with two alternating trial blocks: fixed and jittered. In the fixed block, grating durations will remain constant across all trials other than mismatch trials. In the jittered block, grating durations will be drawn from a normal distribution with a large standard deviation, to introduce variability in timing. Each block will include duration mismatches once every 11 seconds, on average.

The inclusion of jittered blocks is critical for dissociating temporal predictive processing from alternative mechanisms. Under a predictive processing framework, duration mismatches should produce substantially stronger responses in fixed blocks, where stimulus timing is highly predictable. However, Huang et al. (2025) found that neural responses in the visual cortex to oddball temporal intervals were of a similar magnitude in both fixed and jittered temporal contexts. This indicates that responses to temporal oddballs may be explained by intrinsic stimulus-evoked dynamics, instead of temporal prediction or prediction-error signaling (Huang≈ et al., 2025).

By shifting the expected temporal structure within the standard oddball paradigm varying stimulus duration, we aim to investigate how neurons encode and resolve prediction errors related to stimulus duration, potentially engaging temporal mechanisms distinct from those involved in stimulus feature-based mismatches.

All feature-based sessions (Session 1 to 3) will experience 4 temporally based oddballs with equal frequencies: two alternative drifting grating orientation, one drifting grating halt and one omission. These 4 oddballs will last 275 ms, will be shuffled and occur randomly, on average every 11 seconds throughout the 37 min long block. All sessions will experience 4 shared controls blocks:

- Randomized drifting gratings presented at 16 orientations *with* gray periods in between. Each orientation will occur once every 11

seconds to match the occurrence of mismatches in the first experimental block.

- Randomized drifting gratings presented at 16 orientations *without* gray period in between. Each orientation will occur once every 11 seconds to match the occurrence of mismatches in the first experimental block. Open-loop replay of a closed-loop sensory-motor block with all oddball types pre-recorded but uncoupled to the movement of animals.
- Randomized temporal jittered presentation of drifting gratings. Each jittered stimulus will occur every 11 seconds to match the occurrence of jittered mismatch in the experimental block.

## 2. Recording techniques

In sections I to VI, we discussed how neuronal responses to the types of mismatches included for
Session 1 to 4 could be supported by a variety of mechanisms including adaptation, recurrence between pyramidal cells and E/I balance. To properly evaluate the relative contribution of these mechanisms, it is critical to measure the activity of the excitatory and inhibitory neuron populations in all cohorts. Two-photon calcium imaging offers ideal access to different classes of inhibitory neurons in dense networks but lacks the high temporal resolution needed to resolve individual spikes and event timing. Complementary Neuropixels recordings will address this limitation, capturing spike timing with high spatial and temporal resolution for the same stimuli. Combining these two techniques will provide a richer, multi-dimensional dataset, enhancing both data analysis and modeling. Primates will only be recorded with electrophysiological probes due to technical limitations in the ability to record neural activity using two-photon imaging in this species
Given our goal to build integrated models, we aim to record as many neurons as possible within relevant cortical networks. Based on our previous review, **V1** is consistently engaged by the stimuli presented across all cohorts in mice. We also propose to record from LM and a set of distributed areas using both techniques (see **Table 1** and **Figure 10** for mice). Primates will be recorded from V1/V2/V3/MT/MST and prefrontal cortex. This approach will allow for a comprehensive comparison of predictive processing across species and cortical areas.

| Transgenic mice | Stimulus cohort | Minimum number of mice | Areas recorded |
|---|---|---|---|
| Pan-neuronal GCAMP line | 2P Cohort 1 | 5 | V1 + LM. 4 planes in each. Layer I, Layer II/III, Layer 4, Layer 5 |
| Pan-neuronal GCAMP line | 2P Cohort 2 | 5 | V1 + LM. 4 planes in each. Layer I, Layer II/III, Layer 4, Layer 5 |
| SOM-optotagging | Neuropixel Cohort 1 | 5 | V1<br>LM<br>M1<br>M2,<br>PL/IL/ACA<br>RL+LGN |
| SOM-optotagging | Neuropixel Cohort 2 | 5 | V1<br>LM<br>M1<br>M2,<br>PL/IL/ACA<br>RL+LGN |

**Table 1 - Mice recorded for Project 1 by the OpenScope program.** See **Figure 10** for Neuropixels geometry.

For two-photon imaging, we will use a pan-neuronal line to record from the majority of neuron types across cortical layers. For the Neuropixels recordings, we propose to insert probes in V1 and LM as well as motor areas, prefrontal cortex and LGN.

## 3. Recording sessions

Recording sessions will be organized across 4 cohorts, two using two-photon imaging and two entirely separate cohorts of mice using neuropixel probes (see **Figure 9B**). Each recording session will be one of four different stimuli designs (see **Figure 9A**). Each mouse brain will be chronically recorded across either 8 sessions (two-photon imaging) or 4 sessions (neuropixel recording). Each session will present habituated stimuli, as well as blocks of stimuli containing oddballs and control blocks. Session type 1 to 3 will differ only in their sensory context but will share oddball types and control blocks. In addition, session 4 will introduce jittered oddballs instead of feature oddballs. Individual animals will experience all 4 contexts in different orders across 2 cohorts for each data modality. Each stimulus session will be implemented in Bonsai and immediately open source (https://github.com/AllenNeuralDynamics/openscope-community-predictive-processing).

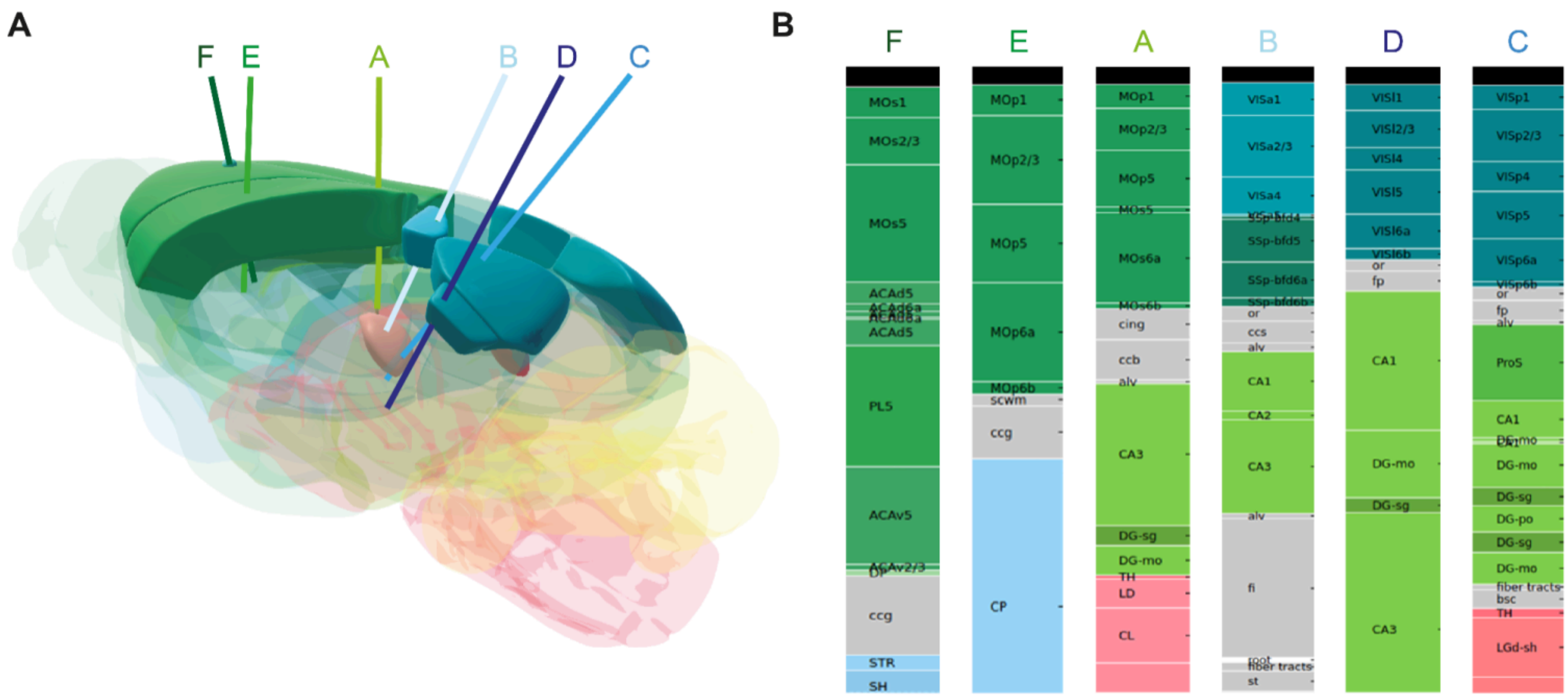

**Figure 10 - Neuropixels probes. A.** Geometry of intended Neuropixels probe trajectories in mice, modeled using PinPoint (Birman et al., 2023). **B.** Examples of areas to be recorded with each probe. Note that exact insertion positions will vary across mice and recording days.

## 4. Multi-lab collaboration

In addition to recordings performed by the OpenScope program, we propose a multi-lab collaboration where individual labs will share sub-components of the stimulus sets but have the flexibility to vary the targeting of brain areas and their recording methods. Those complementary datasets are listed on **Table 2**. This approach will expand the coverage of neuronal activity across different experimental conditions. The following are currently planned:

### Bastos lab : Primate data recordings

Following on the cross-species considerations discussed in section 2 and in the background, we propose to run the same studies in parallel in mice along with collaborating institutions that will provide the non-human primate data (as in (Westerberg et al., 2025), see **Table 2**). For a better quality in visual tasks (both passive and active), we use eye-tracking systems to ensure they are paying attention and control eye movements. For the visual flow experiments in monkeys, we propose to have monkeys control their movement using eye-tracking while being head-fixed (headposted) for habituation, training and invasive recordings (INTAN electrophysiology interface and diagnostic biochips deep laminar electrodes). Other than that, the methodology and data analysis between primates and mice will be as similar as possible, including use of optogenetics to identify pan-inhibitory interneurons (Dimidschstein et al., 2016).

**Table 2 - Complementary datasets proposed to be collected by collaborating laboratories**

| Laboratory | Animal details | Technique | Stimulus | Areas recorded | Details on neuronal recordings |
|---|---|---|---|---|---|
| Allen Institute OpenScope program | Mice | Multi-plane two-photon + Neuropixels | Cohort 1 + 2 | V1, LM, M1, M2, Prefrontal cortex, LGN | See table 1 |
| Najafi lab | Mice | Single plane two-photon imaging | Temporal jitter context | Prefrontal and Premotor cortices, likely some sub-cortical areas too | |
| Podgorski lab | Mice | SLAP2 voltage imaging | Sensory-motor context cohort | V1 | Single pyramidal cell imaging (somas + dendrites) |
| Ruediger lab | Mice | Neuropixel recordings | Temporal jitter context | To be determined | |
| Bastos lab | Primates | Diagnostic Biochips (128 channels per probe) recordings | Session 1,2,3,4 | V1, V2, V3, MT/MST and prefrontal cortex | As described in Westerberg et al., 2025 |
| Oweiss lab | Mice | Two-photon optogenetics+voltage imaging | Standard oddballs session<br>BCI control | V1 | Single pyramidal cell (somas + dendrites)<br>Small ensembles (BCI) |

### Najafi lab and Ruediger lab: Temporal jittered data recordings across inhibitory cell types and visual areas.

The temporal mismatch condition will be extended with more variation of the oddball conditions. These variations will allow us to assess how neurons encode and resolve timing prediction errors within a stimulus sequence, distinguishing between local adaptation, recurrent processing, and hierarchical feedback contributions. Importantly, this data will complement the observations of feature-based mismatch responses (in particular session 3), enabling a direct comparison between feature-based mismatch responses and temporal mismatch responses. Understanding these distinctions is crucial, as it provides insight into how the brain integrates different types of prediction errors to optimize perception and behavior in dynamic environments.

Animals will be exposed to a sequence of drifting gratings of the same orientation. To introduce interval mismatches, the inter-stimulus intervals (ISIs) between some gratings will be either shorter or longer than expected. The session will consist of two alternating trial blocks: fixed and jittered. In the fixed block, ISIs will remain constant across all trials. In the jittered block, ISIs will be drawn from a normal distribution with a large standard deviation, introducing variability in timing. Each block will include 15% interval mismatches once every 11 seconds, on average.

### Podgorski lab: Dendritic recording with voltage imaging

A subset of laboratories will record dendritic activity using voltage imaging in individual excitatory neurons (see **Table 2**). It is important to note that the current scale of voltage imaging is more amenable to single session recordings. In addition, these experiments should be designed to cover mismatch learning from start to finish. Remarkably, standard oddballs (**Session 1**) have been shown to trigger very fast learning (Hamm et al., 2021b; Bastos et al., 2023). We will therefore leverage Session 1 design in those experiments. We expect dendritic data to be highly valuable toward a more detailed investigation of within-neuron mechanisms. For example, the contribution of adaptation-like models could be better uncovered with fast dynamics recordings across the dendritic tree. This is because individual button and dendritic branch dynamics are more directly available. In addition, voltage imaging across the dendritic tree of a single pyramidal cell will uncover the impact of inhibitory input. SOM and PV cells have drastically different projection target geometry onto pyramidal cells. We therefore aim to better define the contribution of both adaptation and inhibitory activity.

### Oweiss lab: Calcium and voltage imaging with BCI and optogenetics

In addition to the standard mismatch condition (**Session 1**), the sensorimotor mismatch condition 2 will be extended to randomly decouple the visual flow from the running disk rotation within single trials (see **Table 2**). When decoupled, the visual flow will be modulated using the decoded $Ca^{2+}$ activity of a 'rewarded' neural population, selected by the experimenter in real time in a BCI paradigm. In this setting, we seek to investigate how the nervous system could adapt to—and possibly predict – unexpected perturbation in sensory feedback online. We expect that 'within-movement' feedback corrections will gradually improve over trials, despite the unexpected dynamics associated with the random perturbations. Both $Ca^{2+}$ and voltage imaging data will provide valuable insights into how both supra and subthreshold membrane dynamics mirror these effects, both within and across neuron mechanisms and across multiple timescales. Furthermore, it will provide critical data to assess the extent to which recurrent excitation and/or feedback inhibition at the circuit level could be shaping these response dynamics without being confounded by overt movement-related neural dynamics. Finally, it will be the first to assess the impact of neuromodulatory reward signals (e.g., dopamine) on shaping predictive processing at the dendritic, somatic and population levels (Chueh et al., 2025).

# Analysis Plan

Our review in **Sections II** to **VI** highlighted the presence of mismatch responses throughout the cortical network, spanning multiple areas and cellular populations, including excitatory neurons and inhibitory subtypes. These responses involve dynamic contributions from both dendritic and somatic compartments. Consequently, our analysis must disentangle these relative contributions within a tightly integrated network, across multiple types of mismatches.

A key assumption in our analysis is that different types of mismatches may recruit distinct relative contributions from computational primitives (see **Review Summary**). To test this assumption, we must measure the precise dynamic properties of individual compartments across neuronal types, areas and layers. Our goal is to compare the relative timing and strength of predictive responses, complemented by decoding analyses to extract instantaneous prediction strengths emerging across the network. Neuropixels recordings will enable decoding with millisecond precision, such that the first occurrences of mismatch encoding across circuit components (brain regions, cortical layers, neuronal subtypes, and neuronal compartments) can be identified, while imaging experiments will provide denser recordings to measure the broader impact of these predictions on the overall network.

Modeling these responses will be a key integrative effort, facilitating the unification of multi-modal and multi-species datasets. First, analytical metrics derived from real physiological data can be designed and iteratively refined using simulated neuronal activity from cortical models, where the ground truth is known. Second, modeling will enable the multi-modal integration of these datasets by leveraging the relative strengths of various techniques to constrain model parameters. Simulated models will vary in complexity to evaluate our ability

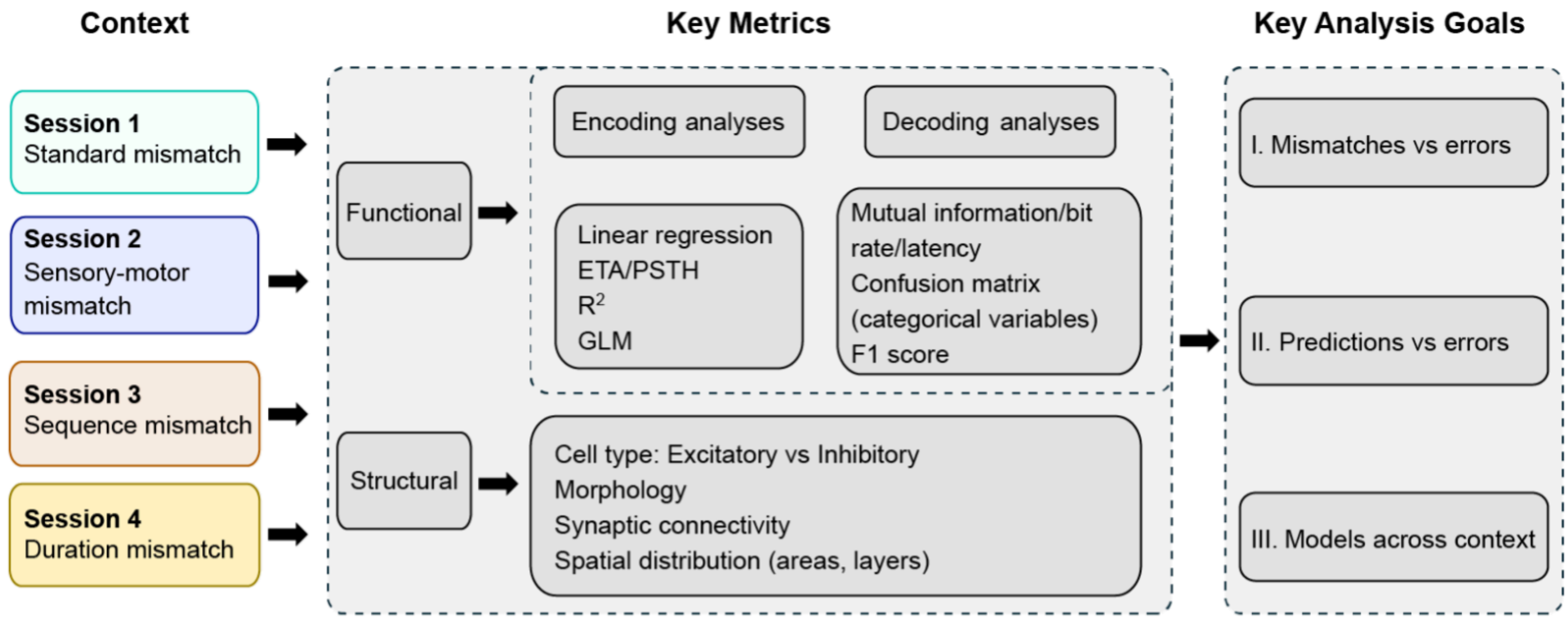

**Figure 11 - General framework for the analysis plan.** Data from the four recording contexts (Session 1: Standard mismatch; Session 2: Sensory-motor mismatch; Session 3: Sequence mismatch; Session 4: Duration mismatch) are analyzed along two complementary dimensions. Functional analyses include extracting encoding and decoding metrics. Structural analyses consist of characterizing responses by cell type, morphology, connectivity, and spatial distribution across cortical areas and layers. Together, these metrics serve three key analysis goals: (I) determining whether mismatch responses represent multiplicative, additive, or subtractive novelty signals; (II) distinguishing whether neurons encode detailed predictions about upcoming stimuli or deviations from expected stimulus ensembles; and (III) testing whether a common neural mechanism underlies mismatch responses across different types of predictions.

to disentangle mechanisms such as adaptation, E/I balance, and other underlying processes.

The analysis can be organized to address three main scientific hypotheses: I) whether mismatch responses are “additive”, “subtractive”, or “multiplicative” in nature; II) whether mismatch responses contain detailed, temporally specific predictions or expectations about the stimulus ensemble; III) whether there exists a common neural mechanism underlying different kinds of mismatch responses. Here, we provide further details about the data analysis and hypothesis testing that this experiment makes possible.

Throughout all hypotheses, we will leverage a shared set of metrics computed on all datasets (see **Figure 11**). **Encoding metrics** should include measures used to evaluate deterministic models, like linear and logistic regressions, such as accuracy, mean square error, and the coefficient of determination R2, or for probabilistic models such as generalized linear models (GLMs). **Decoding metrics** should include measures from pattern clustering and/or classification, for e.g., Mahalanobis distance, confusion matrix (categorical variables) or F1 score, mutual information, or bit rate/latency (for BCIs). In addition, analysis of response distribution across anatomical location and cell types will be used to test all hypotheses.

## I. What kind of information is encoded by mismatch responses?

A. *Multiplicative novelty:* Stimulus-specific enhancement for novel / unpredicted stimuli

B. *Additive novelty:* A generalized "alert" signal that encodes novelty per se

C. *Subtractive novelty:* The difference between the expected vs. actual stimulus

D. *No effect:* In particular, this empirical outcome could constitute a form of rejection of the hypothesis that predictive computation was involved in the experimental conditions tested

***Analysis #1:*** For each neuron and each mismatch stimulus, construct either the event-triggered average (ETA; for $Ca^{++}$ imaging data) or peri-stimulus time histogram (PSTH; for Neuropixel data):

- Significant mismatch responses will be determined in each neuron by comparing activity evoked by a given mismatch stimulus to that same stimulus when it appears during the appropriate control setting. For session 1, this will be a comparison to the spaced randomized control. For session 2, this will be a comparison to the open loop pre-recorded sequence. For session 3, this will be a comparison to the contiguous randomized sequence control. For session 4, this will be the response to a time interval presented as an oddball to the same time interval in random order.
-
- The significance of mismatch responses will be rigorously tested using bootstrap resampling, to avoid making the assumption of normal statistics for each neuron (which is often a poor assumption). Neurons with $p < 0.01$ will be considered "mismatch" neurons.
- Assuming that mismatches occur at random times on an interval [$ITI_{min}$ , $ITI_{max}$], then the ETA from $t = -ITI_{min}$ to $t = 0$ serves as a baseline response.
- *Absolute response measure*: integrated neural activity over a time window shifted by a standard latency (~50-100 ms).
- *Relative response measure:* integrated neural activity minus baseline activity (use a longer time window for baseline for better SNR, but then scale the integral to compare to the activity at $t > 0$).

***Analysis #2:*** Compare the mismatch response in the novel vs. control conditions:

A. Make a scatter plot of responses in the two conditions and carry out a linear fit. Here are possible interpretations of this analysis, keeping in mind that the data may exhibit combinations of these outcomes:

- multiplicative novelty coding = slope of linear fit > 1
- additive novelty coding = offset of linear fit > 0
- subtractive novelty coding = slope of linear fit is not statistically different from zero (or extensive deviation for a subset of neurons)
- no effect = neurons on the identity line

***Analysis #3:*** Compare responses to different mismatch stimuli in the novel condition (for Sessions 1 and 2):

- Calculate the relative response to the four different mismatch stimuli
- If neurons encode subtractive novelty, then the following will be true:

i. R(downward, 90° shift) > R(45° shift), because this is a bigger change in orientation

ii. R(halt) < R(90°) and R(45°), because the halt involves a smaller change in velocity

- Other possibilities: i) make some index that captures this relationship for individual neurons, ii) calculate the fraction of neurons fulfilling these conditions and compare them to a shuffle test, iii) assess the effects of depth and subregion on fraction of neurons showing mismatch responses, and compare between types (different sessions).

***Analysis #4:*** Calculate decoding performance / information encoded for mismatch stimuli and novelty *per se*:

- What fraction of neurons encode significant info about novelty per se?
  - a large fraction indicates a major, distributed encoding of novelty per se
- What fraction of neurons encode significant info about individual mismatch stimuli?
  - a large fraction indicates a major, distributed encoding of the identity of novel stimuli
- Calculate decoding performance vs. N neurons, extrapolate to large N:
  - extrapolation → ~1 indicates strong encoding (expected for individual stimuli, but unclear for novelty *per se*)
- Compare decoding performance of novelty *per se* vs. performance for individual stimuli:
  - similar performance indicates a strong encoding of novelty *per se*
  - lower performance for novelty indicates a weak or secondary encoding of novelty
- Scatter plot of info encoded for novelty vs. individual stimuli:
  - high correlation indicates a joint encoding of novelty and stimulus identity
  - low correlation indicates a separate encoding of novelty and stimulus identity.

## II. Distinguish between two categories of prediction made by neurons:

A. ***Detailed predictions*** about the identity of the upcoming stimulus

B. Deviation of stimulus probability from the expected *stimulus ensemble*, often described in the literature as “adaptation”. This empirical outcome could be interpreted as a form of refutation of the hypothesis that predictive computation was involved in the experimental conditions tested.

***Analysis #1:*** Compare the response to the same mismatch stimulus in all three conditions for the sensorimotor mismatch (session 2):

- Is the mismatch response > for closed loop vs. open loop
  - YES indicates that the neuron encodes a detailed prediction (as only the closed loop condition allows a detailed prediction)
- Is the mismatch response > control vs. open loop
  - YES indicates that the neuron encodes deviation from the expected ensemble (as a blank is differs more from the mismatch grating than the vertically oriented grating present in the closed loop condition)

***Analysis #2:*** Calculate decoding performance / info encoded for individual mismatch stimuli vs. for novelty *per se*.

- Use population decoder to identify the occurrence of an individual mismatch stimulus (target) versus all the other neural activity; start with a linear decoder (support vector machine):
  - this quantifies the fidelity for encoding the identity of each of 4 mismatch stimuli

- In a complementary fashion, calculate the mutual information each neuron represents about an individual mismatch stimulus versus all other neural activity
- Similarly, calculate decoding performance and information for a comparison of neural activity during any mismatch stimulus vs all other neural activity;
  - this quantifies the fidelity for encoding stimulus novelty *per se*
- If significantly more information is encoded in the closed loop condition vs. open loop
  - YES indicates encodes of a detailed prediction
- If significantly more information is encoded in the control condition vs. open loop
  - YES indicates encoding of a deviation from the expected ensemble

### ***Analysis #3:*** Emergence of Prediction Signals in Single Neurons and Neural Populations

When new, arbitrary correlations are created by the experimenter, the brain must, in principle, learn these new correlations. This can be demonstrated by showing several kinds of changes in neural responses to the same stimuli over time. These changes may occur within a single recording session, which is often interpreted as a form of adaptation, or across recording sessions, which is typically interpreted as learning.

*Key Hypothesis Tests:*

- Predictive coding vs. static tuning: Do individual neurons or neural populations show changes in their response to the same oddball stimuli?
  - YES indicates evidence of predictive computation
  - NO indicates evidence of static or previously learned tuning to stimuli
- "Predictive" Activity: Do neurons or populations of neurons exhibit activity that systematically depends on what the upcoming stimulus is (as can be demonstrated by changing stimulus contingencies)?
  - YES suggests that the neural activity was in part encoding the identity of the upcoming stimulus
  - NO indicates that the neural activity encodes the identity of the current stimulus
- "Pattern completion" activity: Do neurons or populations of neurons exhibit activity during stimulus omission that depends systematically on the preceding stimulus?
  - YES indicates a form of predictive computation, in which predictions are embodied, in part, by specific neural activity driven by events that predict an upcoming stimulus (rather than by the stimulus itself)
  - NO indicates that a response to the omission itself
- Latent component dynamics: Do identified latent variables exhibit systematic changes over trials?
  - YES indicates evidence of predictive computation revealed only at the population level
- Neural dimensionality reduction: Does the manifold structure of mismatch responses shift toward a more compact, lower-dimensional space with repeated exposure?
  - YES indicates a structure of predictive computation that is consistent with theories about efficient coding and/or maximization of coding capacity
- Conjunctive vs. disentangled representation: Does the visualized geometric structure of population activity embedded in a 3D space; e.g., using unsupervised UMAP (Uniform Manifold Approximation and Projection), show distinct, possibly orthogonal, trajectories that could reveal disentangled coding schemes for different signals (e.g., for stimulus evoked responses vs. prediction errors)?
  - YES indicates that the population neural code can simultaneously represent information about the stimulus as well as its predictive context

*Single Neuron Analysis:* Determine whether individual neurons exhibit changes in their responses with repeated oddball presentations, indicative of learning.

- Trial-by-Trial Response Analysis: Measure the amplitude and timing of neuronal responses to each oddball stimulus across trials.
- Model Fitting: Apply exponential or linear decay models to these responses to measure trends over time.
- Statistical Validation: Use bootstrap resampling to evaluate the significance of observed changes.
- Time Points for Analysis: Pre-Oddball Baseline Period: A period before the oddball onset (e.g., -200 ms to stimulus onset at 0 ms) to establish baseline activity levels. Oddball Response Window: A post stimulus onset interval (e.g., 0 to 300 ms) capturing the immediate neuronal response to the oddball stimulus.

*Population Latent Analysis:* Identify latent patterns within neural populations that correspond to predictions and prediction error signals.

- Tensor Component Analysis (TCA): Decompose multi-dimensional neural data to uncover components with trial-dependent dynamics.
- Time Points for Analysis: Pre-Oddball Baseline: A period before oddball onset (e.g., -200 ms to stimulus onset at 0 ms) to establish baseline population activity levels. Oddball Response Window: The duration of the oddball stimulus presentation (e.g., 0 to 300 ms) capturing immediate population responses to the oddball stimulus. Post-Oddball Period: A post stimulus offset interval (e.g., 300 ms to 600 ms) to monitor any sustained or delayed responses. Inter-Trial Intervals: Periods between oddball trials to evaluate baseline stability and potential anticipatory activity.

*Cross-Day Analysis:* Monitor the activity of individual neurons or neural populations over time to identify changes in prediction error signaling and learning processes.

## III. Mismatch responses across different types of predictions

These experiments test mismatch responses resulting from different kinds of predictions: i) repetition vs. oddball (session 1), sensorimotor mismatch (session 2), and temporal sequence prediction (session 3 and 4). Are there different circuit mechanisms for these four kinds of prediction?

In particular, sensorimotor prediction requires a corollary discharge of the motor command, so it requires feedback from outside V1. While there is evidence for feedback from higher-level cortex for oddball responses, reduced oddball responses seem to remain after blocking this feedback. Temporal sequence prediction could, in principle, be carried out by recurrent neural circuits within V1, but it is likely that feedback from higher cortex could enhance or extend these predictions.

Importantly, if the outlined paradigms show the same essential distribution of feature-based mismatch responses across areas and layers, then this would argue against the hypothesis for distinct mechanisms.

***Analysis #1:*** Map the locations of neurons showing significant mismatch responses using two-photon imaging and neuropixels recordings.

- For spatial analyses, we will focus on the firing rate (using a deconvolution approach for $Ca^{++}$ imaging) averaged over all timepoints (e.g., 0 to 275 ms) for each trial. For each cohort, we will map the density of mismatch neurons as a function of region, layer, and cell-type. We will compare the percentage of mismatch responses (over all responsive neurons; each mouse as one observation) using a mixed ANOVA with paradigm (paradigm 1, 2, or 3) as a between subjects variable and region and layer as within subjects variables. Sex and mouse age will be covariates. We will carry out a separate analysis for each method (two-photon vs neuropixel) and cell-type (two-photon imaging of PYRs and interneurons subtypes).
- Using PSTHs, compute the variability (standard deviation) of spike times relative to stimulus onset, as well as peak latency; compare to different models and across experimental conditions.
- Use dimensionality reduction techniques (principal components analysis (PCA), t-distributed stochastic neighbor embedding (t-SNE), UMAP, *etc.*) to visualize population activity across units and identify functional clusters.
- Characterize how different coding subspaces are oriented relative to each other in neural state space by computing the joint angles (Rule et al., 2020).
- Another approach would be to examine how much the coding direction of one variable aligns with the direction of another variable.

***Analysis #2:*** Compare responses for the *same* neurons between sensorimotor (session 2) and temporal sequence (session 3) mismatches.

- Is the mismatch response stronger for sensorimotor than temporal sequence prediction?
    - YES suggests different neural circuits for these two kinds of prediction
    - NO suggests common circuitry may explain data
- Make a scatter plot of mismatch response in sensorimotor vs. temporal sequence prediction
    - data scattering all over the plane suggests different neural circuits for these two kinds of prediction
    - data falling near a line suggests that additional circuitry for sensorimotor prediction “feeds into” common circuits
- Are there more examples of ‘pure mismatch responses’ (i.e. no baseline activity) in sensorimotor prediction vs. others
    - YES suggests different neural circuits for these different kinds of prediction

**Analysis #3:** Compare responses for the *same* neurons between the oddball (session 1) and sequence (session 3) mismatches.

- For comparing magnitudes of mismatch responses, the average firing rate for each neuron showing a significant mismatch response will be averaged over trials, and then layers and regions. We will compare these values using a mixed ANOVA with paradigm (session 1, 2, 3 or 4) as a between-subjects variable and region and layer as within-subjects variables.
- Is the mismatch response stronger for repetition than temporal sequence prediction?
    - YES suggests different neural circuits for these two kinds of prediction

- NO suggests common circuitry may explain data
- Make a scatter plot of mismatch response in oddball vs. temporal sequence prediction
- data scattering all over the plane suggests different neural circuits for these two kinds of prediction
- data falling near a line suggests that additional circuitry for oddball prediction “feeds into” common circuits

***Analysis #4:*** Analysis of recording from inhibitory interneurons.

- Are inhibitory neurons more strongly activated in session 2?
    - YES suggests that there is feedback from higher cortical areas
- Is inhibitory activity stronger in closed loop vs. open loop (session 2)?
    - YES inhibitory activity may reflect a sensory prediction
- Similar analyses for sessions 1 and 3

***Analysis #5:*** Temporal Mismatch Analysis (session 4).

- Test whether baseline activity and/or visual evoked responses under control conditions are different than for temporally deviant visual stimuli
    - YES indicates neurons encode specific temporal predictions about the time of occurrence of stimuli
- Assess how distinct classes of interneurons contribute to predictive timing by examining their responses to temporally based mismatches when the stimulus duration deviates from the control condition

**Analysis #6:** Test various prediction models across session types.

- Quantify learning effects as a function of region and layer. Measure the response amplitude before and after repeated presentations of the same stimulus within a recording session.
- Analyze changes in neural responses within a recording session (e.g., occurring over periods of seconds to minutes) to detect patterns likely to reflect short-term memory processes. Compute autocorrelations and cross-correlations across spike trains.
- Train deep learning models using self-supervised learning (e.g., to predict future activity from past activity) to extract latent feature representations of the neural data. Analyze the accuracy of stimulus decoders trained on the representations extracted from different areas and using different temporal windows.
- Analyze changes in neural activity patterns across learning days to detect patterns likely to reflect longer-term experience-dependent plasticity processes.
- Use information theory criteria and cross-validation techniques to compare the goodness-of-fit of different models. Validate models using separate test datasets, including ones obtained from different laboratories.

# Methods

This perspective was developed through an innovative and open collaborative process, engaging a global network of over 50 scientists.

## Collaborative Writing Process

The drafting process began with the creation of a shared Google Document which was seeded with an initial outline. A publicly accessible link to this document was then disseminated via social media and direct communications, along with detailed contribution guidelines to encourage broad participation. To lay a solid foundation, two supplementary documents were started, summarizing key experimental and computational publications relevant to predictive processing. These high-level summaries provided a structured knowledge base for starting the review. Participants were invited to contribute text, comments, and references following a set of general guidelines designed to ensure respectful engagement and constructive discourse. The document was open with full editing access, with no restrictions, throughout the entire process of writing the review. A weekly Zoom meeting was scheduled every Monday at 9 AM PST (12 PM EST, 6 PM CEST) to facilitate real-time discussions. These meetings occurred over a period of 10 months. Additionally, a Slack channel was created for asynchronous communication, enabling contributors to exchange ideas, address specific sections of the manuscript, and organize discussions around emerging themes.

The majority of discussions occurred through Google Doc comments. Over the course of the collaboration, approximately 1,900 comments were created in the document. Due to the limit imposed by Google Docs on both resolved and open comments, the document had to be migrated through three consecutive versions to accommodate ongoing discussions. Comments were systematically reviewed and resolved once consensus was reached. In some cases, primary authors of cited publications were invited to review and confirm the accuracy of specific text sections, ensuring fidelity to the original research findings.

## Citation Management

To streamline reference management, the Paperpile extension was used within Google Docs, allowing contributors to insert citations efficiently. Participants who could not use Paperpile were instructed to include references in a standardized format for later integration.

## In-person Workshop

An in-person workshop was held at MIT on August 8, 2024. The workshop, titled *Attending to Errors in Predictive Coding: A Collaborative Community Experiment through the OpenScope Program*, gathered experimentalists and theorists to discuss two competing hypotheses on predictive coding mechanisms: a cellular hypothesis and a dendritic hypothesis. The workshop was also used to encourage participation in the review, which was in its early stages.

The workshop was structured into three sessions:

- Session 1: Presentations on the broader context of predictive coding and specific theoretical predictions.
- Session 2: Presentations from experimentalists on key data from their labs and discussions of how their findings relate to theoretical models.
- Session 3: Presentations of concrete experimental proposals, followed by an open discussion.

## Authorship and Attribution

A transparent, opt-in authorship model was implemented. Contributors at any scientific career level who provided substantive input—either through direct text additions or thoughtful commentary—were invited to request their name be added to the author list. All authorship requests were approved by a panel of four scientists who initiated the GAC workshop. Approvals were done continuously throughout the process. Then, at the end of the review process, anyone who had requested authorship but whose substantive contributions were not immediately apparent to the panel were directly asked to support their authorship request by briefly describing or pointing to their contributions. In most cases, this led to acceptance of the suggested authorship, and in a few cases it led to voluntary withdrawal. Notably, while the majority of contributors to our shared document opted for authorship, some did not. Finally, authors are listed in alphabetical order so as not to over-emphasize the contribution of a given contributor. Overall, we believe this approach allowed for a balance between inclusion and fairness. A summary table of contributions from each author is included in **Supplementary Figure 1**.

## Final Experiment Selection and Polling Process

A consensus-driven process was established to finalize experimental proposals. An online voting round was held and Google Forms was used to register these votes anonymously. The poll was designed to finalize the experimental section of the perspective and plan future efforts. It included questions on:

- Preferred journal submission options (first and second choices)
- Selection of primary and secondary experimental proposals for OpenScope data acquisition
- Laboratory interest in analyzing or collecting data for selected projects.

# Discussion

Predictive processing is a broad theoretical framework that unifies a wide range of computational models, theoretical refinements, and empirical findings under the core idea that the brain continuously generates and updates statistical expectations about sensory input.

In our review, we outlined both convergences and divergences between experimental results and modeling work. We highlighted that a major challenge in interpreting experimental data is the diversity of experimental designs, as each laboratory has developed only partially overlapping tasks. As a result, despite decades of research, significant conflicts remain—for example when it comes to the role of specific cortical layers or specific inhibitory neurons. We hypothesized that these divergences may be context dependent, stemming from differences in predictive contexts, and thus designed an experiment to resolve some of the most pressing and readily resolvable conflicts and open questions. Our analysis plan proposes to test many specific hypotheses regarding the nature of possible predictive computations, and offers standards by which these hypotheses can be empirically refuted. Our future public dataset will also provide an opportunity for all interested researchers to investigate whether different phenomena studied in predictive processing rely on overlapping or separate neural circuit components.

To conclude, we felt it would be useful to clarify at which level our work lies. As we stated in the introduction, our focus is on Marr level 3, and thus on understanding how a potential predictive processing algorithm would be implemented at the level of neuronal circuits. This point is important as we do not intend to interrogate the broader set of hypotheses underlying predictive processing. Rather we aim to test specific computational and network primitives that have been proposed over the last decades (see **Figure 8**). These primitives, concretely instantiated in models of predictive processing are directly testable.

The falsifiability (sometimes equated with “testability”) of predictive processing has been repeatedly questioned (Kogo and Trengove, 2015; Cao, 2020), with some suggesting that lack of falsifiability marks a theory as pseudoscientific (Popper, 2002). Below, we detail how our approach and *Experimental Proposal* do not suffer from this issue.

It can be beneficial to clearly delineate the key concepts that are used to describe the scientific interplay between logical reasoning and empirical observations and measurement: A *theory* is a structured set of abstract concepts refined through empirical testing (Kuhn, 1962; Lakatos, 1970; Popper, 2002). A *model* instantiates a theory by formalizing specific assumptions, constraints, and parameters to generate testable predictions (Suppes, 1960; Van Fraassen, 1980; Giere, 1990). A *hypothesis* is a concrete, often quantitative, prediction derived from a model (**Figure 12**).

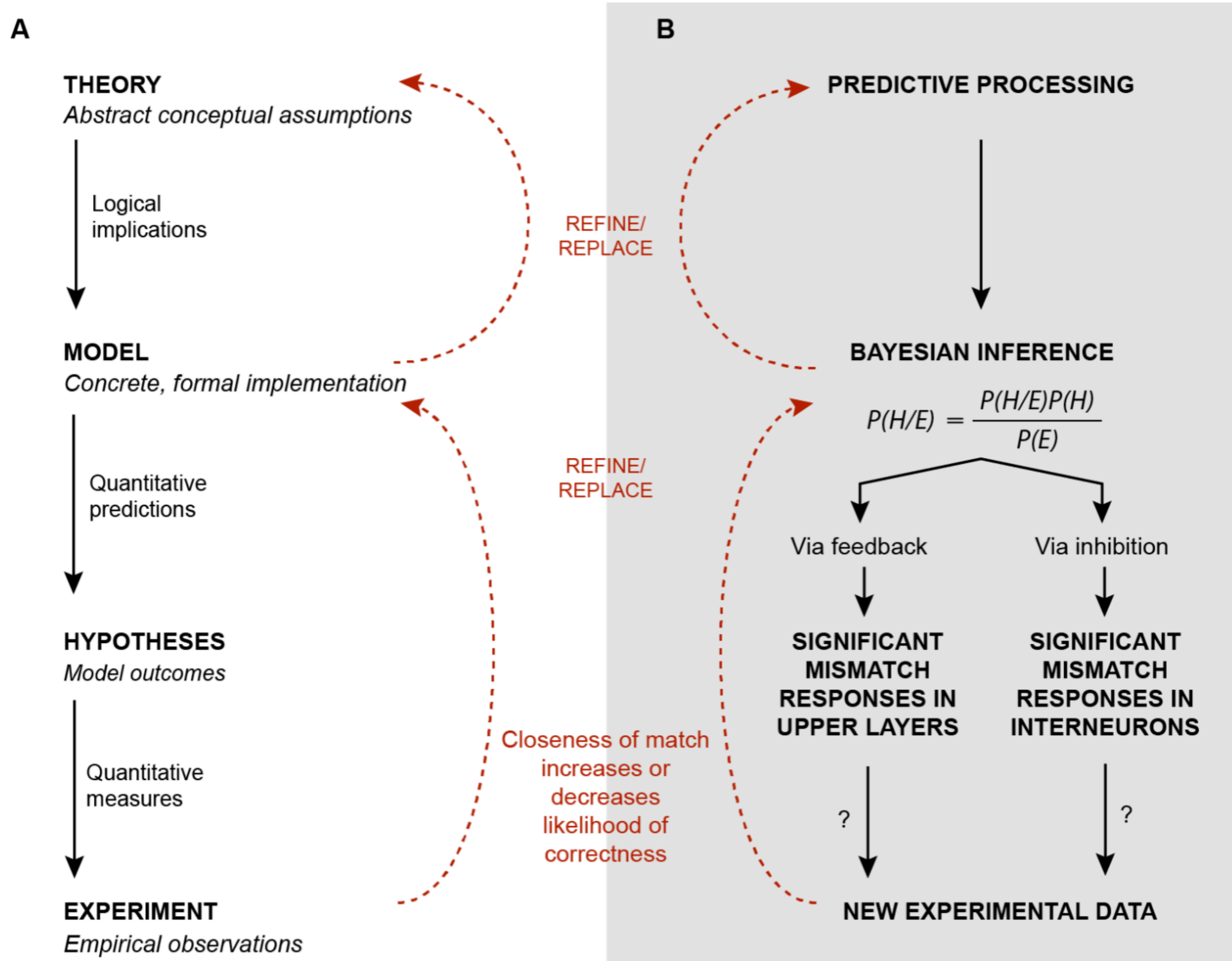

**Figure 12 - The scientific interplay between theory and data. A.** General schematic of the iterative cycle linking theory to experiment. A theory yields logical implications formalized in a model which generates quantitative predictions tested as hypotheses. Quantitative measures from the experiment are then compared to predictions: the closeness of match increases or decreases the likelihood that the theory is correct. Incompatibilities prompt refinement or replacement of the theory, model, or hypotheses (red dashed arrows), reflecting a broadly Bayesian approach to scientific progress rather than binary falsification. **B.** Application of this logic to the proposed experiments. The overarching theory of predictive processing is instantiated through Bayesian inference which generates testable hypotheses about specific circuit mechanisms: for example, that mismatch responses in upper cortical layers arise via feedback or via inhibition. New experimental data from the proposed recordings will update the likelihood of these hypotheses, iteratively refining or replacing the underlying models (red dashed arrows).

As we demonstrate below, the core assumptions of many, if not most, scientific theories are not *directly* testable, but their model-derived implications (predictions) are. Consequently, scientific theories are tested *indirectly* through model predictions and hypotheses(Platt, 1964). Instead of outright rejection (falsification), empirical testing adjusts the likelihood of a theory's correctness based on available data (Ziman, 1991; Jaynes, 2003). Incompatibilities between a theory's predictions and empirical observations generally prompt modifications. When a competing theory better accounts for the empirical evidence, replacement may be warranted. In scientific practice, replacement theories often preserve key explanatory elements of prior theories (Ladyman et al., 2009).

## Problems with Falsifiability as a Criterion

What does it mean for a theory to be "testable"? Is this synonymous with "falsifiable"? Popper's famous falsifiability criterion asserts that no theory can be fully proven, as new data could always refute it (Popper, 2002). However, this principle cuts both ways: falsified theories can later be revived by new evidence. For instance, science transitioned from geocentrism ("the Sun moves around the Earth") to heliocentrism ("the Earth moves around the Sun"), only for special relativity to render both statements valid under different reference frames (Einstein, 1988).

## Refutability - a Multifaceted Test Criterion

In response to these challenges, many philosophers of science have argued that a broader set of criteria—best summarized as *refutability*—provides a more appropriate framework for evaluating scientific theories than falsifiability (Gruenberger, 1964; Toulmin, 1972; Dutch, 1982; Radner and Radner, 1982; Laudan, 1983; Grove, 1985; Langmuir and Hall, 1989; Bunge, 1991; Vollmer, 1993; Fernandez-Beanato, 2020):

1. **Logical Coherence:** Testing for internal consistency is as important as empirical validation. A theory containing contradictions is inherently refuted, as contradiction nullifies logical reasoning. Empirical science has never observed a truly self-contradictory phenomenon (even Schrödinger's cat is mathematically and logically coherent).
2. **Testable Implications:** Many core theoretical assumptions in science are not directly testable but provide testable implications.
3. **Empirical Consistency:** The implications of theories must not contradict empirical observations. If a theory systematically fails to predict data, it must be modified or replaced.
4. **Probabilistic Knowledge:** Scientific insight is not absolute but evolves with data. Science assesses theories based on how well they fit current evidence rather than declaring them definitively true or false. Most progress refines theories rather than eliminating them entirely (Ladyman et al., 2009) (**Figure 12**)

A multi-criterion approach like this one better reflects the nature of scientific progress, which occurs through revision and refinement, rather than the pursuit of falsification. This approach also accounts for the role of quantitative measurements, mathematics, predictive power, coherence, and empirical success in scientific evaluation.

Thus, in accordance with a refutability framework that emphasizes logical coherence, testable implications, empirical consistency, and probabilistic knowledge, our experimental plan is designed to rigorously evaluate predictive processing at the circuit level. Rather than aiming for an outright falsification, our approach embodies a broadly Bayesian perspective where the likelihood of specific computational primitives is continuously updated with empirical evidence.

# Conclusion

In summary, we have reviewed in detail both experimental and theoretical research on predictive processing, organized into the following topics: I) the diversity of error and mismatch types, II) the distribution of error computation across the brain, III) the diversity of predictive neuronal responses, IV) the role of E/I balance and interneurons, V) the role of apical dendrites, VI) synaptic plasticity and learning rules, and VII) the link between single neuron activity and broader neural dynamics. Based on the convergences and divergences identified, we have proposed a detailed experiment designed to address core open questions in our understanding of the circuit underlying predictive processing in the brain. We have proposed a plan for analysing the resulting dataset, and also intend for it to serve as a valuable resource for the broader community to use.

Specifically, we aim to assess detailed, quantitative predictions—such as the differences between stimulus-specific versus general novelty signals predicted by additive, subtractive, or multiplicative models of mismatch responses—using a suite of techniques (e.g., Neuropixels recordings, two-photon imaging, decoding analyses, and latent space analyses). By comparing empirical measurements across various experimental conditions and integrating them with iterative modeling efforts, we will not only test the predictions derived from the predictive processing framework but also evaluate its internal consistency and empirical success relative to alternative theories. This multifaceted approach, grounded in rigorous statistical validation and hypothesis-driven analysis, reflects the progressive nature of scientific inquiry: theories are refined and strengthened through continuous, quantitative testing rather than being outright rejected based on single conflicting observations. We expect that the most likely outcome of this line of investigation will be that it will add detail and/or accuracy and rigor to existing theories of predictive processing, as has historically occurred in many other avenues of scientific investigation (**Figure 13**).

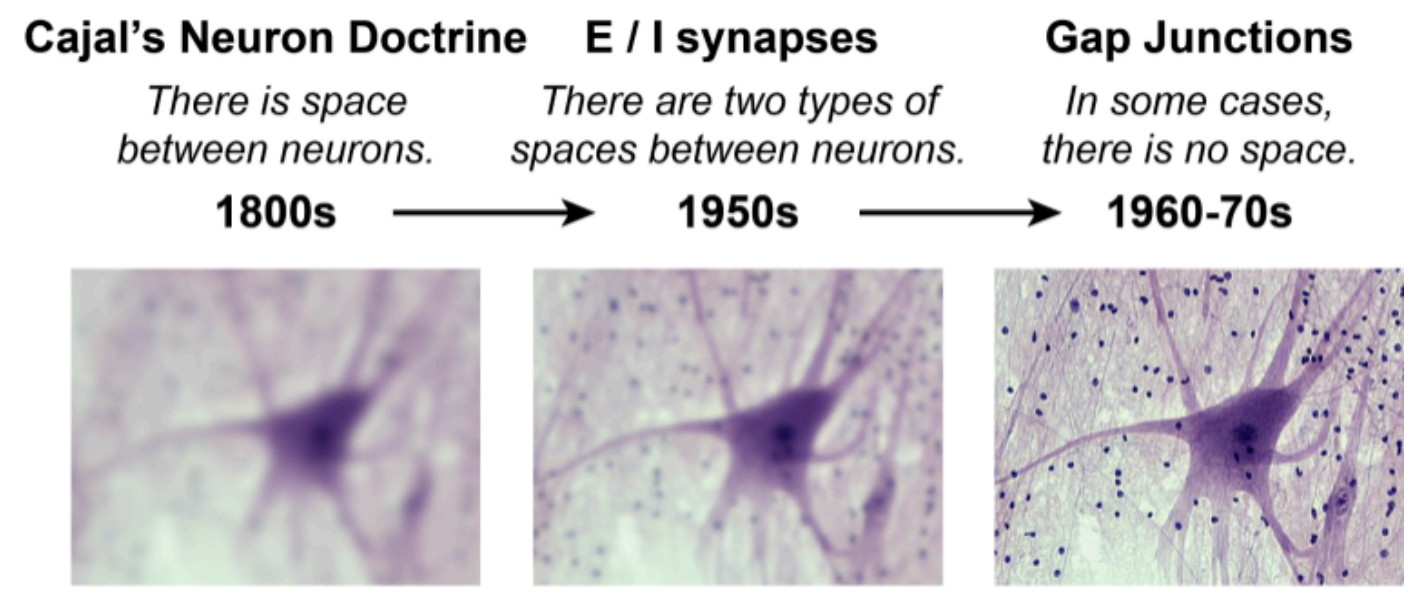

**Figure 13 - Example of scientific progress.** Science often progresses by increasing precision, while keeping the fundamental theoretical structure constant, akin to an image slowly coming into focus. Microscopy image of a spinal cord motor neuron by Fayette A. Reynolds, M.S., via Flickr (Public Domain/CC0).

# Supplementary Figure 1:
# Authorship contribution table

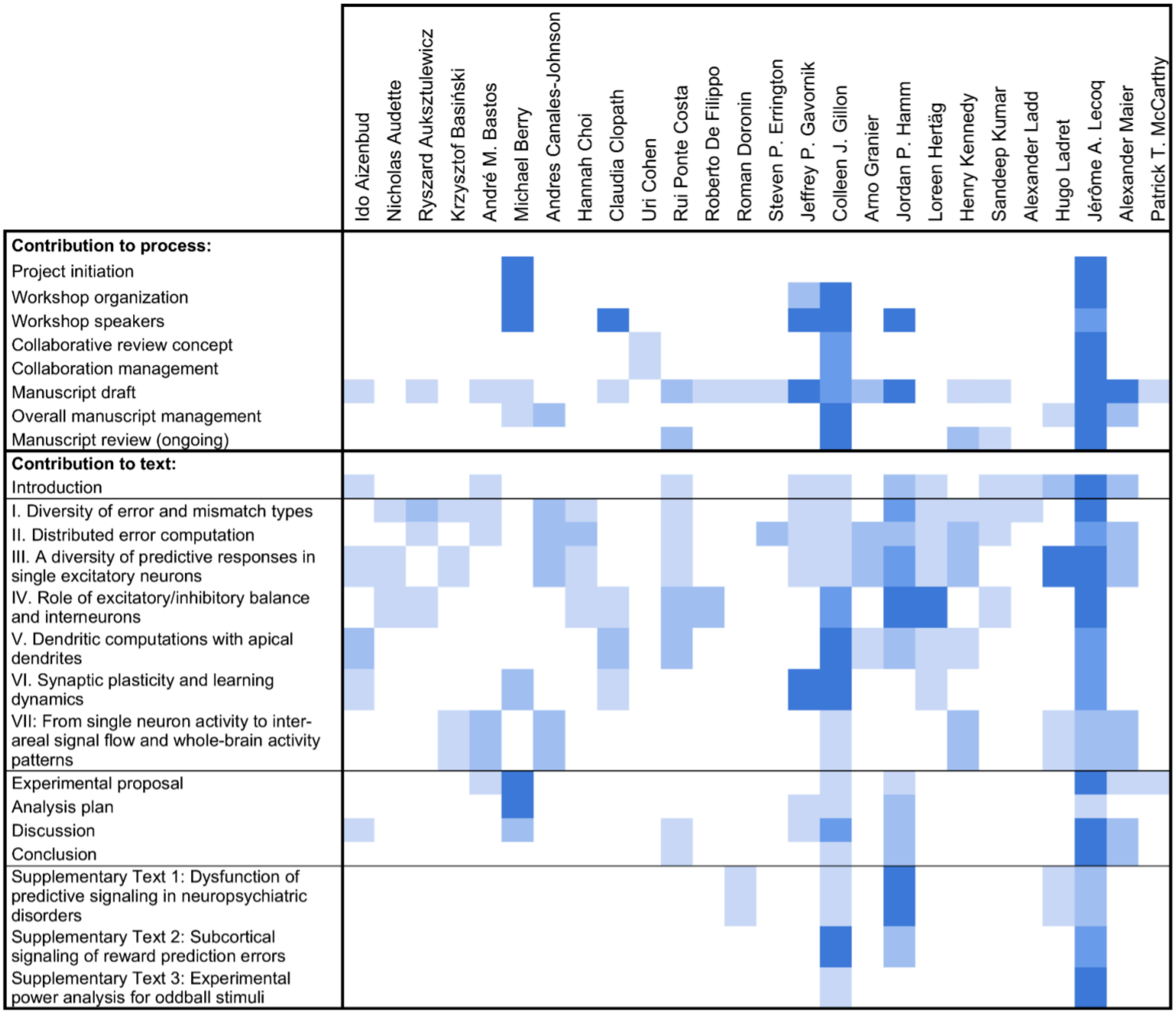

**Legend**

| |
|---|
| Did not contribute |
| For processes, made minor punctual contributions. For text, made small contributions/suggestions, such as contextualizing individual work or correcting mistakes. |
| For processes, contributed to tasks throughout but did not lead. For text, made intermediate contributions, such as linking multiple studies across paragraphs. |
| For processes, led the process to conclusion. For text, made major contributions throughout the task/section, including editing, shaping flow of ideas or improving coherence. |

**Supplementary Figure 1 - Authorship contribution table.**

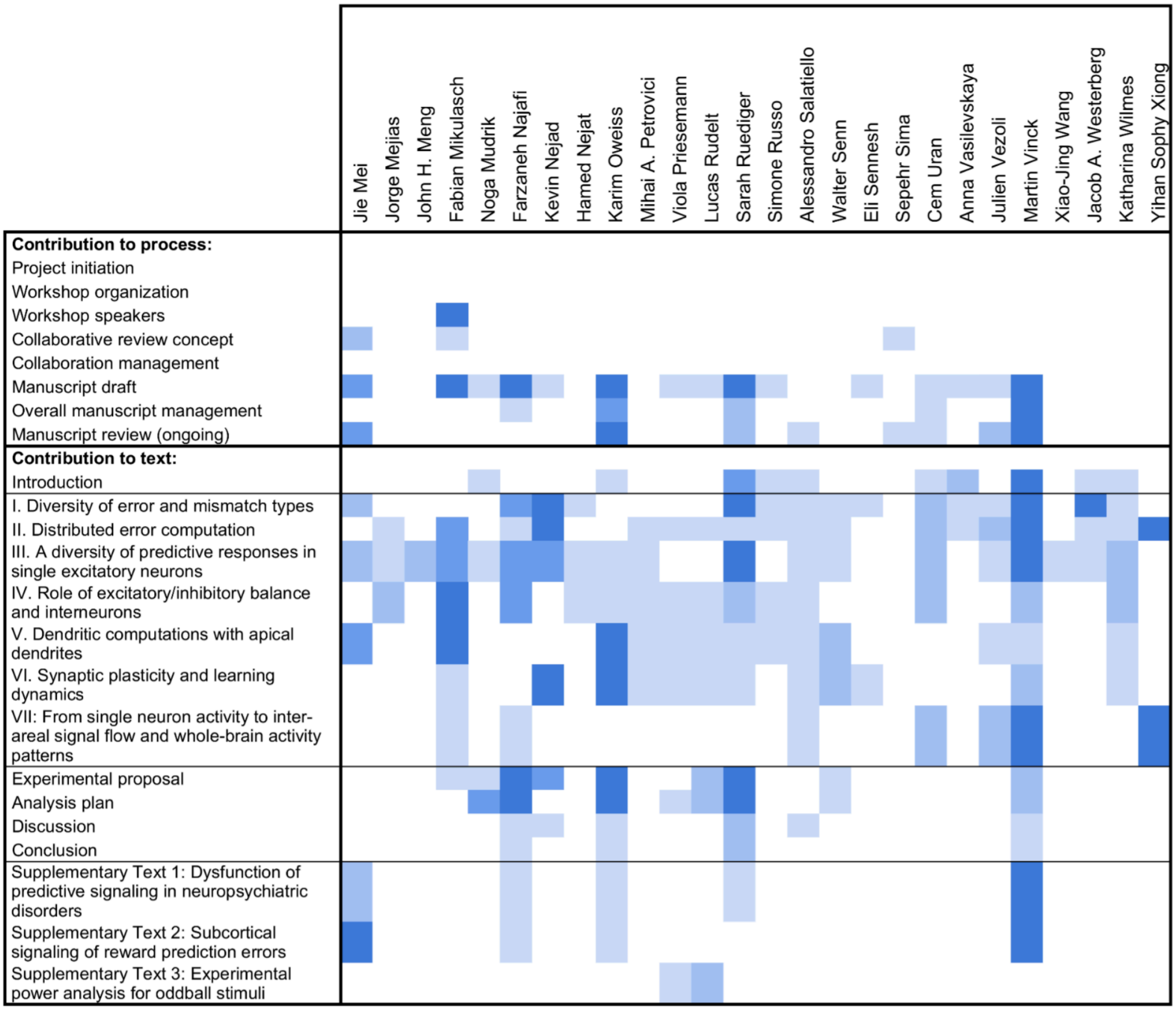

**Legend**

| |
|---|
| Did not contribute |
| For processes, made minor punctual contributions. For text, made small contributions/suggestions, such as contextualizing individual work or correcting mistakes. |
| For processes, contributed to tasks throughout but did not lead. For text, made intermediate contributions, such as linking multiple studies across paragraphs. |
| For processes, led the process to conclusion. For text, made major contributions throughout the task/section, including editing, shaping flow of ideas or improving coherence. |

**Supplementary Figure 1 - Authorship contribution table (continued).**

# Supplementary Text 1: Dysfunction of predictive signaling in neuropsychiatric disorders

Predictive coding is a powerful framework for studying and understanding neuropsychiatric disorders. It carries potential for explaining phenomenology and symptomology for psychotic disorders (Sterzer et al., 2018), autism spectrum disorders (ASD) (Sinha et al., 2014), major depressive disorder (MDD) (Kube et al., 2020), and others. Specifically, hallucinations, whether caused by underlying clinical conditions or external chemical compounds (like psychedelics), can be explained in a predictive coding framework by an over-weighting or under-weighting of priors (Carhart-Harris and Friston, 2019; Corlett et al., 2019; Weilnhammer et al., 2020). Carefully designed cognitive tasks aimed at assessing predictive processing suggest that in schizophrenia, for example, the encoding of priors may be more unstable or imprecise, and thus more susceptible to disruption by unexpected events (Adams et al., 2018).

Neurophysiological evidence also supports the notion of dysfunctional predictive processing in schizophrenia. For example, mismatch negativity, an EEG component thought to reflect a basic sensory prediction error (Friston, 2005), has been consistently shown to be reduced in individuals with schizophrenia (Erickson et al., 2016). Importantly, this reduction is observed in auditory and visual domains alike (Avissar et al., 2018; Mazer et al., 2024), often precedes the onset of psychotic symptomatology (Hamilton et al., 2022), and correlates well with global and cognitive function (Light and Braff, 2005), suggesting that it reflects deficits in information integration impacting both perception and cognition – each core aspects of this disorder.

Disruption in predictive processing has the potential to explain a number of neuroanatomical observations, possibly providing conceptual unity for schizophrenia pathophysiological theories. SOM interneurons are known to play a key role in predictive processing (e.g., (Bair et al., 2003; Hamm and Yuste, 2016; Keller et al., 2020b; Kirchberger et al., 2023; Furutachi et al., 2024; Ross and Hamm, 2024; Cuevas et al., 2025)) and a preponderance of evidence suggests that SOM neurons display the most dramatic molecular and cellular alterations in schizophrenia, across all studied cortical regions (Fung et al., 2010, 2014; Van Derveer et al., 2021; Batiuk et al., 2022). PV interneurons are also altered in the neocortex in schizophrenia, potentially due to reductions in excitatory inputs, particularly in L3 and L4 (Lewis et al., 2012; Dienel et al., 2023). Importantly, SOM and PV neurons are not reduced in number in schizophrenia, but display lower transcript levels suggesting hypoactivation (Dienel et al., 2023). Findings around VIP neurons are notably less consistent, suggest that this population is intact (Tsubomoto et al., 2019; Batiuk et al., 2022) or at least less altered than PV or SOM neurons (Fung et al., 2010; Arbabi et al., 2024).

Given the relatively greater involvement of PV and SOM neurons in cortical feed-forward pathways and recurrent circuitry, as opposed to VIP neurons, which feature prominently in mediating cortical feed-back modulation (Batista-Brito et al., 2018), a unifying interpretation is that signal propagation in feed-forward (Sweet et al., 2004; Schoonover et al., 2024) and recurrent circuits (Hamm et al., 2017) are more altered across neocortex in schizophrenia, as compared with feed-back or top-down modulation, which may be relatively spared. Therefore, schizophrenia may involve a shift in balance toward top-down predictive modulation, and away from bottom-up sensory processing (Javitt, 2009). Concurrent with unstable priors in higher brain regions (Rolls et al., 2008; Adams et al., 2018), this could give rise to illusory percepts and thought disorder. Consistent with this hypothesis, acute exposure to sub-anesthetic ketamine, which has been used as a model of schizophrenia pathophysiology for decades (Javitt, 1987; Javitt et al., 2012), increases top-down suppression of visual cortex activity in mice (Ranson et al., 2019). Such network-based interpretations are altogether concordant with Friston and Firth's hypothesis that the disorder is best understood as a network-level imbalance in fronto-sensory interactions (Friston and Frith, 1995).

Changes in interneuron transcript levels are only one piece of the puzzle for schizophrenia. The genetic landscape explaining schizophrenia risk is highly heterogeneous (Owen et al., 2023), though there is some convergence around excitatory synapses in cortical and hippocampal structures (Trubetskoy et al., 2022). Recent post-mortem work suggests that key excitatory synaptic transcripts may be most disrupted in early visual areas as compared to downstream frontal regions (Schoonover et al., 2024), consistent with the top-down dominant/bottom-up degraded hypothesis described above. However, this idea is less consistent with the pattern of gray matter loss seen in chronic schizophrenia, which tends to impact frontal and temporal regions most dramatically (but not exclusively (Gupta et al., 2015). In fact, some translational work further supports an opposite hypothesis according to which a *reduction* in feed-back modulation to sensory regions from higher brain areas explains the altered corollary discharge and auditory hallucinations seen in the disorder (Ford et al., 2001; Rummell et al., 2023). Both of these perspectives suggest an imbalance in top-down predictive modulation vs bottom-up sensory processing in the disorder, yet they suggest different underlying mechanisms. One potential explanation is that the cause of the imbalance may be changing from early stages (which involve degradations in sensory processing (Javitt, 2009; Javitt and Freedman, 2015) to later stages of the disorder (which show worsening gray matter loss in prefrontal regions) (Vita et al., 2012; Cropley et al., 2017). While the neuromodulatory system is less implicated in the most recent Genome-Wide Association Studies (GWAS) for schizophrenia (Trubetskoy et al., 2022), emerging treatments highlight a potential role for acetylcholine (Kaul et al., 2024), which may help to restabilize the balance of feed-forward vs feed-back circuits via its interactions with cortical interneuron systems (Batista-Brito et al., 2018).

In major depressive disorder (MDD), predictive coding-based theories suggest an explanation in which overly stable priors maintain negative beliefs and become insensitive to contrary (positive) prediction errors (Kube et al., 2020) or overly precise priors predict unreliable outcomes for an individual's actions, such that a patient comes to expect a lack of control (Clark et al., 2018). Behavioral studies (Kube et al., 2019) and neurophysiological evidence support this paradigm, pointing to reduced activation in reward prediction circuits (Pizzagalli et al., 2009; Kumar et al., 2018). Interestingly, psychedelic drugs have recently come into focus given their efficacy in treating MDD, among other neuropsychiatric disorders, with potency comparable to the standard antidepressant escitalopram (Carhart-Harris et al., 2021; Nutt and Carhart-Harris, 2021). A prominent hypothesis suggests that psilocybin, LSD, and other serotonergic psychedelics work to reduce the precision of high-level priors – or, in the case of MDD, negative belief states – that have become pathological, allowing for new information to be accommodated (Carhart-Harris and Friston, 2019). The potential clinical importance of psychedelic compounds is supported by studies in patient samples (Lyons and Carhart-Harris, 2018; Roseman et al., 2018; Ramos and Vicente, 2024; Timmermann et al., 2024), as well as rodent models (Fisher et al., 2024; West et al., 2024). Neurophysiological evidence comes from human neuroimaging studies, showing changes in default mode connectivity during and long after a dose of psychedelics (Carhart-Harris et al., 2016; Siegel et al., 2024). Further, changes in saccadic behavior and visual event-related potentials during an oddball paradigm also concord with the notion that psychedelics work to rebalance predictive processing circuitry (West et al., 2024).

In ASD, it is proposed that top-down predictive modulation of sensorimotor structures is altered (Sinha et al., 2014) due to imprecise or overprecise, but inaccurate, priors (Chao et al., 2024). This account could explain sensory sensitivities commonly seen in the disorder and in animal models, if incoming sensory data is insufficiently predicted or anticipated (Schmitz et al., 2003; Van de Cruys et al., 2014) and thus undergoes more extensive processing (Chao et al 2024). Furthermore, difficulty in interpreting complex social interactions, a characteristic feature of ASD, is well-explained in a predictive processing framework (Keysers et al., 2024). EEG studies show that people with ASD exhibit weaker pre-stimulus anticipatory “prediction potentials” in an oddball task (Grisoni et al., 2019) and diminished corollary discharge (van Laarhoven et al., 2019), also consistent with this hypothesis. In a basic visual oddball paradigm, FMR1-KO mice modelling genetic risk for ASD demonstrate impaired adaptation and enhanced, spatially unrestricted prediction error relative to WT mice; that is, while mismatch responses in WT were restricted to layer 2/3, Fmr1-KO mice exhibited mismatch responses across all layers (Pak et al., 2021).

Importantly, overarching models that posit a given disorder as simply resulting from too much top-down or too little bottom-up input are often met with contrary evidence (Arthur et al., 2023; Pesthy et al., 2023) and altogether fail to capture the complexity of the psychiatric disorder under study. In the case of psychosis, for example, certain symptom clusters (e.g., delusions) appear to reflect an imbalance in predictive processing in favor of predictions, while other symptom clusters (e.g., certain types of hallucinations) appear to reflect the opposite: an imbalance in favor of sensory evidence (Sterzer et al., 2018). Notably, these symptoms not only fall under the same diagnosis, they also manifest in the same individuals, sometimes even simultaneously. Measures like MMN, though reliably reduced in patient populations, are not diagnostic of any specific disorder. It is, for example, tempting to conceptualize schizophrenia, MDD, and ASD as distinct disorders of predictive processing, but reduced MMN has been identified in all three (Lassen et al., 2022). Similarly, the inability to suppress sensory cortical responses to self-generated sounds, a function of predictive processing known as “corollary discharge”, is found in both ASD and schizophrenia (Ford et al., 2001; van Laarhoven et al., 2019).

One possibility is that existing paradigms to measure and study predictive processing in humans (such as MMN or impaired corollary discharge to self-generated vocalizations (see **Section III.2**) are not sufficiently precise on their own to disentangle distinct dysfunctions of predictive circuitry. Alternatively, it’s important to note that these disorders show considerable heterogeneity not just in symptomatology, but in stable measures such as electrophysiology, brain structure, and genetics. Thus, distinct “biotypes” of schizophrenia, for example, may exhibit distinct alterations in predictive processing (Clementz et al., 2016). Going forward, it will be important to add nuance to these models, identifying distinct alterations in predictive processing (e.g., imprecise priors vs weak bottom-up drive) that may manifest across different levels of the cortical hierarchy within the same disorder or within the same individuals, accounting for delusions (at higher levels) and hallucinations (at lower levels) with regionally distinct biological mechanisms (Sterzer et al., 2018).

Insofar as these are essentially disorders of information processing, with significant cortical neuropathology, predictive processing presents itself as a useful framework for studying and conceptualizing psychiatric disorders and the efficacy of various treatments (Sterzer et al., 2018). Psychiatry as a whole stands to benefit from computationally grounded theories, as these provide a structured framework for gathering and interpreting data, and can also be directly tested against and updated in the face of challenging data (Huys et al., 2016).

# Supplementary Text 2: Subcortical signaling of reward prediction errors

In hierarchical predictive coding, a prediction error is defined as a signal that represents the mismatch between a teaching signal and a prediction. Here, predictions are the set of top-down inputs each neuron receives from higher levels in the hierarchy, rather than an organism-level set of beliefs about the environment. Some examples of prediction errors include sensory prediction errors, motor prediction errors, temporal prediction errors, and Reward Prediction Errors (RPEs) (den Ouden et al., 2012). In this box we first highlight RPEs, their role in Temporal Difference (TD) learning and their subcortical neural correlates. We then distinguish and relate RPEs to sensory prediction errors.

## The TD learning framework

In TD learning, a simulated agent seeks to learn a value function. This function is used to estimate the value of different *actions* an agent can take, or *states* it can enter with respect to a future target signal. This framework is frequently employed in tasks where it is helpful to forecast future rewards, because the reward signals are distant in time from the actions and states required to obtain them (Niv and Schoenbaum, 2008). TD learning considers events that unfold over time: when new information is received, like a reward signal, the model compares what it expected with what it received. Mathematically, the model measures the difference between the predicted value at time t and the updated value computed at time t+1 (hence the term "temporal difference"). This signal is then used to update the predicted value of the current state, and propagated backward in time to update the predicted value of actions and states leading up to the discrepancy, such that their value is also better estimated in the future. Notably, updates to actions and states are discounted as the model moves backward in time to reflect their weaker causal relationship to the state or action for which the discrepancy was observed.

## Reward Prediction Errors (RPE)

Although exact definitions vary, an RPE broadly corresponds to the difference between the discounted, predicted future reward at time *t,* and the actual reward obtained at time $t + 1$, combined with an estimate of the discounted, predicted future rewards, updated based on any new information received at time $t + 1$. Future rewards are typically discounted to take into account the fact that rewards appear, behaviourally, to be valued less highly the further away they are in time (Starkweather and Uchida, 2021). The estimate of the value function used to guide actions in TD learning is optimized by minimizing absolute RPEs (Ludvig et al., 2012; Cone et al., 2024). Thus, theoretically, when RPEs reach zero for every state, the value function remains stable and learning is complete.

Putative neuronal correlates of RPEs have typically been identified in subcortical regions. For example, dopamine neurons in the ventral tegmental area increase firing in the presence of a better-than-expected outcome. Conversely, they decrease their firing below baseline in the presence of a worse-than-expected outcome (Schultz, 1998). Importantly, dopamine neurons respond not only to actual rewards or RPEs, but also to stimuli that are predictive of rewards (Waelti et al., 2001; Fiorillo et al., 2003; Frémaux and Gerstner, 2015). The fact that dopamine neurons respond to both RPEs and stimuli predictive of reward has classically been explained via the temporal difference learning framework (Ludvig et al., 2008, 2012; Cone et al., 2024). In addition, dopamine neurons that signal motivational values are heterogeneous and demonstrate different functional characteristics. A study by Matsumoto and Hikosaka identified two populations of dopamine neurons in the substantia nigra pars compacta and ventral tegmental area (Matsumoto and Hikosaka, 2009). One population of neurons showed signed error responses, as it was excited by reward-predicting stimuli and inhibited by punishment-predicting stimuli. In contrast, the second population showed unsigned prediction errors, as it was excited by both. These two types of dopamine neurons were also located in different subregions, suggesting that motivational values are signaled by two functionally and anatomically distinct groups of neurons.

Compared to the types of signed and unsigned prediction errors discussed in section III which relate to the sensory features of the environment, signed prediction errors in the context of rewards indicate whether the outcome of an action is better or worse than expected. Thus, they can help an agent more accurately update its estimate about the value of things it has experienced, and encode more behaviorally relevant memory traces (Haarsma et al., 2021; Rouhani and Niv, 2021). For these reasons, and as described above, signed prediction errors are often used in the context of TD reinforcement learning (Schultz, 2016a; Hoy et al., 2023). In the brain, signed and unsigned prediction errors may be supported by different neuronal populations. Serotonin neurons in the dorsal raphe nucleus have been shown to support unsigned prediction errors and midbrain dopamine neurons to support signed prediction errors (Matias et al., 2017) Nevertheless, the relationship between serotonin neurons and unsigned prediction error signaling remains speculative, as studies employing different experimental techniques and task designs have not observed the involvement of serotonin neurons in unsigned prediction errors (Cohen et al., 2015; Grossman et al., 2022). The locus coeruleus-noradrenaline system may also help drive the signaling of RPE (Su and Cohen, 2022).

### Distinguishing between sensory and reward prediction errors

Sensory and reward prediction errors can be distinguished by:
(1) their definition: Sensory prediction errors result from a mismatch between a sensory teaching signal and a sensory prediction, whereas RPEs occur when there is a mismatch between the predicted and actual rewards.
(2) their impact on the organism: In the context of hierarchical predictive coding, sensory prediction errors are highly distributed, computed from a broad range of comparisons between and within levels of the sensory processing hierarchy. As a result, sensory prediction errors may not always have cognitive correlates. In other words, many locally computed sensory prediction errors may not enter conscious perception. RPEs, on the other hand, involve organism-level expectations about future rewards.
(3) their neuronal correlates: A multitude of subcortical areas, including the striatum, lateral habenula, hypothalamus and amygdala, are involved in the encoding of reward-related information (Hikosaka et al., 2008). Sensory prediction errors of various modalities, as extensively reviewed here, are heavily associated with cortical areas. These results suggest that cortical and subcortical areas play separate roles in sensory and reward prediction errors. This may in part reflect a self-reinforcing bias in the brain areas sensory versus reward prediction error studies choose to focus on. Indeed, recent studies suggest that midbrain dopamine neurons may also support sensory prediction errors (Takahashi et al., 2017; Stalnaker et al., 2019). These two systems must be heavily intertwined, as information from one is frequently relevant to the other. The states that an agent must navigate to receive rewards are often distinguishable based on sensory features. Thus, sensory processing is often required for reward prediction, and errors from each system are likely relevant to the other. Accordingly, subcortical and cortical areas have been shown to work together in signaling sensory cues relevant to rewards (Takakuwa et al., 2017; Baruchin et al., 2023). Links between the two systems are discussed in more detail below.

### Linking sensory and reward prediction errors

Although the noted differences between sensory and reward prediction errors suggest that they are processed differently, experimental studies have established links between the two. When a sensory mismatch occurs, new sensory information in the environment can help shape RPEs. Meanwhile, an RPE can shift attention to different aspects of an environment, leading to changes in sensory perception. This can, in turn, influence which sensory prediction errors will arise.

Subcortical neurons may be involved in both sensory processing and the representation of rewards. For example, subcortical superior colliculus neurons that are involved in earlier stages of sensory processing, receiving direct inputs from the retina are also modulated by rewards. Indeed, Baruchin et al. found that visually responsive neurons in the superficial layers of the superior colliculus showed increased and more readily decodable stimulus responses on trials that followed reward delivery compared to a negative reinforcement (Baruchin et al., 2023).

Neurons typically implicated in RPEs may also encode sensory information, allowing them to process a reward's physical features and thereby help focus and redirect attention toward it. The early response of dopamine neurons to conditioned stimuli can be divided into two components. The first is a sensory component, which is evoked by the physical salience of a stimulus and promotes the detection of rewards. The second is a reward value component, which is linked to the motivational salience of a stimulus (Schultz, 2016b). As discussed by Takakuwa et al., the lack of direct connections from the cortical visual processing stream to the ventral midbrain suggests that the sensory component involves a subcortical visual pathway via the midbrain superior colliculus (Takakuwa et al., 2017).

Studies have also provided evidence of sensory processing in neurons responsible for RPE signaling. For example, error-signaling dopamine neurons respond to changes in the value-neutral sensory properties of an expected reward (Takahashi et al., 2017). In a study by Gonzalez and colleagues, rapid environmental luminance changes evoked dopamine release in the nucleus accumbens. These dopamine signals encoded both the rate and magnitude of luminance changes, facilitating the monitoring of sensory transitions, but not their valence. The authors concluded that the observed dopaminergic responses to sensory stimuli may orient attention to potential reward sources (Kobayashi and Schultz, 2014; Gonzalez et al., 2023).

Taken together, although sensory and reward prediction errors have distinct characteristics, the two processes are likely highly intertwined. Subcortical structures, such as the superior colliculus and dopamine-releasing midbrain regions, may contribute to linking sensory inputs to reward processing. Neurons in these regions exhibit complex functions that extend beyond RPE signaling, e.g., processing physical features of rewards, detecting sensory property changes, and facilitating sensorimotor learning. Given the heterogeneity of subcortical neurons and their projections within and beyond subcortical regions, further research is needed to clarify their contributions to sensory and reward PEs. For this purpose, developing experimental approaches capable of disentangling reward-related features from sensory features will be essential for understanding their distinct roles.

# Supplementary Text 3: Experimental power analysis for oddball stimuli

Understanding the statistical principles underlying experimental design is critical for studying predictive processing in the brain. A key challenge is determining the number of trials required to reliably detect neuronal responses to oddball stimuli. Variability in neuronal responses, background noise, and trial-to-trial adaptation all influence the statistical power needed to detect meaningful activity.

| Publication | **Attinger et al 2017** | **Homann et al 2022** | **Bastos et al 2023** | **Knudstrup et al 2025** | **Westerberg et al 2025** |
|---|---|---|---|---|---|
| Type of stimulus | Drifting gratings coupled to movement | Superimposed Gabor patches in a sequence | Drifting gratings in a sequence | Static gratings in a sequence | Drifting gratings in a sequence |
| Type of oddball | Drifting gratings decoupled from movement | Final image in sequence replaced with novel image | Deviant grating in sequence of repeated gratings | Second grating in sequence replaced by deviant grating | Local: Final grating in sequence different from preceding ones<br>Global: Final grating in sequence different from established sequence |
| Temporal parameters | Oddball: 15 sec at random | Sequence: 4 images, 250 or 300 ms ON<br>Oddball: Every 6 seconds | Sequence: Repeated gratings, drifting at 2 cycles/sec., 500 ms ON, 1 sec OFF<br>Oddball: 12.5% of gratings | Sequence: Repeated gratings, 75 ms to 2 sec ON, 0 or 1.5 sec OFF<br>Oddball: Different frequencies and spacings tested | Sequence: 4 gratings, drifting at 4 cycles/sec., 500ms ON, 500 ms OFF, 4500 ms per sequence |
| Spatial parameters | Vertical, full screen gratings, 0.04 cycles/deg. | Each image: 100 superimposed Gabor patches, 10-20 deg. in size with random orientation and phase. | Full field gratings, 8 grating orientations, 0.08 cycles/deg. | 4 grating orientations, 0.5 cycles/deg. | Full screen gratings, 0.04 cycles/deg. |
| Species | Mouse | Mouse | Mouse | Mouse | Mouse and primate |
| Nb of subjects | 3-6 per group | 5 | 4-9 per group | 14 | 7-9 per group |
| Session duration | 3x 500s | 10min-1h | 6 min | 10 min | 2h |
| Nb of mismatches | 3x 33 | 100 per condition | 10 per condition | 450 standard, 50 deviant | 100 (global) |
| Recording technique | Two-photon | Two-photon | Two-photon | LFP | Neuropixels |
| % responsive neurons | 26-40% | 77% | 10% (PYR) | N/A | 50-62 % (local oddball), 3-9% (global oddball) |
| Test of significance | Mann-Whitney U test between average in response window vs. randomized window | Z-score significance test on change in response to novelty | Paired two-tailed t-test between control and oddball | Non-parametric test on a bootstrapped distribution | Cluster-based permutation test. against control |
| Habituation | 6 sessions, 2h/day | 0 sessions | 3 sessions | 0 sessions | 5 sessions |
| Nb of oddball repeats required | 33 | 100 | 10 | 50 | 144 |
| Oddball rate (0-1) | 0.07 | 0.1666666667 | 0.125 | 0.1 | 0.2 |
| Time per oddball repeated sequence (s) | 495 | 600 | 160 | 400 | 3600 |
| Total oddball time (s) | 495 | 600 | 1440 | 400 | 3600 |

**Supplementary Table 1. Summary table of parameters values extracted from experimental oddball studies.**

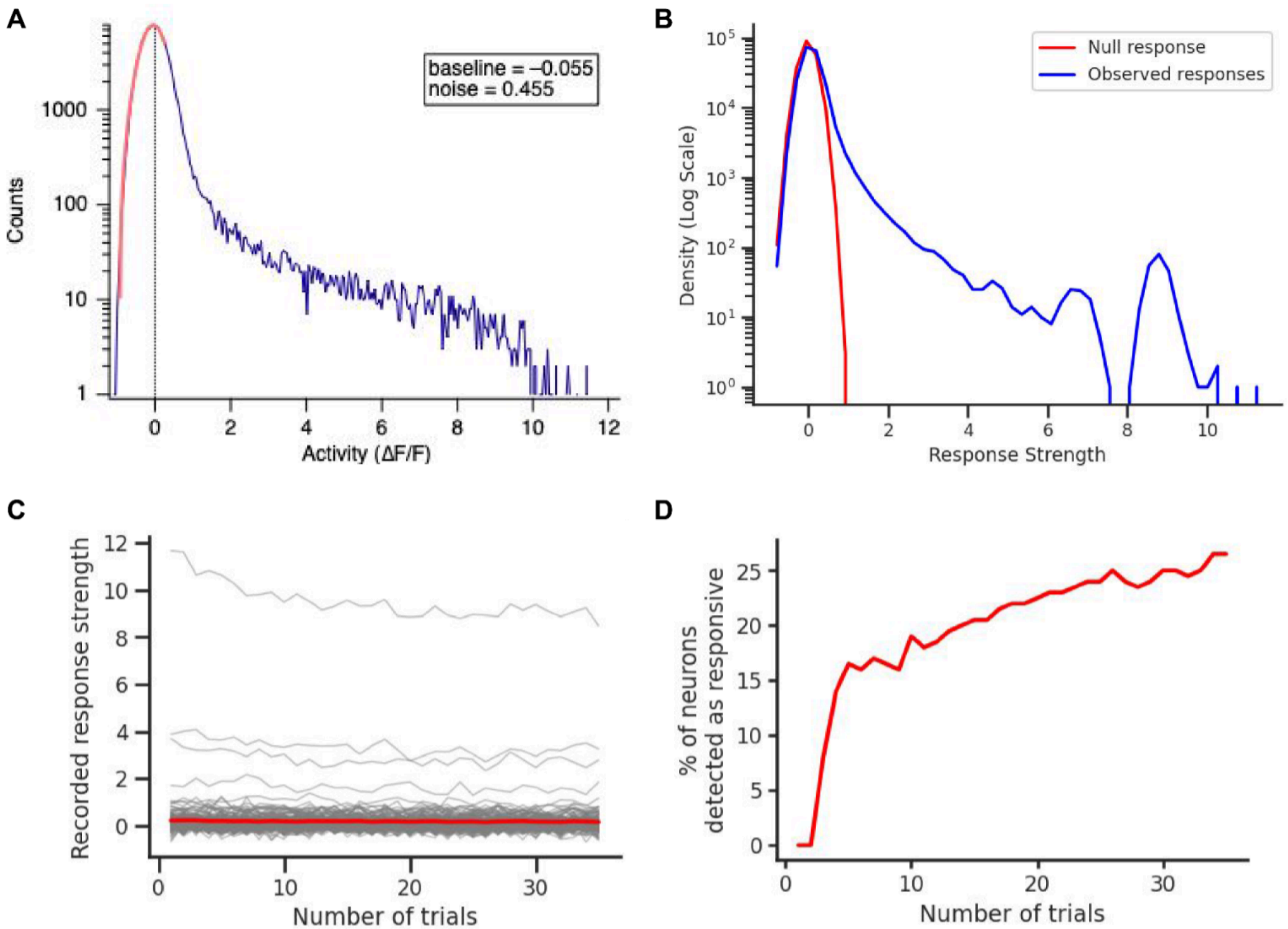

**Supplementary Figure 2 - Simulation of responsive neuron detection rate across trials. A.** Measured distribution of cellular response strength across neurons recorded with two-photon calcium imaging during an oddballs experiment. **B.** Simulated response distribution with parameters (see **Supplementary Table 1**) to qualitatively reproduce response profile recorded in **A**. **C.** Simulated response strength across repeated trials. **D.** Percentage of neurons detected as responsive using a power analysis on simulated data.

To address this challenge, we first summarize parameters used in a few oddball studies in mice (see **Supplementary Table 1**). This table highlights a few key convergences and divergences. The oddball rate across these studies ranged between 0.07 to 0.2. Given the session duration in these experiments, the number of oddballs ranged from 10 to 144 per oddball type. Importantly, studies with fewer oddball numbers operated at shorter timescales: Their oddballs could be predicted from more recent stimulus history and were learned more quickly by mice. Studies with larger numbers of repeats typically involved sequential oddballs with longer temporal history. Across these experiments, recording oddball responses depending on more complex long-term sensory relationships required more repeats.

Based on these observations, we developed a simulation framework to calculate the statistical requirements for optimizing trial numbers in oddball experiments. Neuronal responses can be recorded using methods such as calcium imaging or electrophysiology. In both cases, a measure of cellular responsiveness is compared to background noise fluctuations. These simulations had biologically relevant features, including trial-dependent adaptation and noise, providing a robust approach for estimating the necessary trial counts to achieve reliable detection of neuronal responses. The underlying code can be accessed here: https://colab.research.google.com/drive/1Lnd4kP7pgH9tMW6fySOfCHgsgsrSpKqa?usp=sharing

We simulated a distributed population of neurons with stimulus-evoked responses that followed a log-normal distribution. This distribution captures the presence of a few units with large, easily detected responses and many more units with weaker responses. We qualitatively compared our simulated distribution (see **Supplementary Figure 2B**) with the measured distribution from two-photon imaging experiments (see **Supplementary Figure 2A**).

Using this distribution, we modeled neuronal response decay across repeated trials within the population (gray), reflecting physiological adaptation and reduced sensitivity to repeated oddball stimuli. We then quantified the percentage of neurons detected as responsive as a function of trial number. Detection was assessed using a simple statistical t-test against the null distribution. The resulting curve revealed diminishing returns in detection rates beyond a certain number of trials, highlighting the importance of balancing data collection efforts with statistical power constraints (see **Supplementary Figure 2C-D**).

Next, we examined how the number of recording sessions (or mice) influences detection power. Aggregating data across multiple sessions increased the measure of the percentage of detected neurons (**Supplementary Figure 3**), demonstrating a trade-off between the number of trials per session and the overall sample size. Together, these simulations provide a principled approach for determining the optimal number of trials and sessions needed to reliably detect oddball responses in cortical neurons. Different oddball types might recruit different proportions of neurons with variable effect size. Each oddball type can then be simulated separately by varying input parameters into the model.

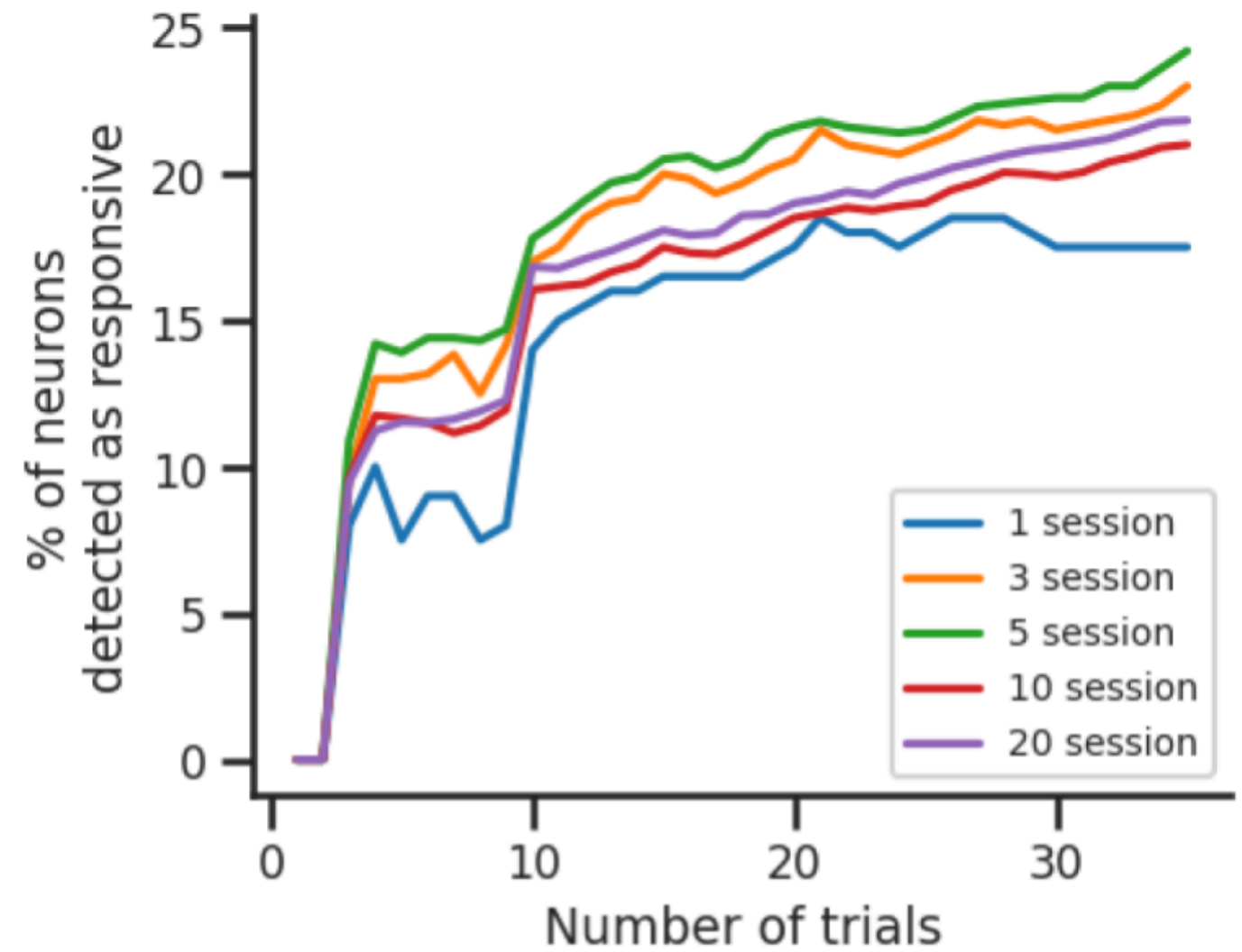

**Supplementary Figure 3 - Simulation of responsive neuron detection rate across sessions.** Percentage of neurons detected as responsive after aggregating data from multiple simulated sessions.